\documentclass[structabstract]{aa}
\usepackage{txfonts}
\usepackage{graphicx}
\usepackage{natbib}
\usepackage{longtable}
\usepackage{subeqnarray}
\usepackage{cases}
\usepackage{ulem}

\usepackage[colorlinks=true,citecolor=blue]{hyperref}

\begin{document}

\title{Ammonia observations towards the Aquila Rift cloud complex}
\author{Kadirya Tursun \inst{1,2}
\and Jarken Esimbek \inst{1,3}
\and Christian Henkel \inst{4,5,1}
\and Xindi Tang \inst{1,3,4}
\and Gang Wu \inst{1,3}
\and Dalei Li \inst{1,3}
\and Jianjun Zhou \inst{1,3}
\and Yuxin He \inst{1,3}
\and Toktarkhan Komesh \inst{1,2,6}
\and Serikbek Sailanbek \inst{1,2,6}
}

\institute{
Xinjiang Astronomical Observatory, Chinese Academy of Sciences, 830011 Urumqi, P. R. China \\
e-mail: kadirya@xao.ac.cn, jarken@xao.ac.cn, tangxindi@xao.ac.cn
\and
University of Chinese Academy of Sciences, 100080 Beijing, P. R. China
\and
Key Laboratory of Radio Astronomy, Chinese Academy of Sciences, 830011 Urumqi, P. R. China
\and
Max-Planck-Institut f\"ur Radioastronomie, Auf dem H\"ugel 69, 53121 Bonn, Germany \\
e-mail: chenkel@mpifr-bonn.mpg.de
\and
Astronomy Department, King Abdulaziz University, P.O. Box 80203, 21589 Jeddah, Saudi Arabia
\and
Department of Solid State Physics and Nonlinear Physics, Faculty of
Physics and Technology, AL-Farabi Kazakh National University, 050040 Almaty, Kazakhstan}

\abstract
{We surveyed the Aquila Rift complex including the Serpens South and W\,40 region in the NH$_3$\,(1,1)
and (2,2) transitions making use of the Nanshan 26-m telescope. Our observations cover an area of
$\sim$$1.5\degr\times2.2\degr$\,(11.4\,pc\,$\times$\,16.7\,pc). The kinetic temperatures of the dense
gas in the Aquila Rift complex obtained from NH$_3$\,(2,2)/(1,1) ratios range from 8.9 to 35.0\,K with
an average of 15.3\,$\pm$\,6.1\,K (errors are standard deviations of the mean). Low gas temperatures
associate with Serpens South ranging from 8.9 to 16.8\,K with an average 12.3\,$\pm$\,1.7\,K, while
dense gas in the W\,40 region shows higher temperatures ranging from 17.7 to 35.0\,K with an average
of 25.1\,$\pm$\,4.9\,K. A comparison of kinetic temperatures derived from para-NH$_3$\,(2,2)/(1,1)
against HiGal dust temperatures indicates that the gas and dust temperatures are in agreement in the
low mass star formation region of Serpens South. In the high mass star formation region W\,40, the
measured gas kinetic temperatures are higher than those of the dust. The turbulent component of the
velocity dispersion of NH$_3$\,(1,1) is found to be positively correlated with the gas kinetic temperature,
which indicates that the dense gas may be heated by dissipation of turbulent energy. For the fractional
total-NH$_3$\,(para+ortho) abundance obtained by a comparison with $Herschel$ infrared continuum data
representing dust emission we find values from 0.1\,$\times$\,$10^{-8}$ to 2.1$\,\times$\,$10^{-7}$ with an average
of 6.9\,($\pm$4.5)\,$\times$\,10$^{-8}$. Serpens South also shows a fractional total-NH$_3$\,(para+ortho) abundance
ranging from 0.2\,$\times$\,$10^{-8}$ to 2.1\,$\times$\,$10^{-7}$ with an average of 8.6\,($\pm$3.8)\,$\times$\,10$^{-8}$.
In W\,40, values are lower, between 0.1 and 4.3\,$\times$\,$10^{-8}$ with an average of 1.6\,($\pm$1.4)$\,\times$\,10$^{-8}$.
Weak velocity gradients demonstrate that the rotational energy is a negligible fraction of the gravitational energy.
In W\,40, gas and dust temperatures are not strongly dependent on the projected distance to the recently formed massive
stars. Overall, the morphology of the mapped region is ring-like, with strong emission at lower and weak emission at
higher Galactic longitudes. However, the presence of a physical connection between both parts remains questionable.}

\maketitle
\section{Introduction}
\label{sect:Introduction}
Systematic studies of dense molecular cores in regions of high mass star formation (HMSF) are
of great importance for our understanding of their physical and chemical properties.
In comparison with low mass star-forming regions, only a few, rather arbitrarily selected cores
associated with HMSF regions have been investigated in some detail \citep{2007ARA&A..45..481Z,2014prpl.conf..149T}.
There is a clear observational dichotomy between low mass star formation (LMSF) and high
mass star formation (HMSF). It is suggested that more than  70\% of HMSF stars are formed
in dense clusters embedded within giant molecular clouds (GMCs) \citep{2003ARA&A..41...57L}.
For a better understanding of massive star formation, it is important to obtain the physical
conditions of HMSF regions. High mass stars form almost exclusively in GMCs while low mass stars can form
in dark clouds as well as in GMCs. High mass stars are predominantly formed in clusters,
while low mass stars may form in smaller molecular complexes as for example in the Taurus
molecular cloud or in isolation. The star formation efficiency is generally higher in HMSF
regions (e.g., \citealt{1986ApJ...301..398M,2003ARA&A..41...57L}). In HMSF regions,
densities and temperatures tend to be higher, and spectral line widths are larger indicating
a higher degree of turbulence. The causal relationship between the presence of young high
mass stars, molecular cloud characteristics, and details of the star formation process,
however, is not well established.

The star formation region in the Aquila Rift is located at a distance of $\sim$436\,$\pm$\,9.2\,pc
(e.g., \citealt{2017ApJ...834..143O,2018ApJ...869L..33O}) in areas extending up to Galactic latitudes
of at least $+10\degr$ and down to $-20\degr$ (e.g., \citealt{2005PASJ...57S...1D}). The Aquila Rift
contains several active star-forming regions: Serpens Main, Serpens South, Serpens MWC297, and W\,40.
Here, we focus on that part of the Aquila Rift complex that harbors two known sites of star formation:
Serpens South and W\,40. Serpens South is a well-known site of star formation, located in the western
part of the Aquila Rift cloud complex, which forms a young embedded stellar cluster \citep{2008ApJ...673L.151G}.
It has a filamentary structure on the cusp of a burst of low-mass star formation. W\,40, located further
to the east in equatorial coordinates, is a site of ongoing high-mass star formation. It still contains
dense molecular cores \citep{2005PASJ...57S...1D}, and includes a blistered H\,{\scriptsize II} region,
powered by a compact OB association that contains pre--main--sequence stars (e.g., \citealt{1978ApJ...222..896Z,
1985ApJ...291..571S,1987A&A...178..237V,2010ApJ...725.2485K,2010AJ....140..968R,2013ApJ...779..113M}).

Aquila is a unique region for studying the physical and chemical conditions of molecular clouds.
A large number of molecular line observations have been performed, such as in CO
\citep{2017ApJ...837..154N,2019ApJS..240....9S,2020ApJ...893...91S},
in NH$_3$ \citep{2013A&A...553A..58L,2014A&A...567A..78L,2016ApJ...833..204F} and in
H$_{2}$CO \citep{2019ApJ...874..172K}. This was complemented by SCUBA-2 450 and 850\,$\mu$m observations
\citep{2016MNRAS.460.4150R}. A few years ago, the whole Aquila complex was extensively studied
by the $Herschel$ Gould Belt Survey\footnote{http://www.herschel.fr/cea/gouldbelt/en/index.php},
yielding more than 500 detections of starless cores imaged in dust emission at 70-500\,$\mu$m
\citep{2010A&A...518L.106K}. However, a systematic spectral survey of the dense gas in the
region, covering and comparing the emission of both Serpens South and the W\,40 complex is
still missing. A basic result of the NH$_3$ studies of \citet{2013A&A...553A..58L,2014A&A...567A..78L},
mostly carried out in a position switching mode, was that they were often finding so far unknown
clouds at the off positions, providing the urgent need for a systematic survey of the entire region.
\citet{2016ApJ...833..204F} presented NH$_3$ measurements of Serpens South ($\sim$4\,pc\,$\times$\,4\,pc)
with the Green Bank Telescope (GBT). Here we present a complementary survey with lower angular resolution,
but twelve times larger area encompassing many Aquila rift clouds with substantial visual extinction.

Ammonia (NH$_3$) is frequently used as the standard molecular cloud thermometer
(e.g., \citealt{1983ARA&A..21..239H,1983A&A...122..164W,1988MNRAS.235..229D}). It
starts to form at an early stage of prestellar core evolution and becomes brightest
during the later stages (e.g., \citealt{1992ApJ...392..551S}). NH$_3$\,(1,1) and (2,2),
both belonging to the para-species of ammonia, have been proved to be an excellent
thermometer at $T_{\rm kin}$\,$<$\,40\,K \citep{1983A&A...122..164W}. They can also serve as a good thermometer for
higher temperatures after some modification of the rotational temperature but with a
reduction in precision \citep{1983A&A...122..164W,1988MNRAS.235..229D,2004A&A...416..191T}.
Moreover, the critical densities of NH$_3$\,(1,1) and (2,2) are about $10^{3}$\,cm$^{-3}$
\citep{1999ARA&A..37..311E,2015PASP..127..299S}, thus providing a proper tracer for dense regions.

In this paper we intend to provide first ammonia maps covering the entire region of the
Aquila Rift cloud complex and to reveal a complete distribution of the dense gas.
The article is organized as follows: In Sect. \ref{sect:Observation} we introduce
our observations and data reduction. Results are highlighted in Sect. \ref{sect:Results}.
We discuss the variation of NH$_3$ abundance and gas temperature in Sect. \ref{sect:discussion}.
Our main conclusions are summarized in Sect. \ref{sect:summary}.

\begin{figure}[t]
\vspace*{0.2mm}
\begin{center}
\includegraphics[width=0.5\textwidth]{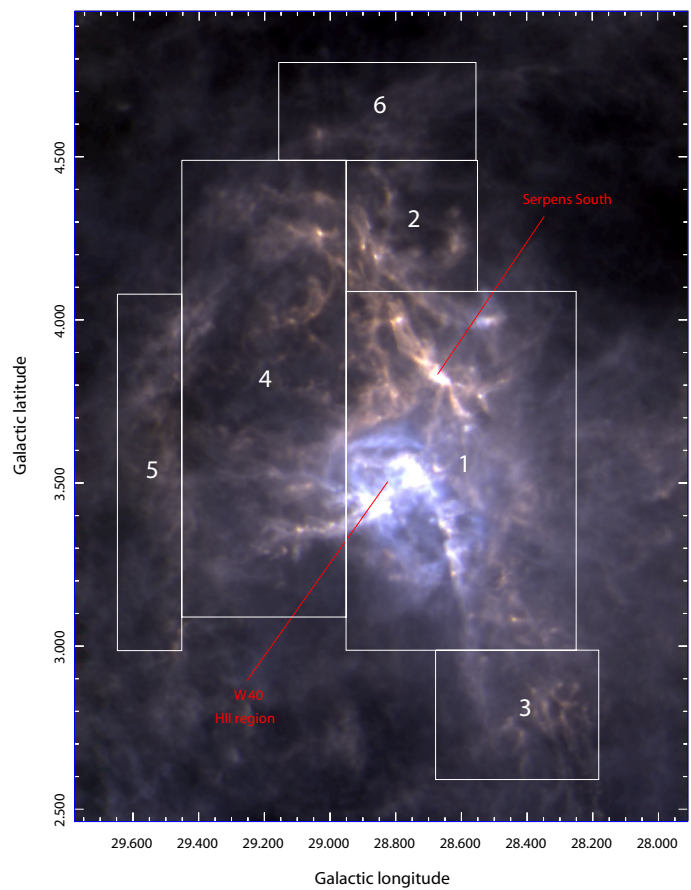}
\end{center}
\caption[]{Color image of the Aquila Rift (red for 500\,$\mu$m, green for 350\,$\mu$m,
and blue for 250\,$\mu$m, all derived from $Herschel$ data \citep{2010A&A...518L..85B,2010A&A...518L.106K}).
The regions subsequently observed in NH$_3$ are marked by six boxes (see Table\,\ref{table:A.1}).}
\label{fg1}
\end{figure}

\section{Observations and data reduction}
\label{sect:Observation}
\subsection{NH$_3$ observations}
\label{sect-2-1}
From March 2017 to August 2018, we observed the NH$_3$\,(1,1) and (2,2)
lines with the Nanshan 26-m radio telescope. A 22.0-24.2\,GHz dual polarization channel
superheterodyne receiver was used. The main parameters of the Nanshan telescope and survey area are
listed in Table\,\ref{table:parameters}. The rest frequency was centered at 23.708\,GHz to
observe NH$_3$\,(1,1) at 23.694\,GHz and (2,2) at 23.723\,GHz, simultaneously. To convert
antenna temperatures $T^\ast_{A}$ into main beam brightness temperatures  $T_{\rm MB}$,
a  beam efficiency of 0.59 has been adopted. The observations were calibrated against
periodically (6\,s) injected signals from a noise diode. We have observed
IRAS\,0033+636 ($\alpha$\,=\,00:36:47.51, $\delta$\,=\,63:29:02.1, J2000) repeatedly during
observations \citep{1996A&A...306..267S}, adopting a main beam brightness temperature $T_{\rm MB}$\,=\,4.5\,K.
Systematic variations and dispersion of brightness temperatures are small (see Appendix\,\ref{Appendix A}).
The standard deviation of the mean is $\sim$$10\%$ (see Fig.\,\ref{FgA.1}). To further check our
calibration stability, we compared our NH$_3$ data with previously taken NH$_3$ data observed
with the GBT in Fig.\,\ref{FgA.2} \citep{2017A&A...606A.133S}. The two data sets obtained from
G035.39-0.33 are in good agreement.

The typical system temperature was $\sim$50\,K ($T^\ast_{A}$ scale) at 23.708\,GHz.
The map was made using the On-The-Fly (OTF) mode with $6\arcmin \times 6\arcmin $ grid size
and $30\arcsec$ sample step. All observations were obtained under excellent weather conditions
and above an elevation of $20\degr$. The mapped region was divided into six areas
(see Fig.\,\ref{fg1}). Main parameters of each area are listed in Table\,\ref{table:A.1}
of Appendix\,\ref{Appendix A}. The whole map covers a region of $\sim$$1.5\degr\times2.2\degr$
\,(11.4\,pc\,$\times$\,16.7\,pc).

\begin{table}[t!]
\centering
\caption{Main parameters of the Nanshan telescope and survey area.}
\label{tbl-1}
\begin{tabular}{l l}
\hline \hline
Telescope & Nanshan 26-m  antenna  \\
Receiver & $K$-band receiver   \\
Mode & On-The-Fly   \\
Molecular lines & NH$_3$\,(1,1) and (2,2)   \\
Rest frequency & 23.708\,GHz \\
Bandwidth & 64\,MHz   \\
Channel number & 8192  \\
Beam size & $\sim$2 arcmin  \\
Velocity resolution & $\sim$0.1 \,km\,s$^{-1}$  \\
Main beam efficiency & 59\% \\
Region & $28.2\degr < l < 29.7\degr$, $2.6\degr < b < 4.8\degr$  \\
Survey area & $1.5\degr \times 2.2\degr$ \\
Serpens South & $28.5\degr < l < 29.1\degr$, $3.6\degr < b < 4.4\degr$  \\
W\,40 & $28.5\degr < l < 29.1\degr$, $3.2\degr < b < 3.6\degr$  \\
\hline
\end{tabular}
\label{table:parameters}
\end{table}

\subsection{Data reduction}
\label{sect-2-2}
\begin{figure}[h]
\vspace*{0.2mm}
\centering
\includegraphics[width=0.45\textwidth]{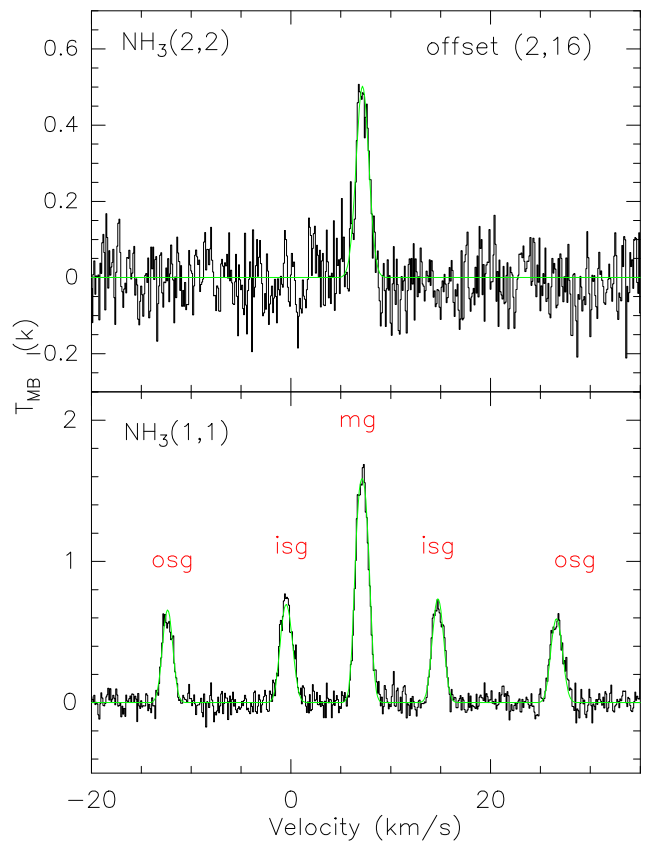}
\caption[]{NH$_3$\,(1,1) and (2,2) spectra at offset ($\Delta l,\,\Delta b$)\,=\,(2$\arcmin$,\,16$\arcmin$) with respect
to the reference position ($l$\,=\,28.59$\degr$, $b$\,=\,3.55$\degr$). In each panel, the black solid
line represents the observed spectrum, the green solid line indicates the NH$_3$\,(1,1) fitting (lower panel)
and Gaussian fitting (upper panel) of the NH$_3$\,(2,2) line (see Sect. \ref{sect-2-2}). The groups of
hyperfine components $``\rm mg", ``\rm isg",$ and $``\rm osg"$  represent the main,
inner-satellite, and outer-satellite groups. The velocity scale is Local Standard of Rest, here and elsewhere.}
\label{Fg2}
\end{figure}

The CLASS and GREG packages of GILDAS\footnote{http://www.iram.fr/IRAMFR/GILDAS/},
and also python plot packages matplotlib \citep{2007CSE.....9...90H} and
APLpy\footnote{http://aplpy.github.com} were used for all the data reduction.
The spectra were resampled in steps of $\sim$1$\arcmin$. To enhance signal to
noise ratios (S/Ns) in individual channels, we smoothed in many but
not all cases (see Sect.\,\ref{sect-3-1}) contiguous channels
to a velocity resolution $\sim$0.2\,km\,s$^{-1}$. A typical rms noise level
(1$\sigma$) is $\sim$0.03\,-\,0.05\,K ($T_{\rm MB}$ scale) for a channel of
$\sim$0.2\,km\,s$^{-1}$ in width (see Table\,\ref{table:A.1}). With respect to
NH$_3$, we chose two fitting methods, `GAUSS' fit and NH$_3$\,(1,1) fit. In order
to convert hyperfine blended line widths to intrinsic line widths in the NH$_3$
inversion spectrum (e.g., \citealt{1998ApJ...504..207B}), we also fitted the averaged
spectra using the GILDAS built-in `NH$_3$(1,1)' fitting method which can fit all 18 hyperfine
components simultaneously. From the NH$_3$(1,1) fit we can obtain integrated intensity,
line center velocity, intrinsic line widths of individual hyperfine structure (hfs)
components, and optical depth. Main beam brightness temperatures $T_{\rm MB}$ are obtained
from `GAUSS' fit. These fitting methods were also used in previous works,
such as by \citet{2012A&A...544A.146W} and by \citet{2018A&A...616A.111W}.
Examples for reduced and calibrated spectra of NH$_3$\,(1,1) and (2,2)
inversion lines are given in Fig.\,\ref{Fg2}.

Because the hyperfine satellite lines of the NH$_3$\,(2,2) transition are mostly weak,
NH$_3$\,(2,2) optical depths are not determined. A single Gaussian profile was fitted to the main
group of NH$_3$\,(2,2) hyperfine components. Physical parameters of dense gas such as rotational
temperature ($T_{\rm rot}$), kinetic temperature ($T_{\rm kin}$), and para-NH$_3$ column
density ($N_{\rm NH_{3}})$ were derived using the method described by \cite{1983ARA&A..21..239H}
and \cite{1986A&A...157..207U} (see Sects.\,\ref{sect-3-2} and \ref{sect-3-3}).

\begin{figure*}[t]
\centerline{\hbox{
\includegraphics[width=9.4cm,height=10.8cm,angle=0]{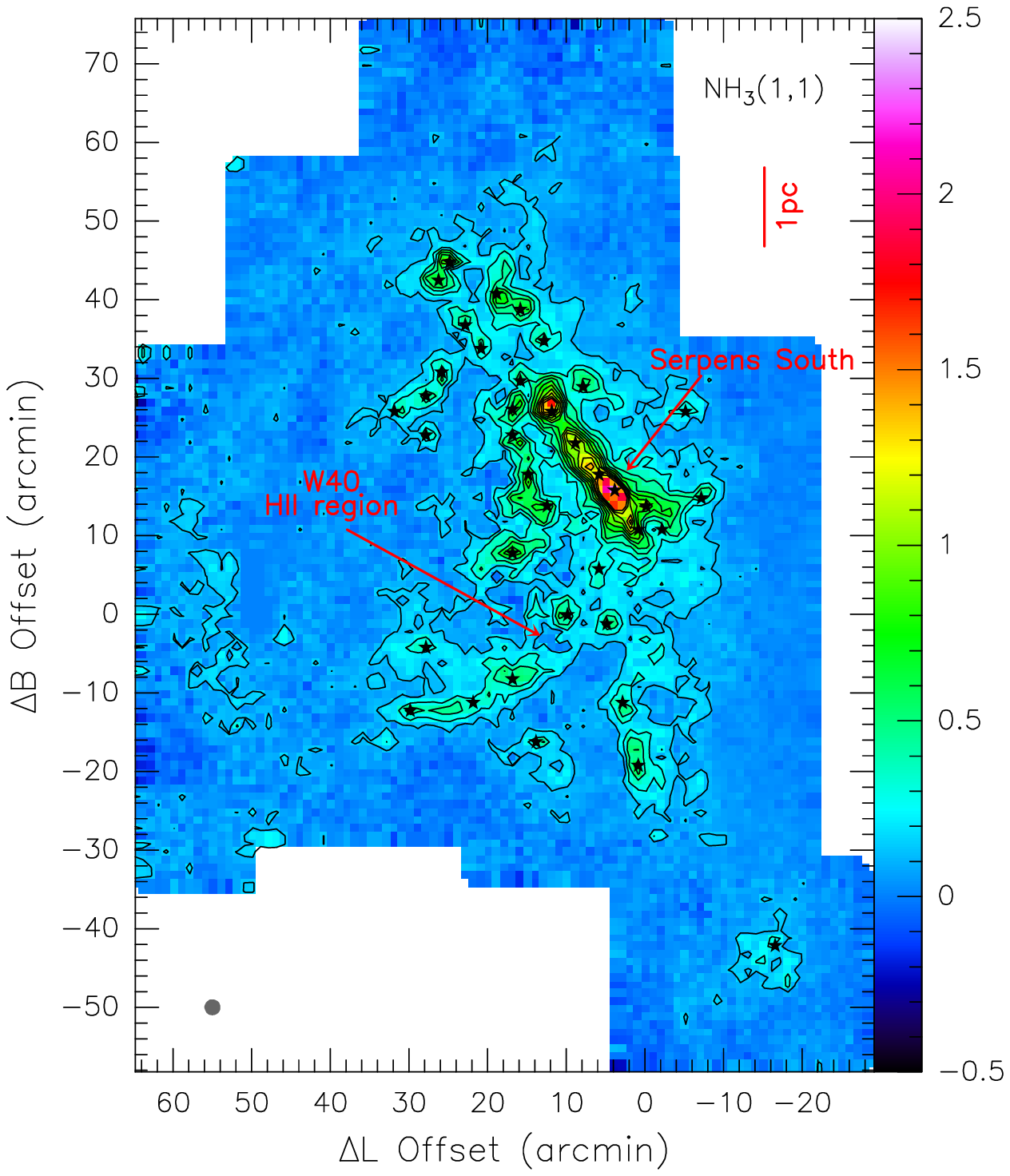}
\includegraphics[width=8.7cm,height=10.7cm,angle=0]{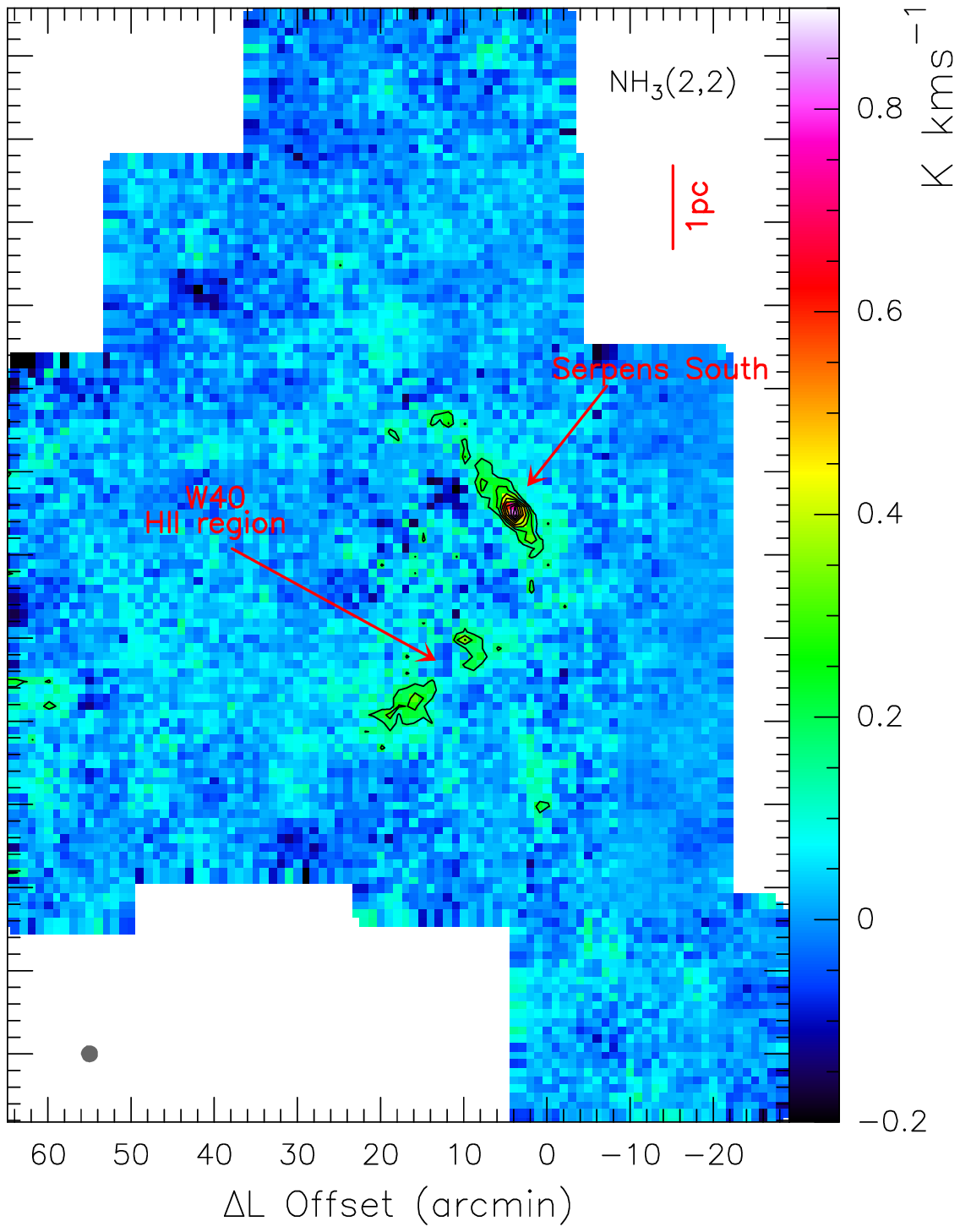}}}
\caption[]{Integrated intensity maps of NH$_3$\,(1,1) (\textit{left}) and (2,2)
(\textit{right}), the reference position is $l$\,=\,28.59$\degr$, $b$\,=\,3.55$\degr$. The integration
range is 4\,<\,$V_{\rm LSR}$\,<\,10\,km\,s$^{-1}$. Contours start at 0.13\,K\,km\,s$^{-1}$ (3$\sigma$)
on a main beam brightness temperature scale and go up in steps of 0.13\,K\,km\,s$^{-1}$.
The unit of the color bars is K\,km\,s$^{-1}$. The half-power beam width is illustrated as a black
filled circle in the lower left corners of the images. The red line in the top right of each map illustrates the 1\,pc scale at a
distance of 436\,pc \citep{2017ApJ...834..143O,2018ApJ...869L..33O}. In the left panel black stars show the
positions of the identified 38 ammonia clumps (see Sect. \ref{sect-3-2}).}
\label{fg3}
\end{figure*}

\section{Results}
\label{sect:Results}
\subsection{NH$_3$ distribution}
\label{sect-3-1}
NH$_3$\,(1,1) and (2,2) velocity-integrated intensity maps of the main groups of hyperfine
components are presented in Fig.\,\ref{fg3}. Intensities were integrated over the Local
Standard of Rest velocity ($V_{\rm LSR}$) range of 4 to 10 \,km\,s$^{-1}$. NH$_3$\,(1,1)
shows an extended distribution and clearly traces the dense molecular structure
including Serpens South and W\,40.  NH$_3$\,(2,2) is only detected in the densest
regions of Serpens South and W\,40, and shows a much less extended distribution.

What we find in Fig.\,\ref{fg3} is an Aquila Rift morphology that resembles the $Herschel$
color infrared image shown in Fig.\,\ref{fg1} in many aspects, but also shows clear differences
(see also Fig.\,1 of \citealt{2010A&A...518L.106K} for an H$_{2}$ column density map of the
region, based on $Herschel$ data). In the Serpens South region, there is a dominant
ridge of strong NH$_3$\,(1,1) emission from the southwest to the northeast (position angle
$\sim$ 30$\degr$ in Galactic coordinates) containing several cores with a total length of
about 15$\arcmin$. This is also seen in the color $Herschel$ infrared image, but the latter
is dominated by relatively hot dust associated with the W\,40 complex, where NH$_3$ emission
is present at a lower level. Weaker NH$_3$ emission extending further to the south and north
follows the color $Herschel$ infrared image. Our mapped area with highest Galactic longitudes,
region 5 in Fig.\,\ref{fg1}, showing only weak emission, is also seen in our NH$_3$\,(1,1) map.
Overall, the NH$_3$ distribution (Fig.\,\ref{fg3}) shows a ring-like morphology, with the center of this
ring located in region\,4 (see Fig.\,\ref{fg1} and Sect.\,\ref{sect-4-5}). With a radius of
25$\arcmin$ to 30$\arcmin$ (about 3.5\,pc) it is roughly circular and exhibits comparatively strong
dust and NH$_3$ emission at its low and weak emission at its high Galactic longitude side.

Previous SPIRE/PACS observations from the $Herschel$ Gould Belt survey towards the Aquila cloud complex
\citep{2015A&A...584A..91K} identified 446 candidate prestellar cores and 58 protostellar cores in the
Aquila region. Using this catalogue, we find 362 prestellar cores and 49 protostellar cores in our observed
area and show their distribution in Fig.\,\ref{FgB.1}. The two data sets of pre- and protostellar cores on
the one side and the NH$_3$ data on the other match each other very well. All protostellar cores are associated
with regions of notable ($>$0.13\,K\,km\,s$^{-1}$) NH$_3$ emission, while prestellar cores are even found in the
weak high-longitude region of the large ring-like structure.

\begin{figure*}[t]
\centering
\includegraphics[width=0.47\textwidth]{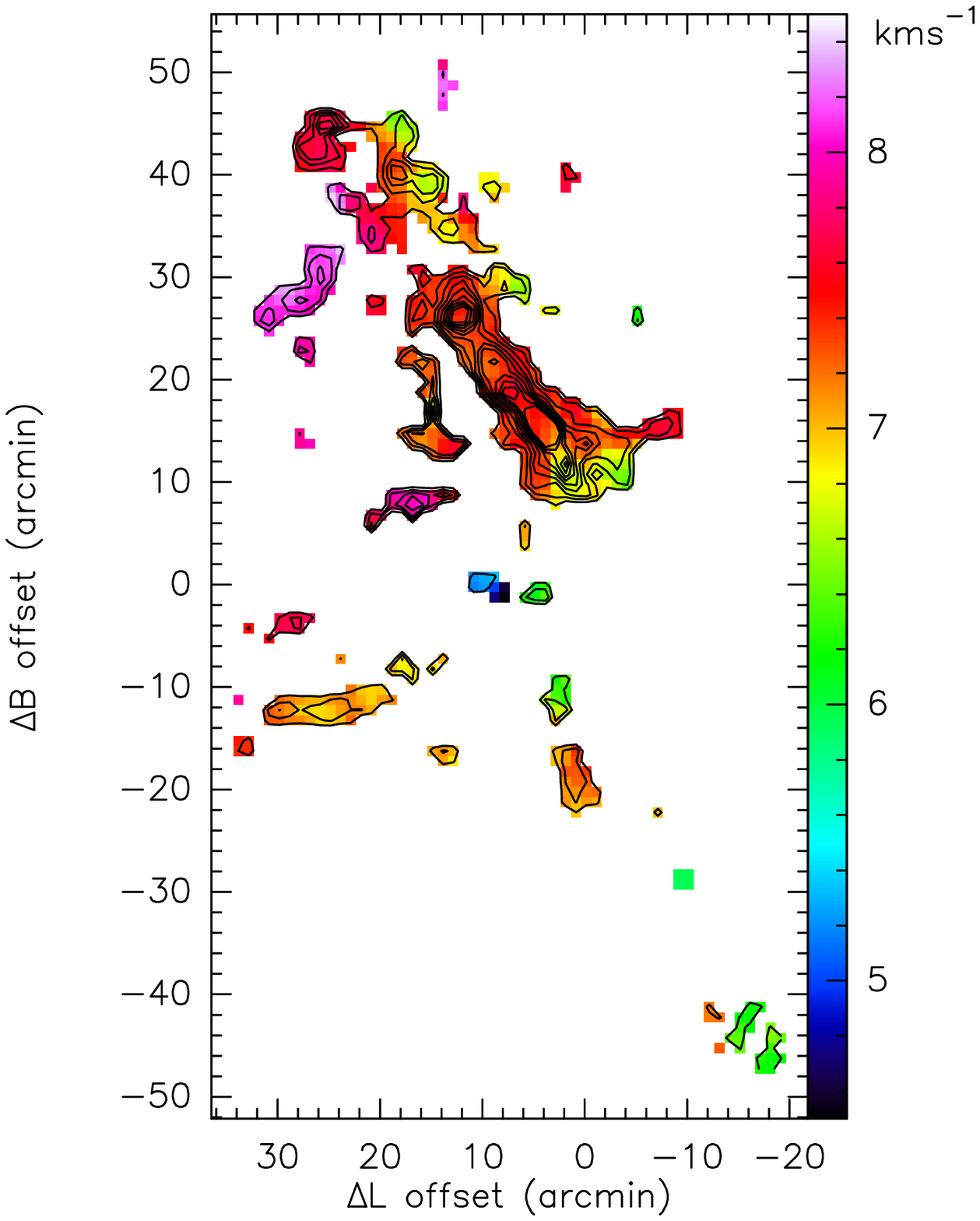}
\includegraphics[width=0.47\textwidth]{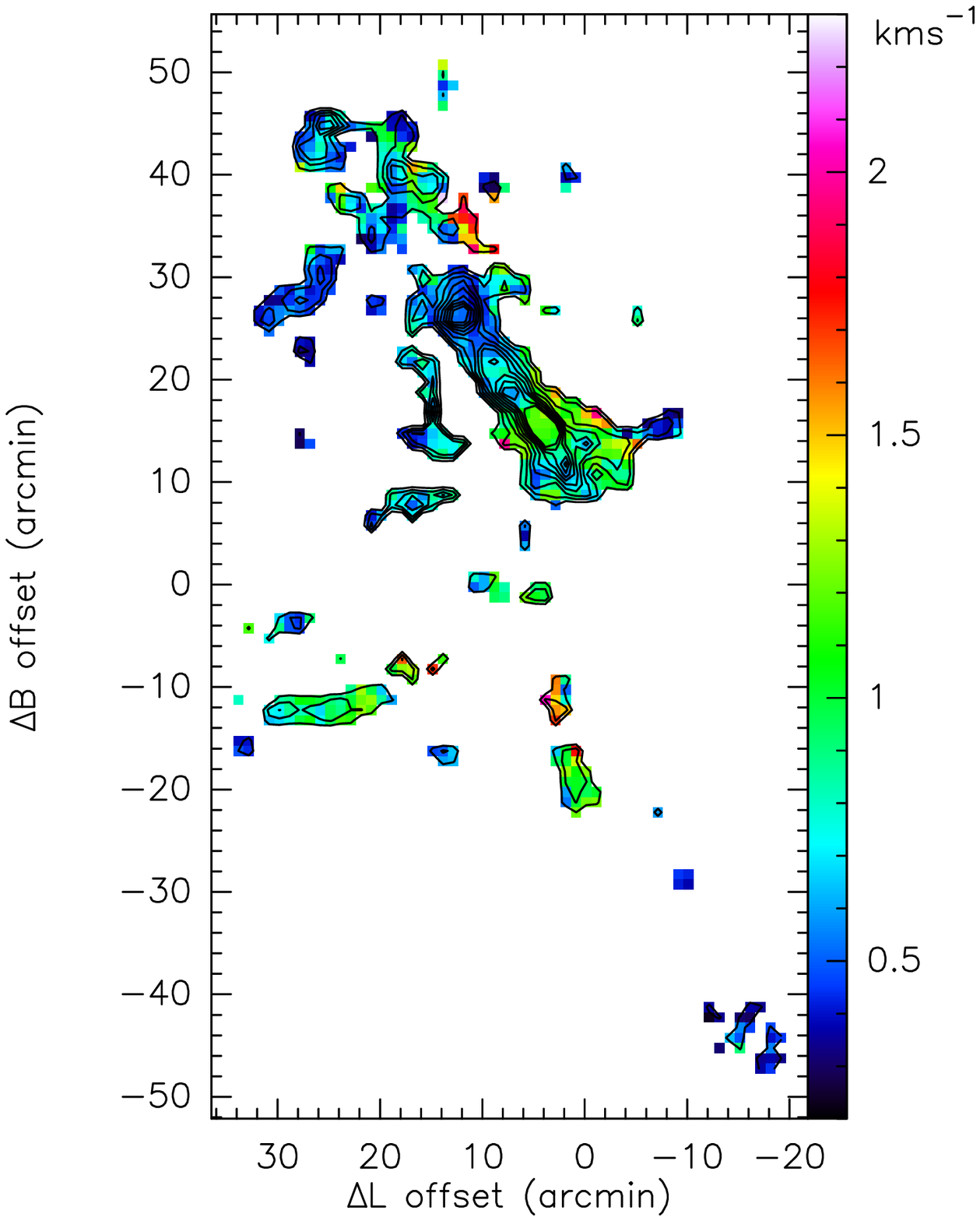}
\caption{Para-NH$_3$\,(1,1) intensity-weighted mean velocity (moment\,1, \textit{left}) and
FWHM line width (moment\,2, \textit{right}) maps. The reference position is $l$\,=\,28.59$\degr$, $b$\,=\,3.55$\degr$.
These intensity-weighted mean velocities and FWHM line widths referring to the individual hf components are derived
from the GILDAS built-in `NH$_3$\,(1,1)' fitting method (Sect.\,\ref{sect-2-2}). The considered velocity range is
4\,<\,$V_{\rm LSR}$\,<\,10\,km\,s$^{-1}$ for each panel. Contours of integrated intensity start at 0.23\,K\,km\,s$^{-1}$ (5$\sigma$)
on a main beam brightness temperature scale and go up in steps of 0.23\,K\,km\,s$^{-1}$.}
\label{fg4}
\end{figure*}

Using the `NH$_3$(1,1)' fit procedure (see Sect.\,\ref{sect-2-2}),
the NH$_3$\,(1,1) intensity-weighted mean velocity (moment\,1) and velocity dispersion (moment\,2)
maps are presented in Fig.\,\ref{fg4}. They visualize the kinematics of the Aquila Rift derived from
NH$_3$\,(1,1). The two images are overlaid with NH$_3$\,(1,1) integrated intensity contours as in
Fig.\,\ref{fg3}. Here we use a higher threshold of 5$\sigma$ for the lowest contours to provide
reliable results. Previous NH$_3$ observations with the GBT (beam size $\sim$30$\arcsec$;
\citealt{2016ApJ...833..204F}) toward the Serpens South indicated narrow line widths, typically of order 0.5\,km\,s$^{-1}$,
but with minima near 0.15\,km\,s$^{-1}$, so that we used in this case a velocity resolution of 0.1\,km\,s$^{-1}$
to fit the NH$_3$ spectral lines. The intensity-weighted mean velocity map (see Fig.\,\ref{fg4} left panel)
reveals that the main group of hf components shows a velocity range from 4.5 to 8.4 \,km\,s$^{-1}$.
The dominant ridge of NH$_3$ emission mostly indicates velocities in excess of 7\,km\,s$^{-1}$. Only
at its southern edge and to the west of its northern edge velocities are smaller (see also Fig. 4
of \citealt{2016ApJ...833..204F}). Velocities in regions 5 and 6 can not be shown because the
signal-to-noise (S/N) rations are low. In these regions with weak NH$_3$\,(1,1) line emission we
obtain averaged radial velocities of 8.2\,$\pm$\,0.8 and 7.1\,$\pm$\,0.5\,km\,s$^{-1}$, respectively,
which is consistent with those encountered in the regions with stronger emission.

From the FWHM line widths (moment\,2) map (see Fig.\,\ref{fg4} right panel) we can see that
the small region with widest lines is located in the northwest of Serpens South, while another
small spot with relatively high dispersion is encountered in the south. Elsewhere rather low
and also uniform intrinsic (for individual hf components) dispersions ($\sigma \sim 0.5$ \,km\,s$^{-1}$)
are present. Irregular and larger dispersions are obtained in region 2 (see Fig.\,\ref{fg1} for location and
extent of this region). However, this may be merely a consequence of low S/N ratios. We study the FWHM line
widths of the NH$_3$\,(1,1) main lines with a peak line flux threshold of 5$\sigma$ as summarized
in Fig.\,\ref{fg5}a. The line width refers to the individual hf components. Apparently the line width
distribution of the NH$_3$\,(1,1) gas has, in agreement with \citet{2016ApJ...833..204F},
an outstanding peak around 0.5\,km\,s$^{-1}$ .

\subsection{Dense clump identification}
\label{sect-3-2}
Based on the distribution of the integrated intensity of the NH$_3$\,(1,1) line in Fig.\,\ref{fg3},
we use the Clumpfind2d algorithm \citep{1994ApJ...428..693W} to identify dense core clumps. First,
the root mean square of the integrated intensity map was derived based on a background noise estimate
of 0.04 K in a 0.2\,km\,s$^{-1}$  wide channel from regions with weak emission. Then the level range
from a threshold of 5 to 37$\sigma$ was set with increments of 4$\sigma$ (where $\sigma$
is the rms noise level). 38 potential dense clumps associated with the Aquila Rift were
identified within the integrated intensity map after eliminating a fake clump located at the
boundary of the map. In order to further confirm the authenticity of the condensations
identified by Clumpfind2d, we also tried different parameter settings and found that the
location of the dense clumps remained the same and also that many false structure
results appeared when using a lower threshold of $\leq3\sigma$. Adopting the output from
the Clumpfind2d algorithm, it should be noted that the clumps identified around Aquila
are located mostly along the dominant NH$_3$ ridge with some additional sources further in the north
and south (see Fig.\,\ref{fg3}). The black stars in Fig.\,\ref{fg3} (left panel) show the positions
of the identified 38 clumps. Among these 38 clumps both NH$_3$\,(1,1) and (2,2)
emission lines are detected in 17 clumps. Observed parameters and calculated model parameters
are given in Tables\,\ref{table:B.1} to \ref{table:B.3}. NH$_3$\,(1,1) and (2,2)
spectral lines towards the 17 clumps are shown in Fig.\,\ref{FgB.2}. For the other 21
clumps NH$_3$\,(1,1) spectra  are shown in Fig.\,\ref{FgB.3}. Clumps are closely associated
with pre- and protostellar cores. While all protostellar cores appear near the clumps, the
prestellar cores exhibit a more widespread distribution (see also Fig.\,\ref{FgB.1} and Sect.\,\ref{sect-3-1}).

\subsection{Kinetic temperature}
\label{sect-3-3}
\begin{figure*}
\centering
\includegraphics[width=0.33\textwidth,height=0.255\textwidth]{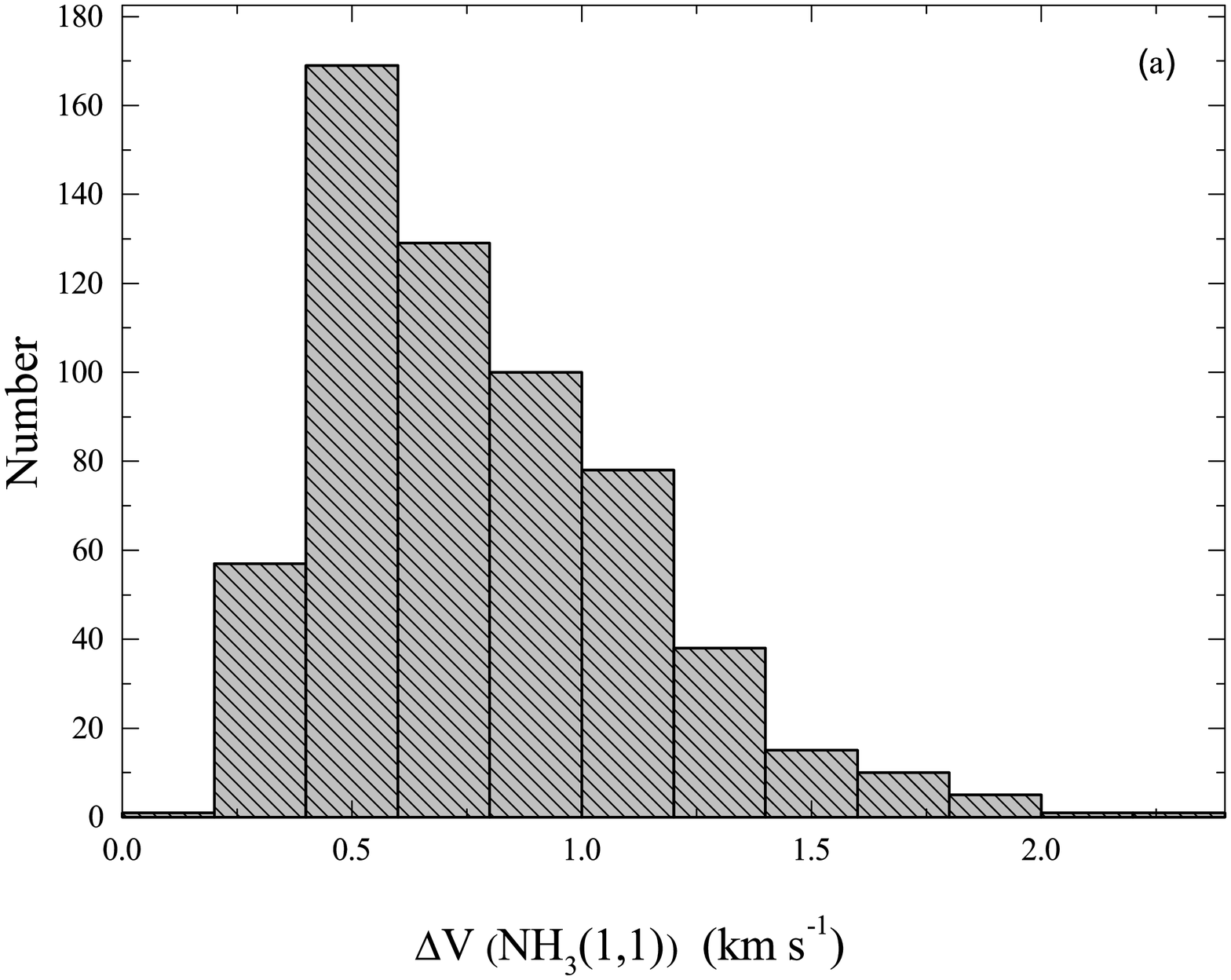}
\includegraphics[width=0.33\textwidth,height=0.255\textwidth]{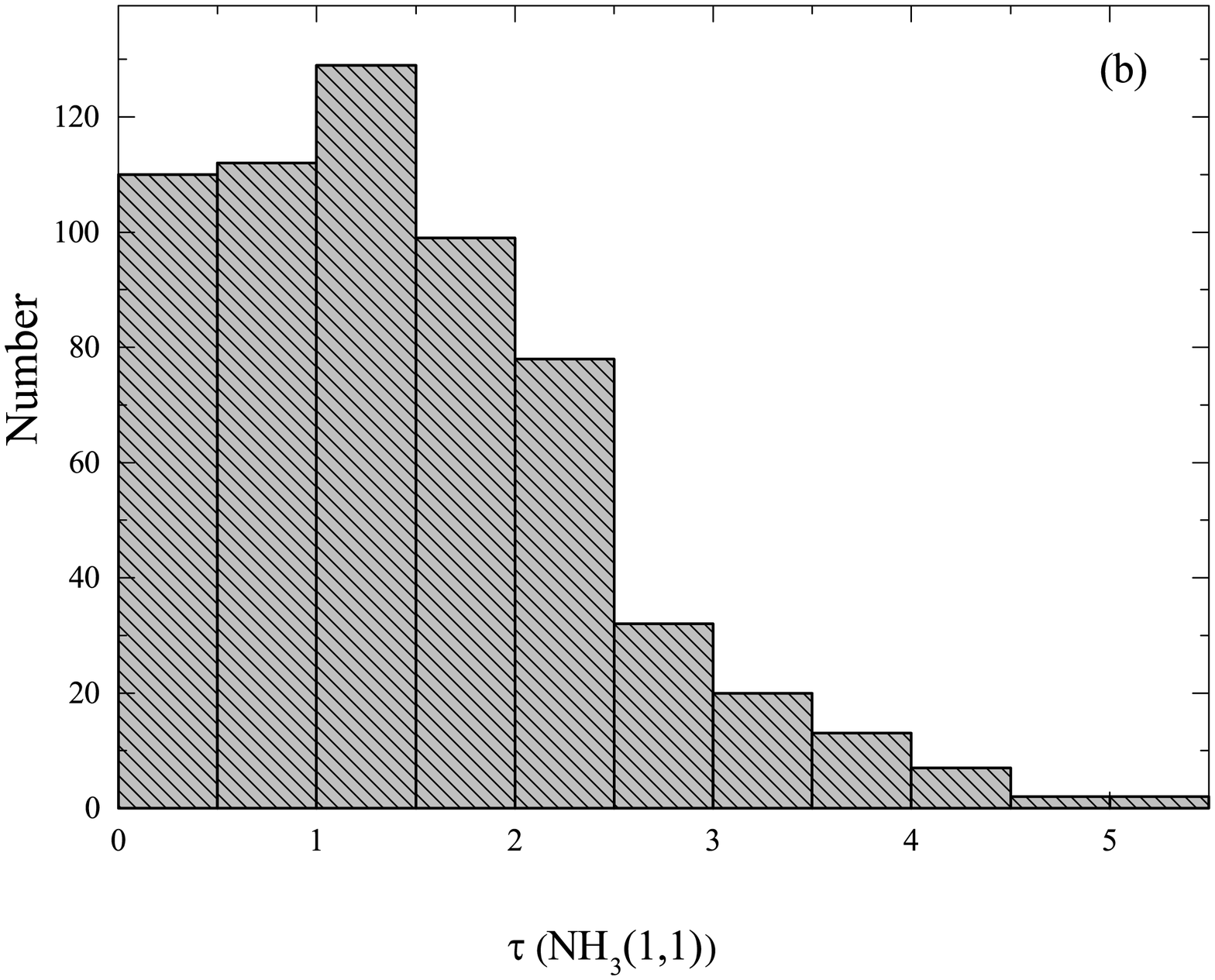}
\includegraphics[width=0.33\textwidth,height=0.255\textwidth]{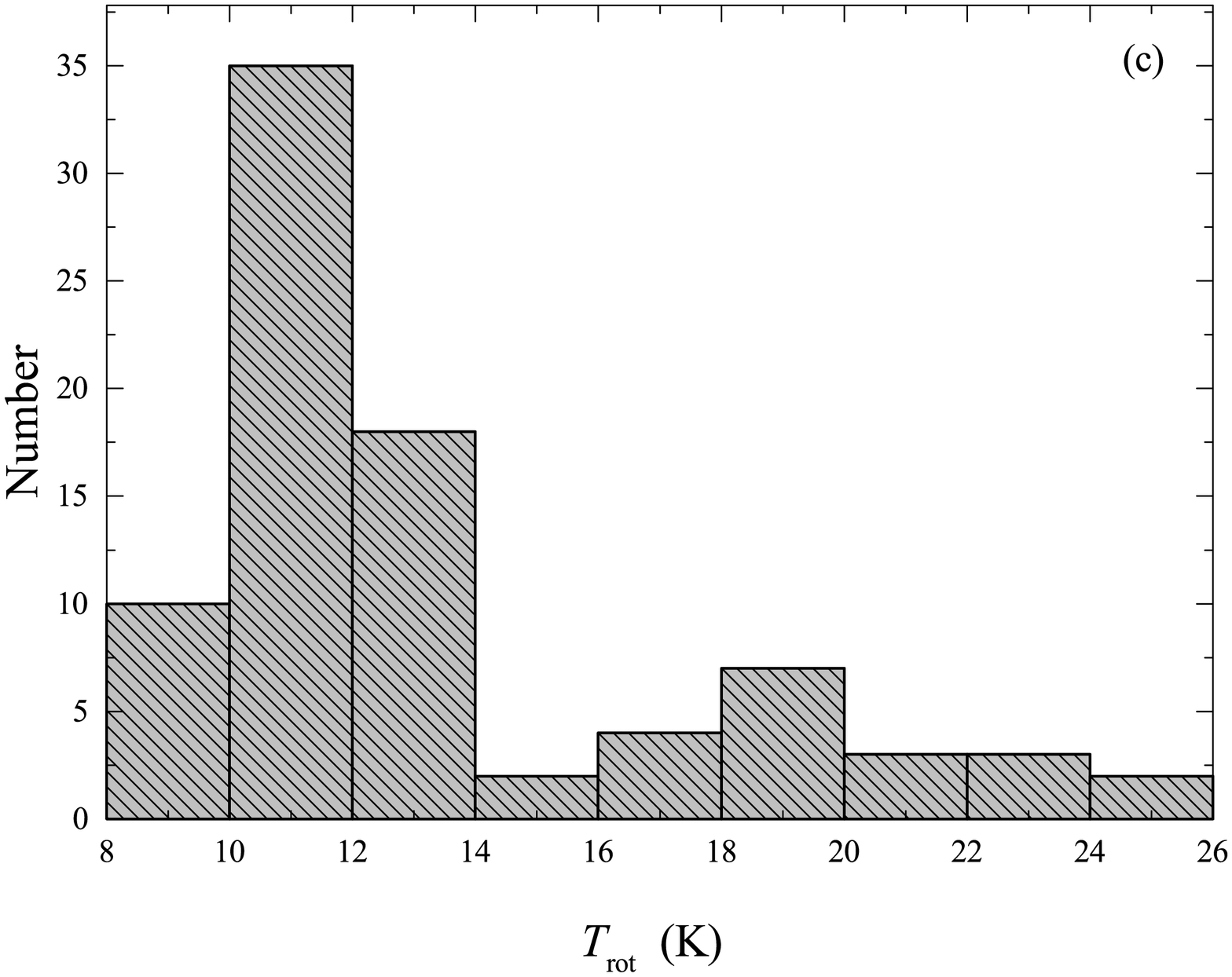}
\includegraphics[width=0.33\textwidth,height=0.255\textwidth]{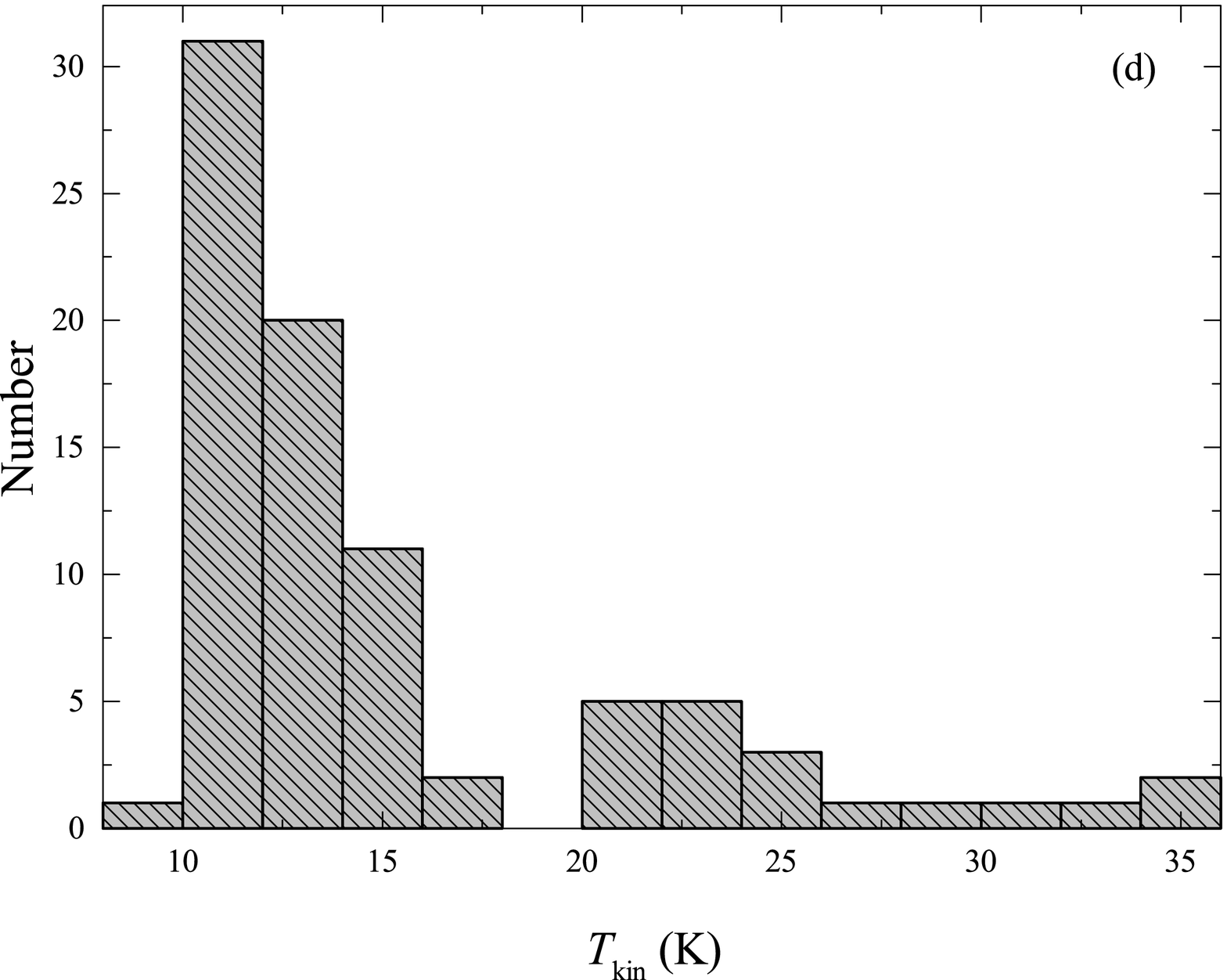}
\includegraphics[width=0.33\textwidth,height=0.255\textwidth]{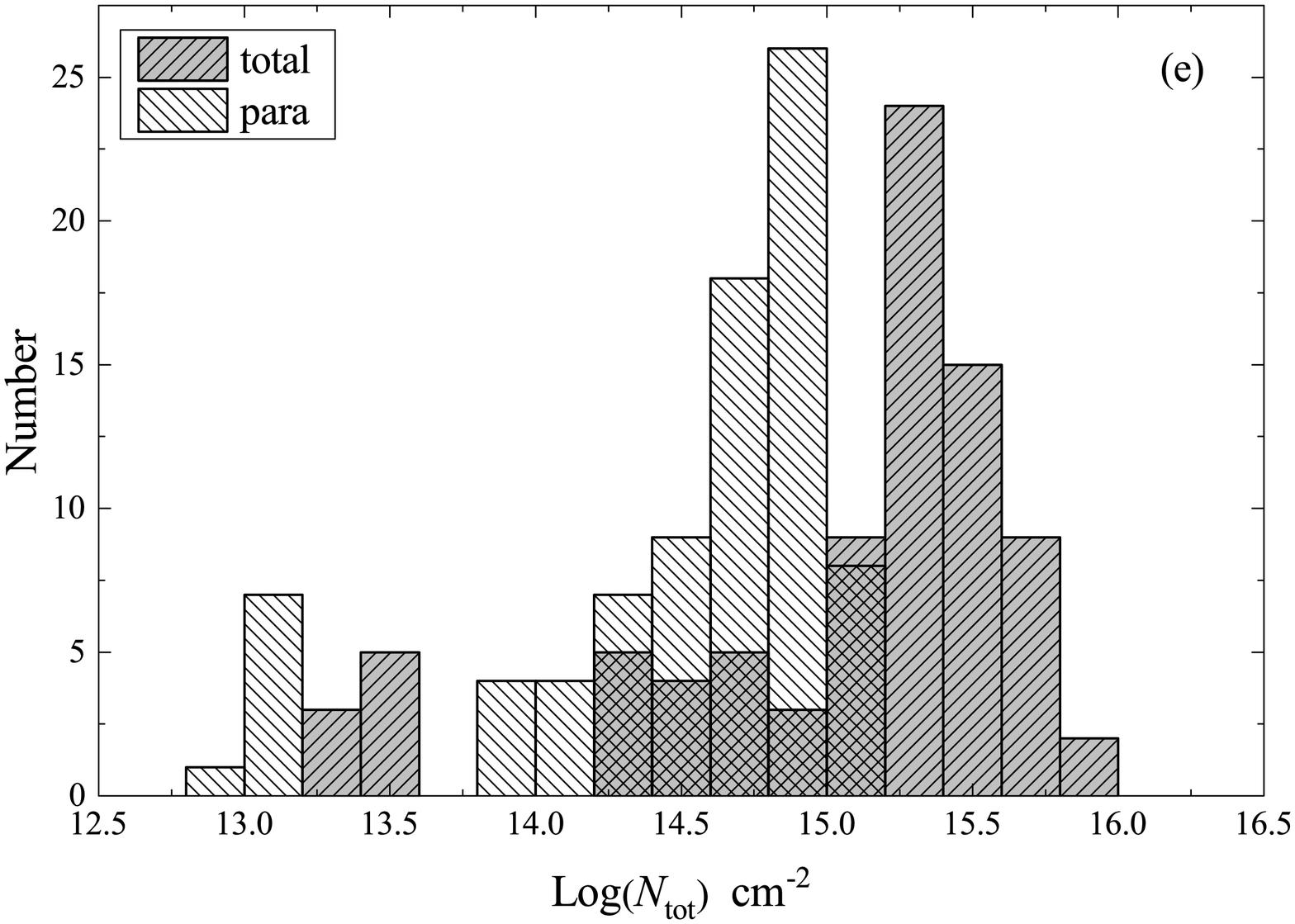}
\includegraphics[width=0.33\textwidth,height=0.255\textwidth]{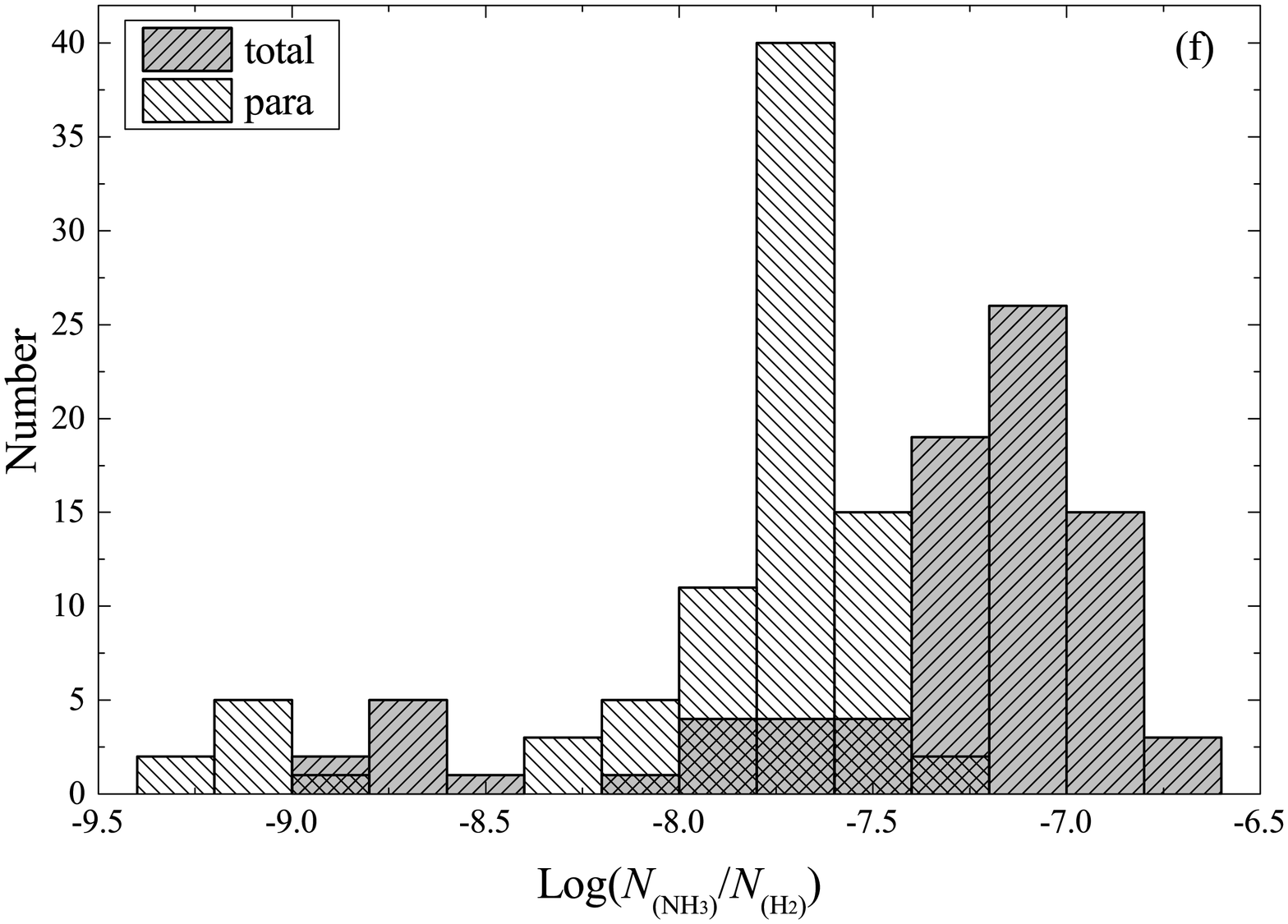}
\caption[]{Histograms of physical parameters derived from NH$_3$: (a) intrinsic FWHM line widths of individual NH$_3$\,(1,1) hyperfine structure components with a peak line flux threshold of 5$\sigma$, (b) peak optical depths of the main group of hf components $\tau_m (1,1)$ for those positions with NH$_{3}$\,(1,1) signal-to-noise ratios\,$>$\,5$\sigma$. These line widths and peak optical depths are derived from the GILDAS built-in `NH$_3$\,(1,1)' fitting method. (c) rotational temperature $T_{\rm rot}$, where both the NH$_{3}$\,(1,1) and (2,2) lines were detected, at levels of at least 5$\sigma$, (d) kinetic temperature $T_{\rm kin}$, for those positions with NH$_{3}$\,(2,2) line detections and\,$>$\,5$\sigma$ features, (e) para and total (para+ortho) column densities of NH$_3$, for those positions exhibiting NH$_{3}$\,(2,2) line detections and $>$\,5$\sigma$ features,
assuming an ortho- to para-ratio in thermal equilibrium at the presently obtained kinetic temperature. (f) para-NH$_3$ fractional abundance $\chi$ = [para-NH$_3$]/[H$_2$], and total-NH$_3$\,(para+ortho) fractional abundance $\chi$ = [total-NH$_3$]/[H$_2$], toward the positions with NH$_{3}$\,(2,2) line detections and $>$\,5$\sigma$ features, again adopting ortho- to para NH$_3$ ratios assuming thermal equilibrium.}
\label{fg5}
\end{figure*}

The relative population of the $K$ = 1 and 2 ladders of NH$_3$ are highly sensitive to collisional processes,
since they are not directly connected radiatively. This allows us to use them as a thermometer of the gas kinetic
temperature. The rotation temperature of NH$_3$\,(1,1) and (2,2) has been obtained by the method described in
\citet{1983ARA&A..21..239H}, which is
\begin{eqnarray}
\label{Eq1}
T_{\mbox{\tiny rot}}=\frac{-41.5}{\ln\left( \frac{-0.282}{\tau_m (1,1)}\ln
\left(1-\frac{T_{\rm MB}(2,2)}{T_{\rm MB}(1,1)}\left( 1- \rm exp(-\tau_m (1,1))\right)
 \right) \right)}\ \  {\rm K,}
\end{eqnarray}
where $\tau_m $ is the peak optical depth of the (1,1) main group of hf components derived
using the GILDAS built-in `NH$_3$\,(1,1)' fitting method. The main beam brightness
temperatures $T_{\rm MB}$ of the (1,1) and (2,2) inversion transitions are derived using
the GILDAS built-in `GAUSS' fitting. A histogram of the peak optical depth
of the (1,1) main group of hyperfine components, $\tau_m (1,1)$, for those positions with
NH$_3$\,(1,1) signal-to-noise ratios\,$>$\,5$\sigma$ is summarized in Fig.\,\ref{fg5}b.
Obviously, the optical depth distribution of the NH$_3$\,(1,1) gas peaks around 1.2.
The NH$_3$\,(1,1) main beam brightness temperatures ($T_{\rm MB}$ $<$ 2\,K)
are higher than those of the NH$_3$\,(2,2) line ($T_{\rm MB}$ $<$ 0.5\,K), so that the
NH$_3$\,(2,2) lines can be considered to be optically thin.

In the most prominent regions, where both the NH$_3$\,(1,1) and (2,2) lines were detected,
at levels of at least 5$\sigma$, the rotational temperature lies between 8.6 to 25.3\,K with
an average of 13.4\,$\pm$\,4.1\,K (errors are standard deviations of the mean throughout the article).
The median and mean values are 11.8 and 13.4\,K, respectively. The statistical distribution of
$T_{\rm rot}$ is summarized in Fig.\,\ref{fg5}c and shows a typical value of about 10\,K.

Following \citet{2004A&A...416..191T} to connect rotational with kinetic temperatures, we used
\begin{eqnarray}
\label{Eq2}
T_{\mbox{\tiny kin}}=\frac{T_{\mbox{\tiny rot}}(1,2)}{1-\frac{T_{\mbox{\tiny rot}}(1,2)}
{42}\ln\left( 1+1.1 \, {\rm exp}\left(\frac{-16}{T_{\mbox{\tiny rot}}(1,2)}\right)\right)}\ \ {\rm K,}
\end{eqnarray}
where the energy gap between the (1,1) and (2,2) states is $\Delta E_{12}$\,=\,42\,K.
\citet{2004A&A...416..191T} ran different Monte Carlo models involving the NH$_3$\,($J,K$)\,=\,(1,1),
(2,1), and (2,2) inversion doublets and an $n(r)$ = $n_0 / (1 + (r/r_{o})^{2.5})$ density
distribution to compare their observationally determined approximately constant rotational
temperatures with modeled kinetic temperatures in dense quiescent molecular clouds.
Eq.\,(\ref{Eq2}) is derived from fitting $T_{\rm kin}$ in the range of 5 to 20\,K, and most of
our sources can be found in this interval.

The kinetic temperatures of the dense gas in the Aquila Rift complex obtained from NH$_3$\,(2,2)/(1,1)
ratios wherever the (2,2) line has been detected at a\,$>$\,5$\sigma$ level range from 8.9 to 35.0\,K
with an average of 15.3\,$\pm$\,6.1\,K. The median and mean values are 12.7 and 15.3 K, respectively.
The distribution of the NH$_3$ kinetic temperatures is presented in Fig.\,\ref{fg5}d and shows a
typical value of about 12\,K.

The gas kinetic temperatures derived from the NH$_3$(2,2)/(1,1) map are shown in
Fig.\,\ref{fg6}, left panel. We find that dense gas temperatures from para-NH$_3$ in
Serpens South are predominantly cold ranging from 8.9 to 16.8\,K with an average of
12.3\,$\pm$\,1.7\,K, while W\,40 at lower Galactic latitudes, which represents a young stellar
cluster associated with an H\,{\scriptsize II} region, shows values from 17.7 to 35.0 K
with an average of 25.1\,$\pm$\,4.9\,K (see Table\,\ref{table:parameters} for the relevant
areas representing Serpens South and W\,40). The kinetic temperatures in the dense gas around
W\,40 and in the lower right part of our map are high ($\sim$25\,K; see Fig.\,\ref{fg6} left panel),
which is twice higher than that ($\sim$12\,K) in the low mass star formation region of Serpens South.
For those 21 clumps with only upper limits to the NH$_3$\,(2,2) lines (see Sect.\,\ref{sect-3-2}),
resulting upper kinetic temperature limits are $\lesssim 10$\,K.

\subsection{NH$_3$ column density}
\label{sect-3-4}
The total para-NH$_3$ column densities can be calculated from NH$_3$\,(1,1), following \citet{2012A&A...544A.146W}
\begin{eqnarray}
\label{Eq3}
N_{\mbox{\tiny tot}} \approx N(1,1) \Bigg( \frac{1}{3} \, {\rm exp} \left(\frac{23.1}
{T_{\mbox{\tiny rot}}(1,2)}\right)+1+ \frac{5}{3}\,{\rm exp} \left(-\frac{41.2}{T_{\mbox{\tiny rot}}(1,2)}\right)\\  \nonumber
+\frac{14}{3} \, {\rm exp}\left( -\frac{99.4}{T_{\mbox{\tiny rot}}(1,2)}\right)\Bigg)\ \ {\rm cm^{-2}.}
\end{eqnarray}
In many cases the NH$_3$\,(1,1) inversion lines are optically thick, and the optical depth is determined
by the GILDAS built-in `NH$_3$\,(1,1)' fitting method (see Sect.\,\ref{sect-2-2}). The column density in
the $(1,1)$ state is related to the optical depth $\tau_{\rm tot}$, according to \citet{1986A&A...162..199M}, by
\begin{eqnarray}
\label{Eq4}
N(1,1)=\frac{1.65\times10^{14}}{\nu} \,\frac{J(J+1)}{K^{2}}\,\Delta v\,\tau_{\rm tot}\,T_{\rm ex}\ \ {\rm cm^{-2},}
\end{eqnarray}
where $N$ is in cm$^{-2}$, the FWHM line width $\Delta v$ is in \,km\,s$^{-1}$, the line frequency $\nu$
is in GHz, and the excitation temperature $T_{\rm ex}$ is in Kelvin. In the optically thick case,
the excitation temperature $T_{\rm ex}$ is derived from the main beam brightness temperatures $T_{\rm MB}$
and the optical depth $\tau$ by
\begin{eqnarray}
\label{Eq5}
T_{\rm MB} = (T_{\rm ex}-2.7\,K)(1-\textrm{exp}(-\tau))\ \ {\rm K.}
\end{eqnarray}
If $\tau$\,$\ll$ 1, the main beam brightness temperature $T_{\rm MB}$ is assumed to be $T_{\rm ex}$\,$\tau$.

We calculated the ortho-column densities using the ortho-\,to\,para-NH$_3$ abundance
ratios as a function of $T_{\rm kin}$ in Fig.\,3 of \cite{2002PASJ...54..195T}. We then
added it to the para-column density to get the total (para + ortho) column densities of
NH$_3$ (see Tables\,\ref{table2} and \ref{table:B.3}).

The conversion between ortho- and para-NH$_3$ is very slow and may take as long as
10$^{6}$ yr \citep{1969ApJ...157L..13C}. As a result, the excitation temperature between
the ortho- and para-species, the so-called spin temperature, is often believed to reflect the
formation temperature. Assuming thermalization in this sense, the ortho- to para-NH$_3$ ratios
are close to unity at $T_{\rm kin}$\,$\ga$\,30\,K, while they can reach values of four at
$T_{\rm kin}$\,$\sim$\,10\,K \citep{2002PASJ...54..195T}. An even slightly higher range of
uncertainty is introduced by detailed models also introducing uncertainties in the poorly
known spin temperatures of H$_{2}$ and NH$_{4}^{+}$, which play an essential role in the
formation of ammonia. Depending on the detailed circumstances, \citet{2013ApJ...770L...2F}
find, rather independent of kinetic temperature for values below 30\,K, ammonia ortho- to
para abundance ratios of $\sim$1.5 and even $\sim$0.7. We conclude that the maximum error in our
adopted thermalized ortho-to-para NH$_3$ abundance ratios could be a factor of 4.9/0.7\,=\,7.0,
representing a case where we determine $T_{\rm kin}$\,$\sim$\,8.9\,K (our lowest kinetic temperature),
while the actual ortho- to para-abundance ratio is not 4.9 as expected in case of thermalization,
but 0.7 (see \citealt{2012A&A...543A.145P}). However, even in this worst case scenario our total NH$_3$
abundance would only be overestimated by half an order of magnitude, i.e by 5.9/1.7\,=\,3.5. The higher the
kinetic temperature, the smaller the uncertainty related to the ortho-NH$_3$ column density correction.

The observed NH$_3$ spectra were analyzed in the way described in Eqs.\,(\ref{Eq1}) to Eq.\,(\ref{Eq5}).
In addition, detailed derivations of these equations are give in the Appendix of \citet{2012ApJ...753...50P} and
\citet{2013A&A...553A..58L}. Furthermore, uncertainty estimations of the NH$_3$(1,1) and (2,2) spectra are given in
Appendix\,\ref{Appendix-C}.

As can be seen in Fig.\,\ref{fg6}, central panel, we provide a total column density map of NH$_3$
for those positions exhibiting NH$_3$(2,2) line detections and >5$\sigma$ NH$_3$\,(1,1) features.
The Aquila clumps show a broad distribution of total-NH$_3$ column densities
from 0.2\,$\times$\,10$^{14}$ to 6.4\,$\times$\,$10^{15}$\,cm$^{-2}$ with an average of
2.1\,($\pm1.6$)\,$\times$\,10$^{15}$\,cm$^{-2}$. The total-NH$_3$ column density
range is 0.3\,$\times$\,$10^{14}$ to 6.4\,$\times$\,$10^{15}$\,cm$^{-2}$ with an
average of 2.6\,($\pm$1.4)\,$\times$$10^{15}$\,cm$^{-2}$ in Serpens South, while
in W\,40 total  NH$_3$ column densities vary from 0.2 to
7.6\,$\times$\,$10^{14}$\,cm$^{-2}$ with an average of 2.6\,($\pm$2.1)\,$\times$\,$10^{14}$\,cm$^{-2}$.
The distribution of total column densities of NH$_3$ is
presented as a histogram in Fig.\,\ref{fg5}e and has an outstanding peak
around 2.4\,$\times$\,$10^{15}$\,cm$^{-2}$, due to our data from Serpens South.

\section{Discussion}
\label{sect:discussion}
\subsection{Variation of NH$_3$ abundance}
\label{sect-4-1}
Measurements of the NH$_3$ abundance in different star-forming regions have shown a large spread,
e.g. 10$^{-5}$ in dense molecular "hot cores" around newly formed massive stars due to dust grain evaporation
\citep{1987A&A...173..352M}, 10$^{-8}$ in quiescent dark clouds \citep{1983ApJ...270..589B} and 10$^{-9}$ in
the Orion Bar Photon Dominated Region (PDR; \citealt{2003A&A...408..231B,2003A&A...402L..69L}) because it is
extremely affected by a high UV flux. A few times 10$^{-10}$ in the Large Magellanic Cloud (LMC) and in M82 may
characterize an environment with low metallicity and a high UV radiation field (\citealt{2001ApJ...554L.143W,2010ApJ...710..105O}).

Our total-NH$_3$ column densities $N({\rm NH_{3}})$ are compared with the column densities
of H$_{2}$ derived from the $Herschel$ infrared continuum data representing dust emission
\citep{2010A&A...518L.102A,2015A&A...584A..91K}, for which we smoothed the data to our
beam size of $2\arcmin$. The NH$_3$ column densities are consistent with studies
reported towards other Gould Belt star-forming regions \citep{2017ApJ...843...63F}, where
the logarithm of the para-NH$_3$ column density (log $N$(para-NH$_3$)) varies from 13.0 to 15.5.

\begin{figure*}
\centerline{\hbox{
\includegraphics[width=6.7cm,height=8.78cm,angle=0]{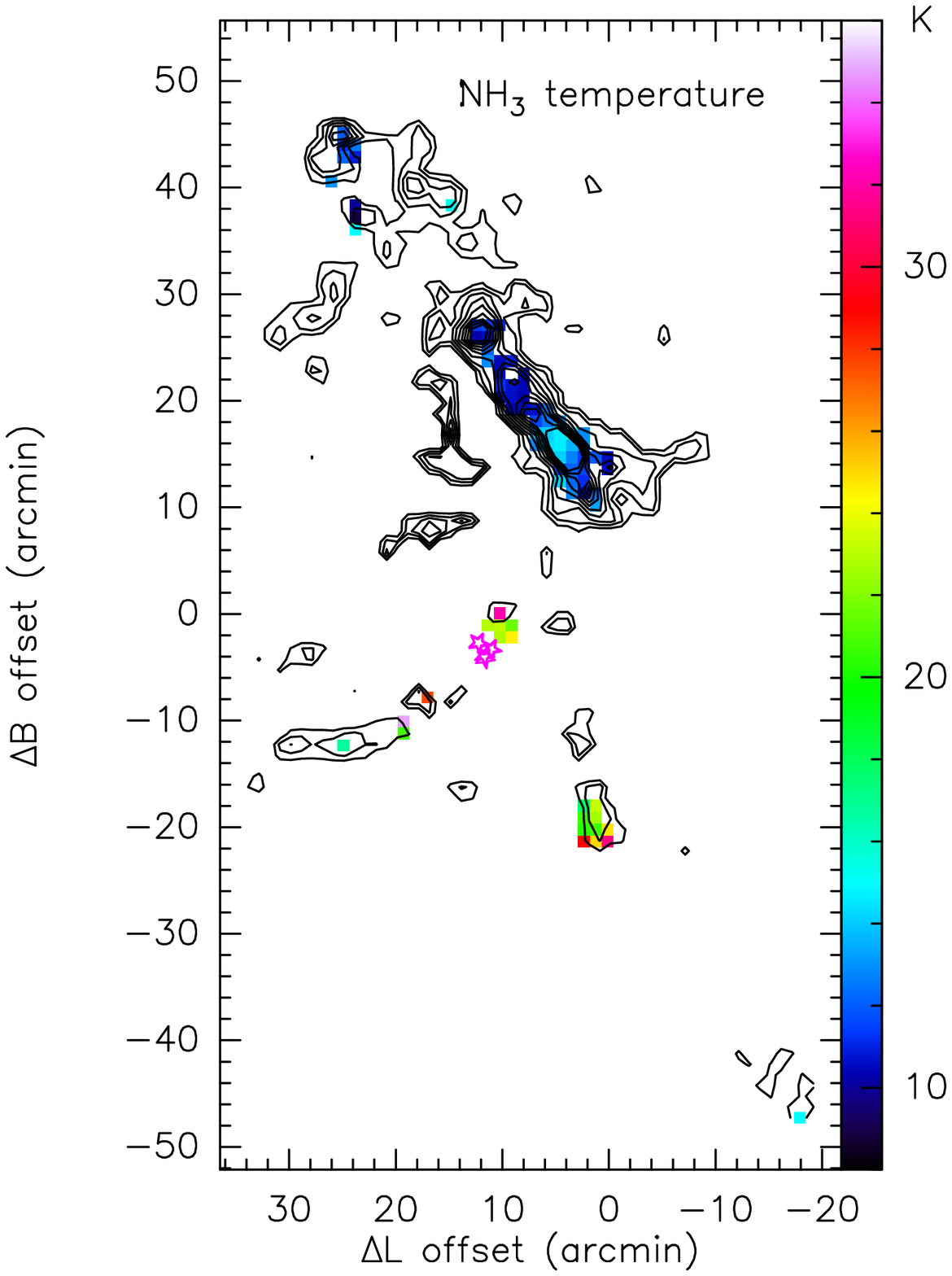}
\includegraphics[width=5.7cm,height=8.83cm,angle=0]{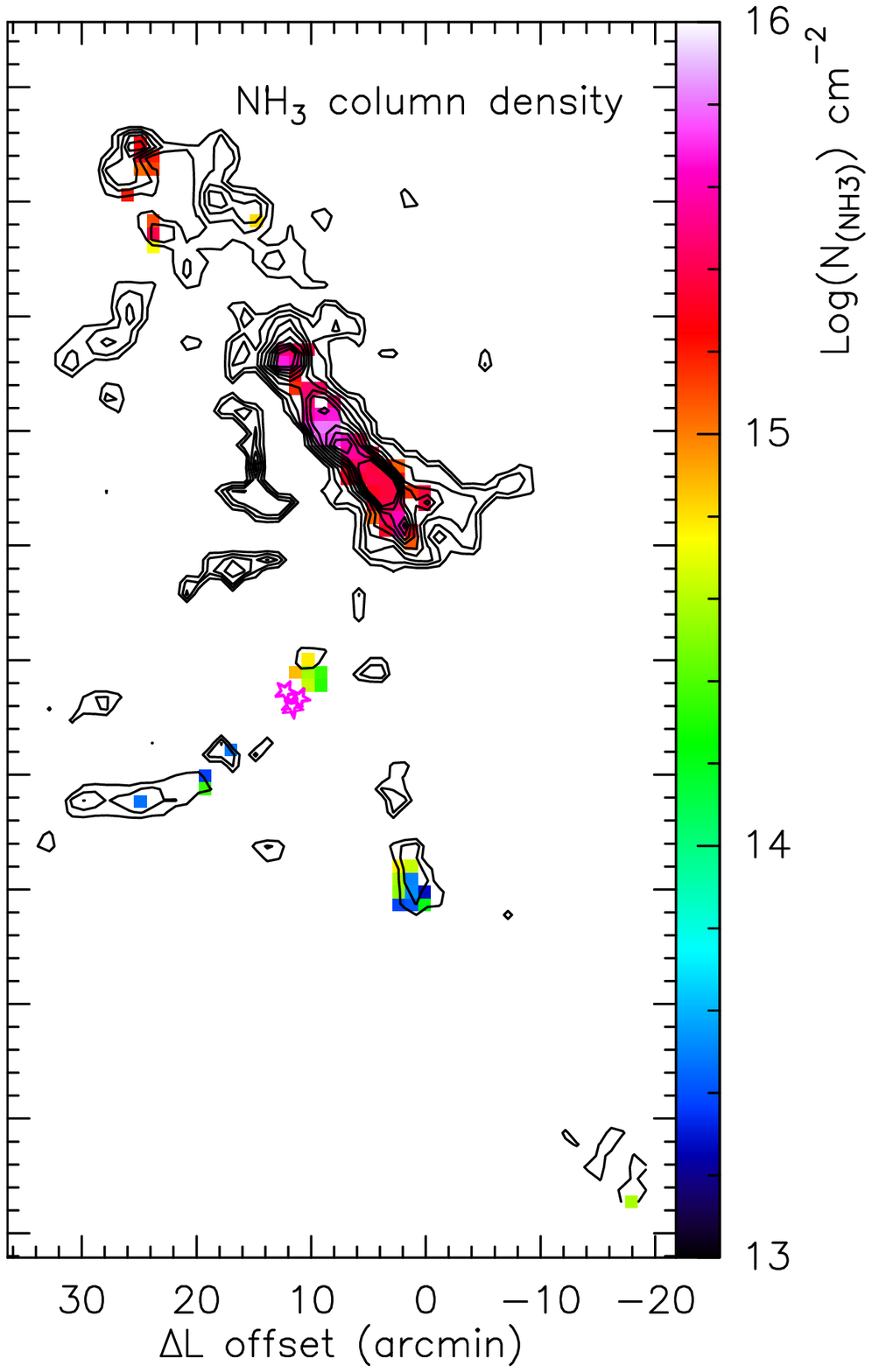}
\includegraphics[width=5.7cm,height=8.80cm,angle=0]{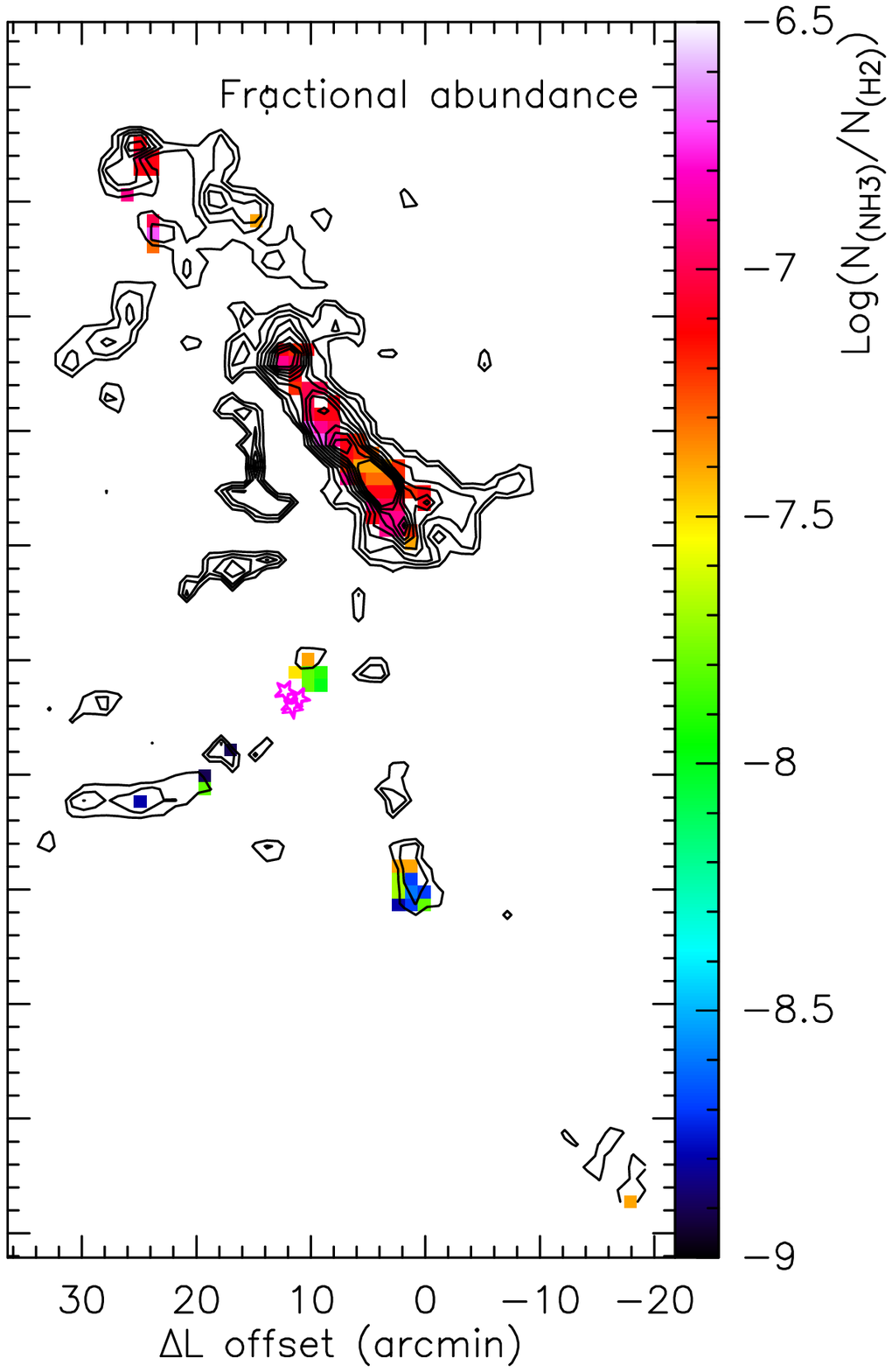}}}
\caption[]{Maps of NH$_3$ kinetic temperature in units of Kelvin (\textit{left}), the logarithm of the
total-NH$_3$ column density in units of cm$^{-2}$(\textit{middle}), and the corresponding logarithm of the
fractional abundance (\textit{right}). The reference position is $l$\,=\,28.59$\degr$, $b$\,=\,3.55$\degr$.
Contours of integrated NH$_3$\,(1,1) intensity are the same as in Fig.\,\ref{fg4} and cover the velocity
range 4\,<$V_{\rm LSR}$\,<\,10\,km\,s$^{-1}$. Contours start at 0.23\,K\,km\,s$^{-1}$ (5$\sigma$) on a main beam
brightness temperature scale and go up in steps of 0.23\,K\,km\,s$^{-1}$. Magenta stars show the locations of the
OB association (OS1a, IRS1, IRS1a, IRS2, IRS2a, IRS2b, IRS3, and IRS3a) in W\,40.}
\label{fg6}
\end{figure*}

The fractional total-NH$_3$ abundance map ($\chi$\,(total-NH$_3$)\,=\,(total-$N$\,(NH$_3$))/$N$(H$_2$))
is shown in Fig.\,\ref{fg6}, right panel. The relative total
NH$_3$ abundances $N$(total-NH$_3$)/$N$(H$_{2}$) range from 0.1\,$\times$\,$10^{-8}$ to
2.1\,$\times$\,$10^{-7}$ with an average of 6.9($\pm$4.5)\,$\times$\,10$^{-8}$, and the relative
para-NH$_3$ abundances $N$(para-NH$_3$)/$N$(H$_{2}$) range from 0.1 to 4.3\,$\times$\,$10^{-8}$
with an average of 1.8\,($\pm$0.9)\,$\times$\,10$^{-8}$ in our entire observed region (see Fig.\,
\ref{fg5}f). Previous observations of NH$_3$ in high-mass star-forming clumps suggest a median
value of $N$(para-NH$_3$)/$N$(H$_{2}$) of 2.5\,$\times\,10^{-8}$ \citep{2015MNRAS.452.4029U}. It
addition, averaged $N$(para-NH$_3$)/$N$(H$_{2}$) values of 1.2\,$\times$\,$10^{-7}$, 4.6\,$\times$\,$10^{-8}$,
and 1.5\,$\times$\,$10^{-8}$, were obtained by \citet{2011ApJ...741..110D}, \citet{2012A&A...544A.146W},
and \citet{2019MNRAS.483.5355M} in clumps of the Bolocam Galactic Plane Survey (BGPS), the APEX
Telescope Large Area Survey of the GALaxy (ATLASGAL), and the Hi-GAL survey, respectively.
Fractional abundances of $\sim$2--3\,$\times$\,$10^{-8}$ were derived for protostellar and starless cores
in Perseus, Taurus-Auriga, and infrared dark clouds \citep{2006A&A...455..577T,2009ApJ...696..298F,
2013A&A...552A..40C}.

The peak of our fractional para-NH$_3$ abundance distribution lies slightly above the
10$^{-8}$ range (see Fig.\,\ref{fg5}f). \citet{2016ApJ...833..204F} found a factor of several
variation in the para-NH$_3$ abundance across Serpens South, with the lowest detected abundances,
$\chi$(para-NH$_3$) $\sim$ 5\,$\times\,10^{-9}$, and highest, $\sim$\,2\,$\times\,10^{-8}$. We find
para-NH$_3$ abundances from 0.1 to 4.3\,$\times\,10^{-8}$ with an average of 2.2($\pm$0.8)\,$\times$\,10$^{-8}$
in the same region. We also find that Serpens South shows total-NH$_3$ fractional abundances ranging
from 0.2\,$\times\,10^{-8}$ to 2.1\,$\times\,10^{-7}$ with an average of 8.6\,($\pm$3.8)\,$\times$\,10$^{-8}$.
In W\,40 the values are lower, between 0.1 and 4.3\,$\times\,10^{-8}$ with an average of 1.6\,($\pm$1.4)$\,\times$\,10$^{-8}$,
while we obtained fractional para-NH$_3$ abundances range from 0.1 to 1.9\,$\times10^{-8}$ with an average of
0.7($\pm$0.5)\,$\times$\,10$^{-8}$. These results imply that in W\,40, toward the positions with NH$_3$(2,2)
line detections and >5$\sigma$ NH$_3$(1,1) features, total-NH$_3$ abundances are a factor of $\sim$5 lower
than in Serpens South. Different stages of star formation apparently lead to different fractional NH$_3$
abundances. The lower total-NH$_3$ fractional abundance in W\,40 compared to Serpens South is
likely due to the fact that W\,40 is strongly affected by FUV photons originating from the H\,{\scriptsize II} region.
Ammonia is a particularly sensitive molecular species with respect to this kind of radiation (e.g., \citealt{2001ApJ...554L.143W}).

\begin{figure*}[t]
\centering
\includegraphics[width=0.331\textwidth]{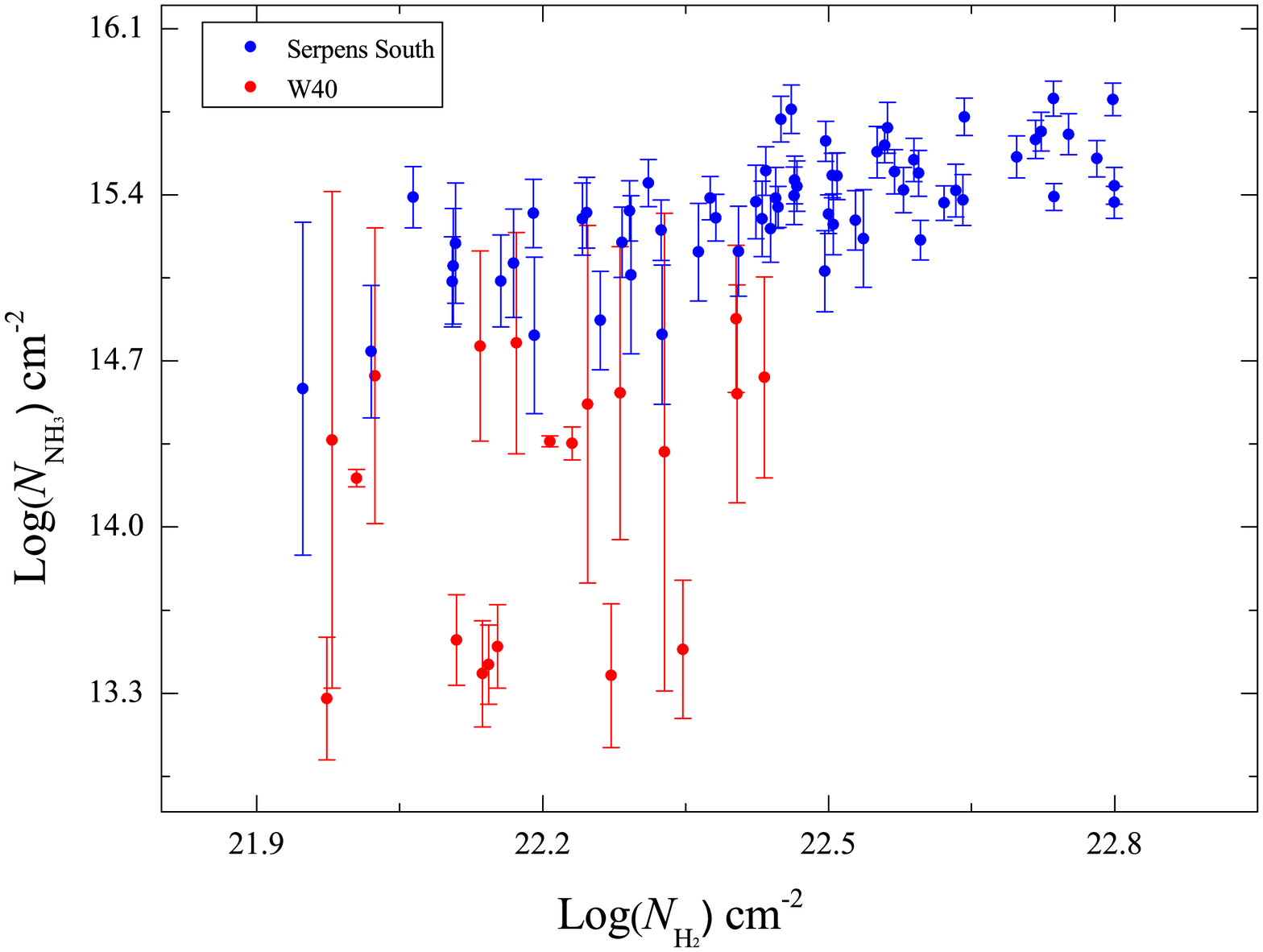}
\includegraphics[width=0.33\textwidth]{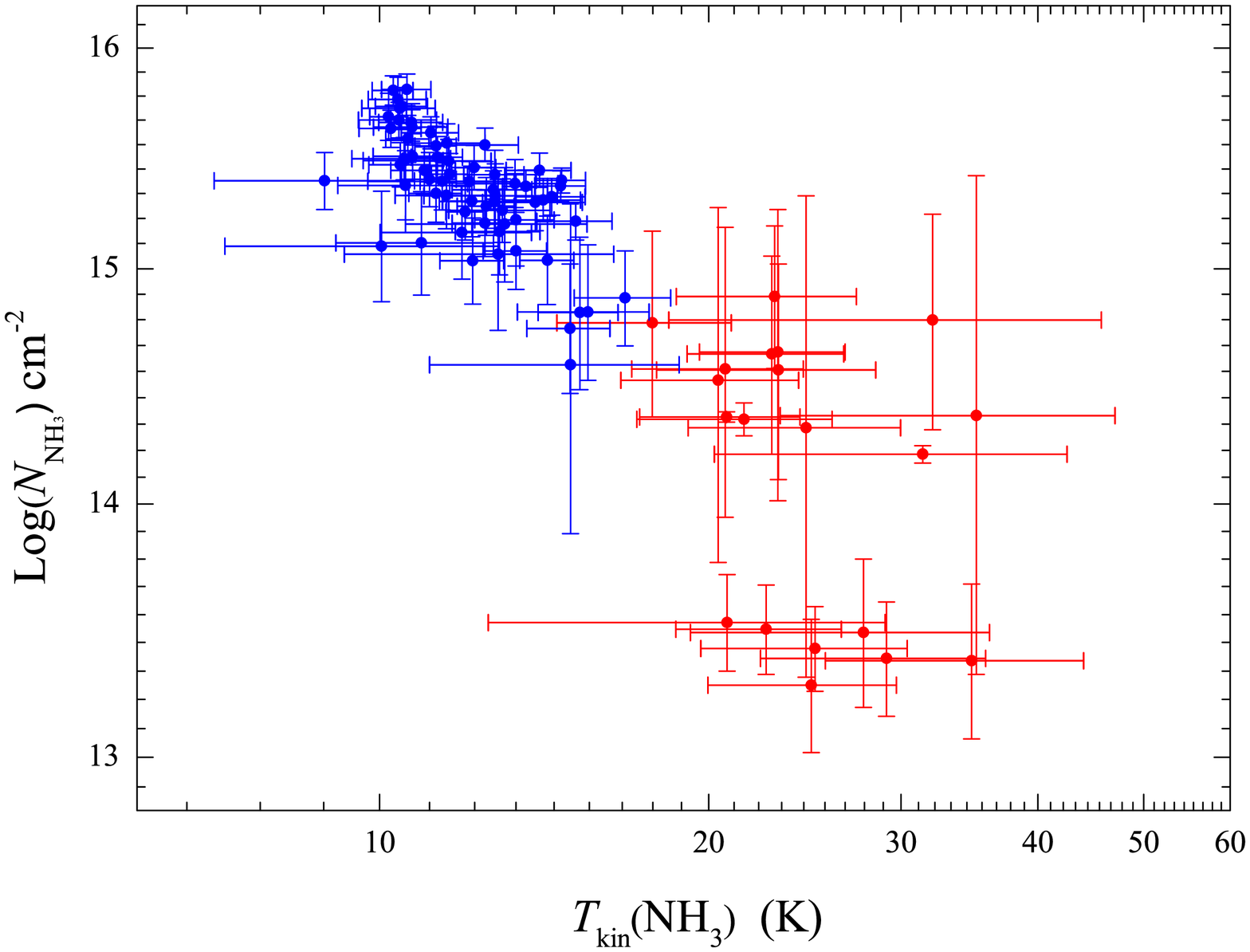}
\includegraphics[width=0.33\textwidth]{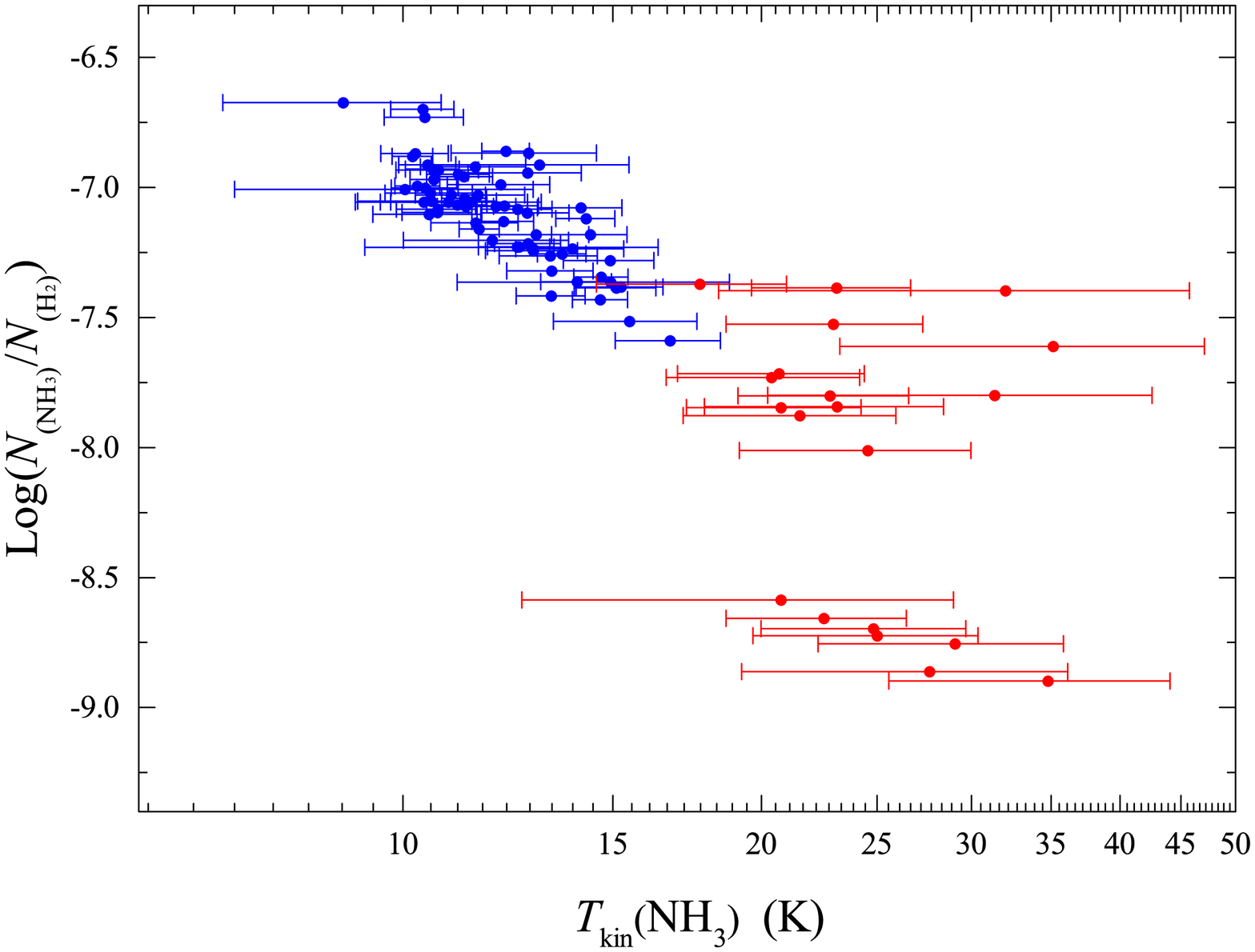}
\caption[]{Column densities derived from total-$N$(NH$_3$) for Serpens South (blue points)
and W\,40 (red points) vs. $N$(H$_{2}$) column densities (\textit{left}), total column densities
of $N$(NH$_3$) vs. kinetic temperature (\textit{middle}), and total fractional NH$_3$ abundance,
$N$(total-NH$_3$)/$N$(H$_{2}$), vs. kinetic temperature $T_{\rm kin}$(NH$_3$) (\textit{right}).}
\label{fg7}
\end{figure*}

The total column densities of NH$_3$, and its fractional abundances,
$N$(total-NH$_3$)/$N$(H$_{2}$), as a function of H$_{2}$ column density
and kinetic temperature $T_{\rm kin}$(NH$_3$) are shown in  Fig.\,\ref{fg7}.
The total-NH$_3$ column densities increase with H$_{2}$ column densities in
Serpens South, while there is no clear functional relation between total-NH$_3$
column densities and H$_{2}$ column densities in W\,40. However, in W\,40, low $N$(total-NH$_3$) values
are only found in case of low H$_{2}$ column densities (see Fig.\,\ref{fg7}, left panel).
$N$(H$_{2}$) varies from 0.9 to 2.7\,$\times$\,$10^{22}$\,cm$^{-2}$ with an average
of 1.7\,($\pm$0.5)\,$\times$\,$10^{22}$\,cm$^{-2}$ in W\,40. Serpens South is characterized by $N$(H$_{2}$) ranging
from 0.9 to 6.3\,$\times$\,$10^{22}$\,cm$^{-2}$ with an average of 3.2\,($\pm$1.4)\,$\times$\,$10^{22}$\,cm$^{-2}$
(see Table\,\ref{table2}). The total NH$_3$ column densities and fractional abundances show a trend inversely
proportional to kinetic temperature in Serpens South (see Fig.\,\ref{fg7} middle,
right) and the entire surveyed region. However, no clear correlation with kinetic temperature
is seen in W\,40 alone.

\subsection{Comparison of gas and dust temperatures}
\label{sect-4-2}
\begin{figure*}[t]
\centering
\includegraphics[width=0.486\textwidth]{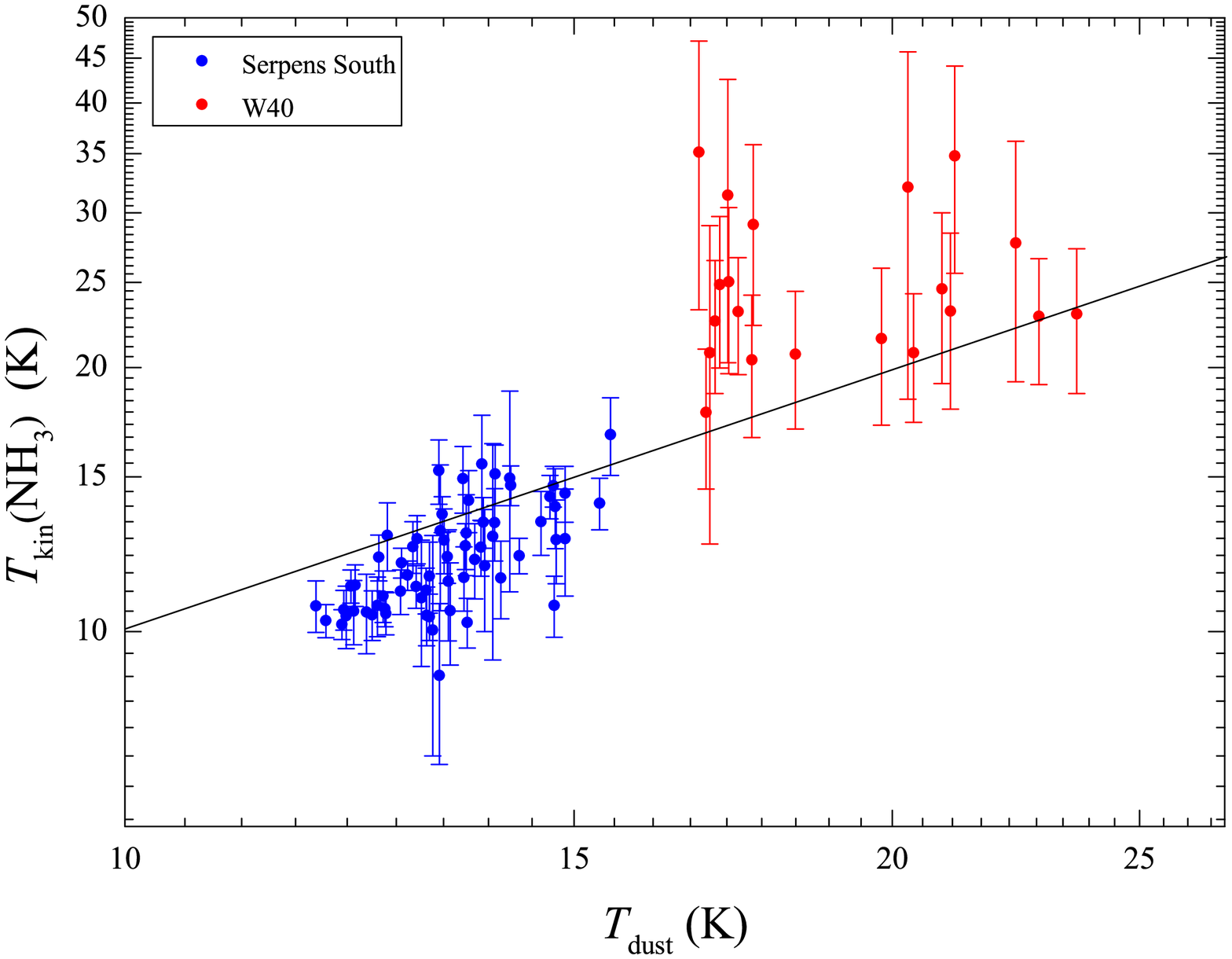}
\includegraphics[width=0.486\textwidth]{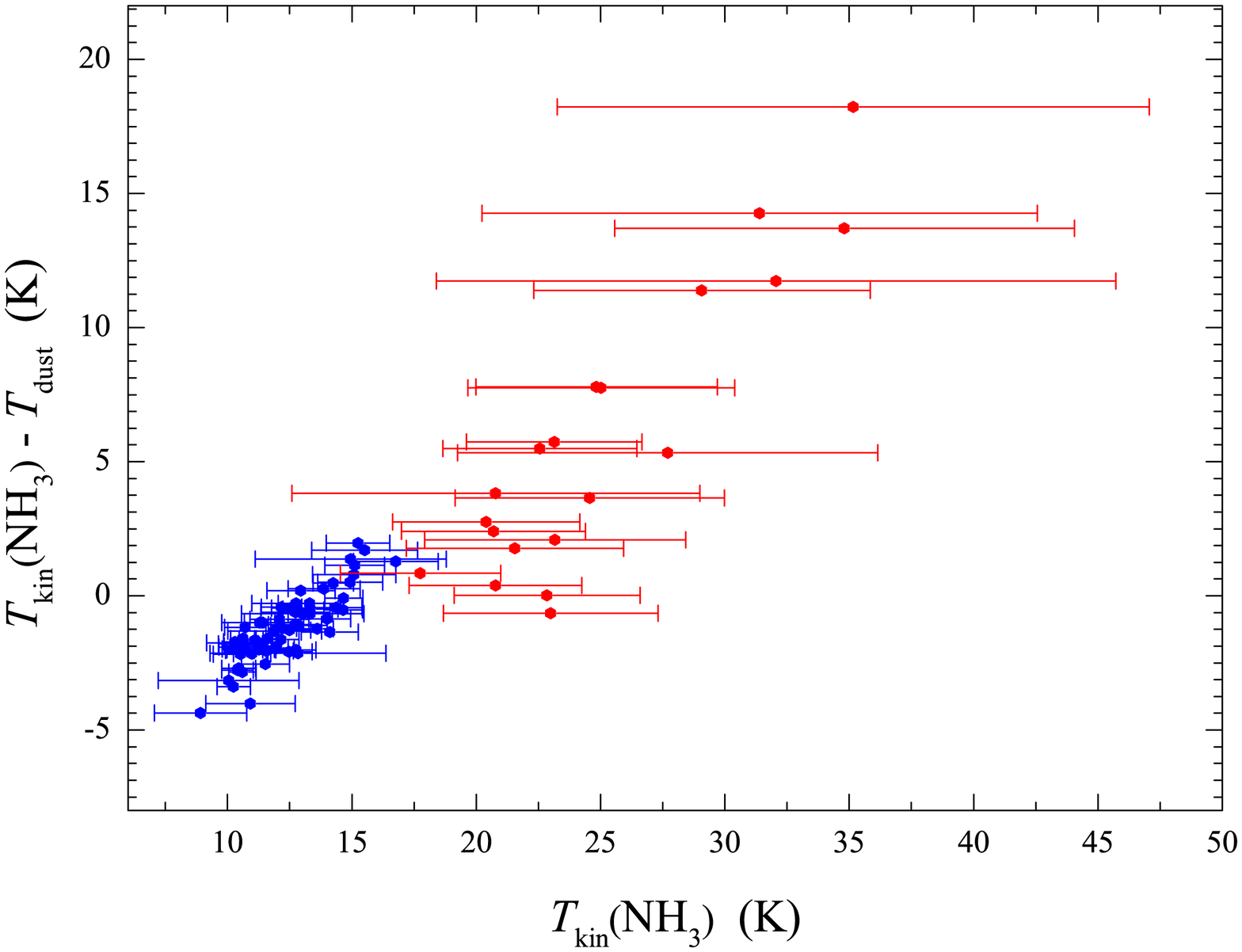}
\caption[]{Comparison of gas kinetic temperature derived from NH$_3$\,(2,2)/(1,1) ratios for
Serpens South (blue points) and W\,40 (red points) against dust temperature. In the left panel,
the black line indicates identical gas and dust temperature.}
\label{fg8}
\end{figure*}

A comparison of gas kinetic temperatures derived from para-NH$_3$\,(2,2)/(1,1) against HiGal dust
temperatures \citep{2010A&A...518L..85B} is shown in Fig.\,\ref{fg8}. The statistical
distribution of $T_{\rm rot}$ shows a typical value of about 10\,K. This is compatible with the
characteristic kinetic temperature of local quiescent molecular gas, as indicated, for example,
by \citet{1983ApJ...266..309M} using CO and by \citet{1983ApJ...266..309M} analyzing ammonia
observations. The gas kinetic temperatures in the Serpens South region
are similar to other active star-forming regions found by \citet{2017ApJ...843...63F},
such as Barnard\,18 in Taurus ($T_{\rm kin}$\,=\,6\,$\sim$\,14\,K), NGC\,1333 in Perseus
($T_{\rm kin}$\,=\,8\,$\sim$\,21\,K), and L1688 in Ophiuchus ($T_{\rm kin}$\,=\,9\,$\sim$\,25\,K).
The Orion A dense molecular cloud has been measured in NH$_3$\,(1,1) and (2,2) with the GBT
\citep{2017ApJ...843...63F}. The typical gas kinetic temperature obtained from NH$_3$\,(2,2)/(1,1)
is 20\,--\,30\,K. Measured gas kinetic temperatures are $>$\,100\,K in Orion\,KL,
$>$\,50\,K in the Orion Bar, $\sim$\,50\,K in Orion South, 20\,--\,30\,K in the north of Orion molecular
cloud\,1 (OMC-1) and $>$\,50\,K in the northeastern part of the OMC-1 region (see Fig.\,5 of \citealt{2018A&A...609A..16T} based on
H$_{2}$CO data). The gas kinetic temperatures in the north of OMC-1 agree well with our W\,40 region. The dust temperatures of
our sample are obtained from spectral energy distribution (SED) fitting to $Herschel$ HiGal data at 70, 160, 250, 350,
and 500 $\mu$m by \citet{2010A&A...518L.102A} and \citet{2015A&A...584A..91K}. The dust
temperatures derived on $Herschel$ scales of $36\arcsec$ are smoothed to our beam size of $2\arcmin$.
In the region observed by us the dust temperatures range from 11.9 to 23.6\,K with an average
of 14.8 $\pm$ 2.8\,K. Overall, the temperatures derived from para-NH$_3$\,(2,2)/(1,1)
tend to show higher temperatures than the HiGal dust temperatures.

Most of the clumps analyzed in this study lie in the optimal range of precise $T_{\rm kin}$
determination when using NH$_3$\,(1,1) and (2,2) lines (see Sects.\,\ref{sect:Introduction} and \ref{sect-3-3}).
Fig.\,\ref{fg8} indicates that there is a large number of cold clumps with $T_{\rm kin}$\,$<$\,20\,K. The gas and
dust are expected to be coupled at densities above about 10$^{4.5}$ or 10$^{5}$ cm$^{-3}$
\citep{2001ApJ...557..736G,2004ApJ...614..252Y}. The temperatures derived from dust and gas are
often in agreement in the active and dense clumps of Galactic disk clouds \citep{2010ApJ...717.1157D,2013A&A...556A..16G,2014ApJ...786..116B,2019MNRAS.483.5355M}.
This is also the case for Serpens South. Low gas temperatures associate with Serpens South ranging
from 8.9 to 16.8\,K with an average of 12.3\,$\pm$\,1.7\,K, which is consistent with the mean value of 11\,$\pm$\,1\,K
found by \citet{2016ApJ...833..204F}. The gas and dust temperatures (mean and standard deviations
$T_{\rm gas,avg}$\,$\sim$\,12.3\,$\pm$\,1.7\,K versus $T_{\rm dust,avg}$\,$\sim$\,13.4\,$\pm$\,0.9\,K)
scatter in Serpens South, but agree reasonably well as can be most directly seen in the right panel
of Fig.\,\ref{fg8} (blue points). In the high mass star formation region W\,40, however, we find that
the measured gas kinetic temperatures are higher than the dust temperatures (mean and standard
deviations $T_{\rm gas,avg}$\,$\sim$\,25.1\,$\pm$\,4.9\,K versus $T_{\rm dust,avg}$\,$\sim$\,19.1\,$\pm$\,2.2\,K),
which indicates that the gas and dust are not well--coupled in W\,40 and that the dust
can cool more efficiently than the gas. This is consistent with the relatively weak NH$_3$ lines associated with
the core region of W\,40, indicating the presence of only small amounts of dense gas. This
illustrates that the interplay between gas and dust cooling/heating is not uniform in the area
covered by our observations. Such a difference between $T_{\rm gas}$ and $T_{\rm dust}$ is also seen in
other regions and appears to be an often encountered property of massive star formation regions.
\citet{2014ApJ...786..116B} and \citet{2015A&A...580A..68K} compare gas and dust temperatures in the massive
star forming infrared dark cloud G32.02+0.05 and the high mass star forming PDR S140, respectively, and find
similar discrepancies between gas and dust temperatures. This likely indicates a lack of coupling
between the gas and dust \citep{2014ApJ...786..116B} or could be due to the clouds being clumpy \citep{2015A&A...580A..68K}.
This may be potential mechanisms relevant to W\,40, where the gas temperature higher than the dust temperature.

\subsection{Thermal and non-thermal motions}
\label{sect-4-3}
Previous observations of NH$_3$ and H$_{2}$CO in Galactic star-forming regions
(e.g.,\citealt{1988A&A...203..367W,1996A&A...308..573M,1999ApJS..125..161J,2006A&A...450..607W,
2011MNRAS.418.1689U,2015MNRAS.452.4029U,2012A&A...544A.146W,2014ApJ...790...84L,
2017A&A...598A..30T,2018A&A...609A..16T,2018A&A...611A...6T}) suggest that the line width
is correlated with kinetic temperature. It implies that the correlation between line width
and kinetic temperature is due to the dissipation of turbulent energy.

\begin{figure}[t]
\vspace*{0.2mm}
\centering
   \includegraphics[width=0.486\textwidth]{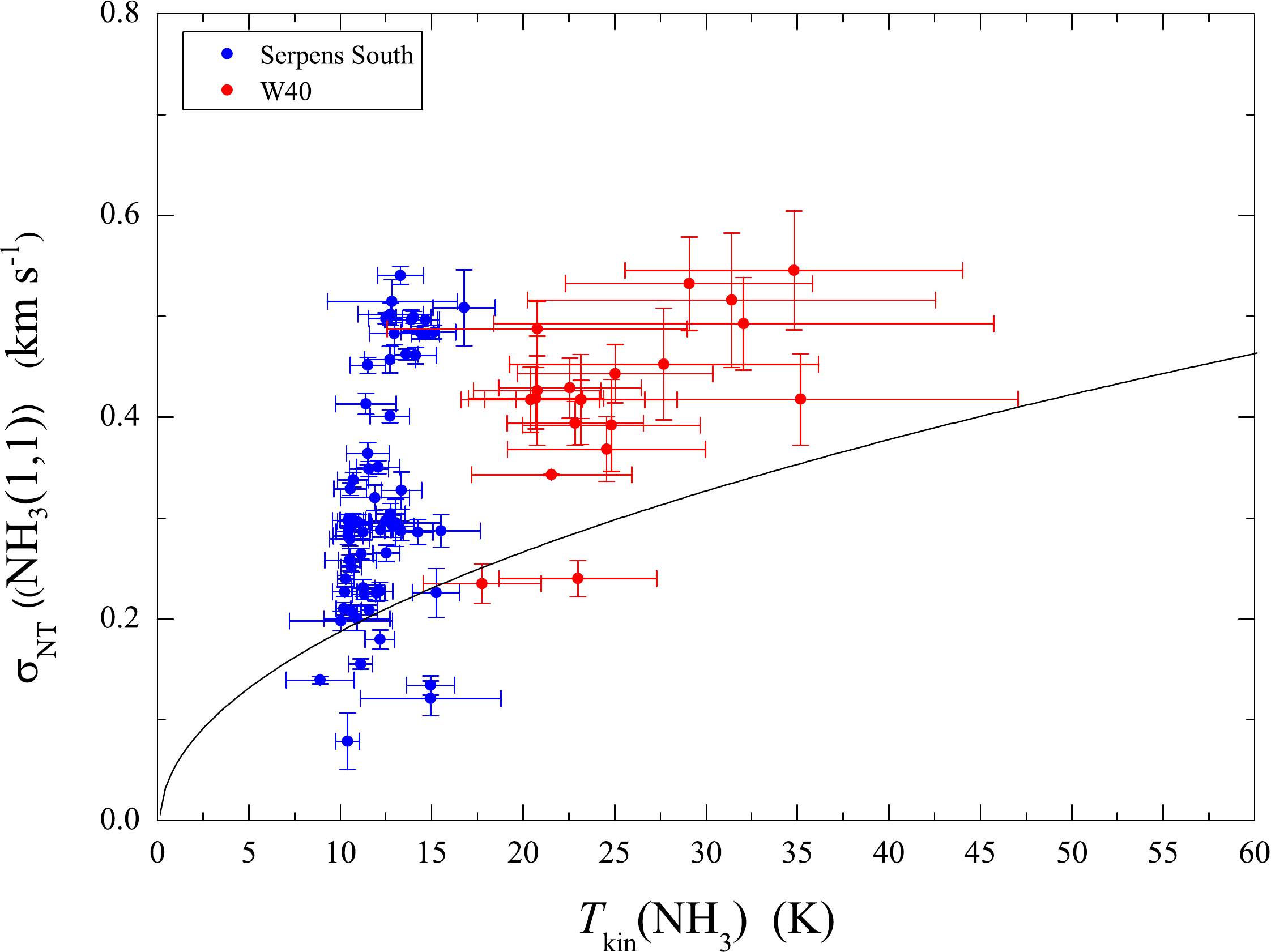}
  \caption{Non-thermal velocity dispersion ($\sigma_{\rm NT}$) vs. gas kinetic temperature derived
  from para-NH$_3$\,(1,1) for Serpens South (blue points) and W\,40 (red points).
  The total dispersion of individual hyperfine structure (hfs) components are derived from
  the GILDAS built-in `NH$_3$(1, 1)' fitting method for the (1,1) line. The black lines represent the thermal sound speed.}
   \label{fg9}
   \end{figure}

Here we examine whether there is a relationship between turbulence and temperature in
our survey area. We computed thermal velocity ($v_{\rm th}$), non-thermal velocity dispersion
($\sigma_{\rm NT}$), and thermal sound speed ($c_{\rm s}$). $v_{\rm th}$,
$\sigma_{\rm NT}$, and $c_{\rm s}$ are defined in Appendix A of \citet{2014A&A...567A..78L}. For $v_{\rm th}$ we have
\begin{eqnarray}
v_{\rm th}=\sqrt{\frac{2k_{\rm B}T_{\rm kin}}{m}}\ \ \,\rm km\,s^{-1},
\end{eqnarray}
where $k_{\rm B}$ is the Boltzmann constant and $m$ is the mass of a particle. $T_{\rm kin}$
is the kinetic temperature of the gas. For NH$_3$ , it is
\begin{eqnarray}
v_{\rm th}=0.03 \sqrt{T_{\rm kin}}\ \ \,\rm km\,s^{-1}.
\end{eqnarray}
The thermal contribution to the observed line with $v_{\rm obs}$ is related to the so far discussed full width to half maximum (FWHM) line width by (see Appendix \ref{Appendix-C})
\begin{eqnarray}
\label{Eq8}
v_{\rm obs} = \Delta v/2\sqrt{\ln 2}\ \ \,\rm km\,s^{-1}.
\end{eqnarray}
The \boldmath $v_{\rm obs}$ \unboldmath can be divided into a thermal and a turbulent part by
\begin{eqnarray}
v_{\rm obs} = \sqrt{v_{\rm th}^2 + v_{\rm turb}^2}\ \ \,\rm km\,s^{-1},
\end{eqnarray}
where \boldmath $\Delta v$ \unboldmath is the FWHM line width of the NH$_3$(1, 1) line,
obtained from the NH$_3$(1, 1) fit in CLASS (see Sect. \ref{sect-2-2}) and $\sigma_{\rm NT}$
is the non-thermal velocity dispersion along the line of sight
\begin{eqnarray}
\sigma_{\rm NT} = v_{\rm turb} / \sqrt{2}\ \ \,\rm km\,s^{-1}.
\end{eqnarray}
This value can be compared with the thermal sound speed
\begin{eqnarray}
c_{\rm s} = \sqrt{\frac{k_{\rm B}T_{\rm kin}}{\mu m_{\rm H}}} \ \ \,\rm km\,s^{-1}.
\end{eqnarray}
where $\mu$ = 2.37 is the mean molecular weight for molecular clouds \citep{2016ApJ...819...66D} and $m_{\rm H}$ is
the mass of the hydrogen atom. Comparisons of velocity dispersion and kinetic temperature are
shown in Fig.\,\ref{fg9}. The thermal velocity of NH$_3$\,(1,1) for lines detected at a $>$5$\sigma$ level
ranges from 0.09 to 0.18\,km\,s$^{-1}$ with an average of 0.11\,$\pm$\,0.02\,km\,s$^{-1}$. The non-thermal velocity
dispersion of NH$_3$\,(1,1) ranges from 0.07 to 0.55\,km\,s$^{-1}$ with an average of
0.34\,$\pm$\,0.12\,km\,s$^{-1}$. The derived non-thermal motions of NH$_3$ are
much higher than the thermal line widths of our survey area. This implies that the line broadening of
NH$_3$ is dominated by non-thermal motions in these clumps. The BGPS sources which contain both starless and active
star forming massive cores observed by \citet{2011ApJ...741..110D} show an average of thermal velocity and non-thermal velocity dispersion of 0.12\,$\pm$\,0.02\,km\,s$^{-1}$ and 0.76\,$\pm$\,0.48\,km\,s$^{-1}$, respectively. \citet{2012A&A...544A.146W} determined averages of 0.14\,$\pm$\,0.02\,km\,s$^{-1}$ and 0.90\,$\pm$\,0.40\,km\,s$^{-1}$ for thermal velocity and non-thermal velocity dispersions
in the cold high mass clumps of the ATLASGAL survey. The average thermal velocity of our NH$_3$\,(1,1) data agrees with previous results observed in massive star-forming clumps \citep{2011ApJ...741..110D,2012A&A...544A.146W}, but the non-thermal velocity dispersions are smaller than their values.

We also calculated the thermal to non-thermal pressure ratio
($R_{\rm P}$=$c_{\rm s}^{2}$/$\sigma_{\rm NT}^{2}$; \citealt{2003ApJ...586..286L})
and Mach number (given as $M$=$\sigma_{\rm NT}$/$c_{\rm s}$). The sound speed
ranges from 0.18 to 0.36\,km\,s$^{-1}$ with an average of 0.23\,$\pm$\,0.04\,km\,s$^{-1}$.
The thermal to non-thermal pressure ratio in the gas traced by NH$_3$\,(1,1) ranges from 0.16 to 6.07
with an average of 0.67\,$\pm$\,0.79. The BGPS sources observed by \citet{2011ApJ...741..110D} show
the $R_{\rm P}$ values vary from 0.02 to 5.06 with an average of 0.20\,$\pm$\,0.33. NH$_3$ samples observed by
\citet{2012A&A...544A.146W} show values of 0.01--0.57 with an average of 0.10\,$\pm$\,0.06. Our average value
of thermal to non-thermal pressure ratio is higher than previous results observed by \citep{2011ApJ...741..110D} and \citep{2012A&A...544A.146W}.
We find that the Mach number for NH$_3$\,(1,1) ranges from 0.41 to 2.47 with an average of 1.49\,$\pm$\,0.45.
The BGPS sources observed by \citet{2011ApJ...741..110D} yield a mean Mach number of 3.2\,$\pm$\,1.8. The average value of the Mach
number we derive from NH$_3$ may also be below the result (3.4\,$\pm$\,1.1) of the various stages of high-mass star
formation clumps with strong NH$_3$ emission from the ATLASGAL survey \citep{2012A&A...544A.146W}. Nevertheless,
all this suggests that non-thermal pressure and supersonic non-thermal motions (e.g., turbulence, outflows, shocks,
and/or magnetic fields) are dominant in the dense gas traced by NH$_3$ in the Aquila region.

The derived values of \boldmath $v_{\rm th}$ \unboldmath, $\sigma_{\rm NT}$, $c_{\rm s}$, $R_{\rm P}$,
and Mach number for Serpens South and W\,40 are listed separately in Table\,\ref{table2}.
We have calculated average non-thermal line widths of NH$_3$\,(1,1) for those positions with NH$_3$\,(1,1) >5$\sigma$ features
in the subsamples consisting of Serpens South and W\,40. For NH$_3$\,(1,1), the average non-thermal line
widths $\sigma_{\rm NT}$ are with 0.32 $\pm$ 0.12\,km\,s$^{-1}$ and 0.41 $\pm$ 0.08\,km\,s$^{-1}$,
respectively, quite similar. The average non-thermal line widths of NH$_3$\,(1,1) calculated here are
consistent with those found in Serpens South by \citet{2016ApJ...833..204F}. They suggested that much of the
dense gas in Serpens South has subsonic or trans-sonic non-thermal motions, while the mean $\sigma_{\rm NT}$ across
the region is similar to the expected $\sim$ 0.2\,km\,s$^{-1}$ sound speed at 11\,K.

The similar non-thermal line widths between the W\,40 region and
Serpens South are surprising. Either the entire region is already strongly
affected by the consequences of massive star formation (e.g., through outflows and shocks)
based on activity related to W\,40 or the young low mass stars in Serpens South
are numerous enough to induce turbulent motions at a similar degree as the more massive stars
in W\,40 that may have dissociated or expelled most of the dense molecular gas in their vicinity.
Figure.\,\ref{fg9} shows that turbulent heating considerably contributes to gas temperature in these clumps.

\subsection{Radiative heating}
\label{sect-4-4}
Previous SCUBA-2 450 and 850\,$\mu$m observations of the W\,40 complex in the
Serpens-Aquila region \citep{2016MNRAS.460.4150R} provide evidence for radiative heating.
 The W\,40 complex represents a high-mass star-forming region
dominated by an OB association that is powering an H\,{\scriptsize II} region. OS1a is
the most luminous star in the W\,40 complex \citep{2016MNRAS.460.4150R}, but heating by
the associated nearby (D\,$\leq$\,0.1\,pc) sources IRS1, IRS1a, IRS2, IRS2a, IRS2b, IRS3, and IRS3a
(their positions are indicated in Fig.\,\ref{fg6}) is also likely playing a role.

\begin{figure}[t]
\vspace*{0.2mm}
\centering
\includegraphics[width=0.48\textwidth]{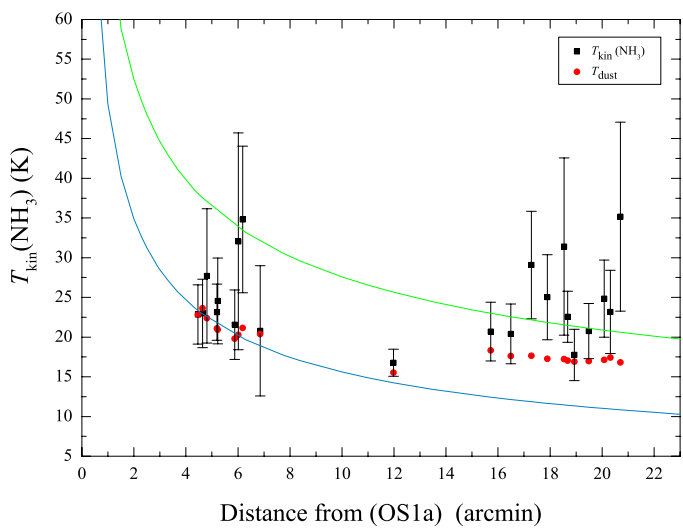}
\caption{Gas kinetic temperature derived from para-NH$_3$ (black points) and dust
temperature (red points) along the W\,40 region (projected distance from its main
stellar source OS1a, $l$\,=\,28.79$\degr$, $b$\,=\,3.49$\degr$). The
blue and green lines are the expected relationships
from a Stefan-Boltzmann law and modified Stefan-Boltzmann law (see Sect. \ref{sect-4-4}),
respectively, assuming OS1a is the dominant source with an approximate luminosity of
$10^{5.25}$\,${\rm L}_{\odot}$ \citep{2010ApJ...725.2485K,2016MNRAS.460.4150R}.}
\label{fg10}
\end{figure}

We investigate the relationship between gas kinetic temperature and projected distance $R$ from
the central part of W\,40 (OS1a, $l$\,=\,28.79$\degr$, $b$\,=\,3.49$\degr$)
in Fig.\,\ref{fg10}. As is shown in the figure and as it was already mentioned before (Sect.\,\ref{sect-4-2}),
dust temperatures are lower than gas temperatures derived from NH$_{3}$ in W\,40.

It is expected that the gas temperature and distance relation from the Stefan-Boltzmann blackbody radiation law is
$T_{\rm kin}$\,=\,100$\,\times$\,($\frac{7.4\times10^{-5}}{R}$)$^{\frac{1}{2}}$\,($\frac{L}{{\rm L}_{\odot}}$)$^{\frac{1}{4}}$\,K,
adopting a molecular cloud distance of 436\,pc \citep{2017ApJ...834..143O,2018ApJ...869L..33O} and assuming that OS1a is the
dominant source with an approximate luminosity of $10^{5.25}$\,${\rm L}_{\odot}$ \citep{2010ApJ...725.2485K,1998ApJ...502..676W}.
For the emissivity of the dust grains smaller than the wavelength at the characteristic blackbody temperature, the radiation law
is adjusted to $T_{\rm kin}$\,=\,100$\,\times$\,($\frac{1.2\times10^{-4}}{R}$)$^{\frac{2}{5}}$($\frac{L}{{\rm L}_{\odot}}$)$^{\frac{1}{5}}$\,K
\citep{1998ApJ...502..676W}.

The two radiation models for gas heating (Stefan-Boltzmann blackbody radiation and its modification related to dust emissivity)
are both not well supported by our para-NH$_3$ data (Fig.\,\ref{fg10}) that exclude positions related to Serpens South.
Contrary to our expectation, temperatures near $\sim$18$\arcmin$ ($\sim$2\,pc) offsets appear to be slightly
higher than those at $\sim$5$\arcmin$ offsets. This holds for temperatures derived by both the gas and dust and is not in
agreement with the declining dust temperatures found by \citet{2016MNRAS.460.4150R}, their Fig.\,15 as a function of distance from
the main stellar source of W\,40. For NH$_3$ this may imply that the dense molecular gas in the vicinity of W\,40 has been destroyed
by UV radiation (Sect.\,\ref{sect-4-1}) and that the projected angular distances in Fig.\,\ref{fg10} may be significantly below the
real distances. In any case, dust grain mantle evaporation as seen in hot cores (e.g. in Orion-KL) and leading to very high ammonia
column densities is not seen. W\,40 might have gone through such a phase of evolution, but UV photons may have destroyed its short-lived
chemical consequences (for models, see e.g., \citealt{1992ApJ...399L..71C}) in the meantime. As discussed in Sect.\,\ref{sect-4-1},
Serpens South shows a fractional total-NH$_3$ abundance with an average of 8.6\,($\pm$3.8)\,$\times$\,10$^{-8}$. In W\,40, values are lower 1.6\,($\pm$1.4)$\,\times$\,10$^{-8}$, which is a factor of $\sim$5 below the result from Serpens South.

The nature of the molecular ridge, not very pronounced in the dust continuum maps of Fig.\,\ref{fg1} but
dominating the maps of NH$_3$ emission in Figs.\,\ref{fg3}, \ref{fg4}, and \ref{fg6} is still poorly defined.
It could represent a PDR similar to the Orion bar northwest of the Trapezium stars \citep{2018A&A...609A..16T}.
Or it could represent swept up material near the edge of an expanding H\,{\scriptsize II} region. Since radial
velocities (see Fig.\,\ref{fg4}) do not hint at a significant difference between W\,40 and Serpens South, such material
would have been swept up mostly along the plane of the sky. Additional molecular surveys addressing the detailed
chemistry of this region could shed more light onto this puzzle.

\subsection{Kinematics of the dense gas}
\label{sect-4-5}
To study the velocity pattern of our observed area in more detail we fitted the velocities of the entire
measured region, Serpens South, and W\,40. There may be weak velocity gradients that
run along the entire observed region, Serpens South, and W\,40 (see left panel of Fig.\,\ref{fg4}).
Following the steps of \citet{1993ApJ...406..528G} and \citet{2018A&A...616A.111W}, we fitted
the velocity adopting a linear form: $V_{LSR} = v_{0} + a\times\Delta l + b\times\Delta b$, where
$v_{0}$ is the systemic velocity of the cloud, $\Delta l$ and $\Delta b$ are offsets in Galactic
longitude and Galactic latitude. The velocity gradient can be derived as
$\nabla v = (a^{2} + b^{2})^{0.5}/D $, where distance $D$ = 436 pc (Sect.\,\ref{sect:Introduction}).
This distance was used for all spectra with peak line fluxes larger than 5$\sigma$. This linear
velocities fitting is based on a solid body approximations for Serpens South and W\,40, Serpens South
alone, and W\,40 alone (see Table\,\ref{table:parameters} for the adopted extent of those regions).
The derived gradients are then 0.16\,$\pm\,0.01$, 0.27\,$\pm\,0.01$, and 0.38\,$\pm\,0.01$\,km\,s$^{-1}$\,pc$^{-1}$
for the observed region shown in Fig.\,\ref{FgD.1}, Serpens South, and W\,40, respectively.

To check the assumption of solid body rotation we study the fitted velocity and velocity residuals
($V_{\rm obs}$\,$-$\,$V_{\rm fit}$) in Appendix \ref{Appendix D} for the region encompassing
both Serpens South and W\,40. The black solid polyline indicates a potential rotation axis ($0.019\Delta l+0.007\Delta b-0.328=0$,
in arcmin) in the fitted velocity map of the entire observed region (Fig.\,\ref{FgD.1} left panel), and
three black parallel lines perpendicular to this rotation axis show the direction of the velocity gradient. Velocities
tend to be lower in the lower right of the left panel of Fig.\,\ref{FgD.1}. For Serpens South alone
(see Table\,\ref{table:parameters} for the extent of the region), the rotation axis is
$0.03\Delta l-0.01\Delta b-0.12=0$, and for W\,40, it is $0.04\Delta l-0.03\Delta b-0.88=0$,
in arcmin. From Fig.\,\ref{FgD.1}, we can see that most of the velocity
residuals are distributed in the range $V_{\rm obs}$ -- $V_{\rm fit}$ = --1 to 1 km\,s$^{-1}$
which is even more clearly shown in Fig.\,\ref{FgD.2}. To be specific, $90\%$ of the
velocity residuals of the observed region shown in Fig.\,\ref{FgD.1}, Serpens South,
and W\,40 are in the range $V_{\rm obs} - V_{\rm fit}$\,<\,0.65, 0.56, and 0.79 km\,s$^{-1}$,
respectively. Larger velocity residuals are mainly located, in terms of Galactic coordinates,
at the western and northeastern part of Serpens South and the southern part of W\,40.

A maximum velocity gradient of 0.38\,$\pm\,0.01$\,km\,s$^{-1}$\,pc$^{-1}$ for W\,40 can be considered as an upper limit
for an estimate of the rotational energy. Taking Serpens South and W\,40 as a rigidly rotating cylinder, the
ratio of rotational to gravitational energy is given \citep{1992ApJ...400..579B,2018A&A...616A.111W} by
\begin{equation}
\begin{aligned}
\beta = \frac{E_{rot}}{E_{grav}}=\frac{1}{2} \frac{ML^2}{12}(1+\frac{3}{2x^2}) \omega ^2 / \frac{3}{2}\frac{GM^2}{L}f(x)  \approx \frac{\omega ^2}{4\pi G\rho}\frac{L^2}{9R^2}\ \,,
\end{aligned}
\end{equation}
where $G$ is the gravitational constant and $\rho$, $L$, $R$, and $\omega$ are the density, height, radius, and the angular velocity of the cylinder.
This can be simplified to
\begin{equation}
\begin{aligned}
\beta = \frac{3.0 \times 10^{-3}  \omega_{-14}^{2}}{n_{4}}  \frac{L^2}{9R^2}\ \,,
\end{aligned}
\end{equation}
where, $\omega_{-14}$ is the angular velocity in units of $10^{-14} s^{-1}$ and the gas density $n_{4}$ is in units of
$10^{4}$ cm$^{-3}$ \citep{2013A&A...553A..58L}. For the entire observed region, Serpens South, and W\,40, we adopt angular
velocities, which are equal to the velocity gradients calculated before, 0.51, 0.88, and $1.23 \times 10^{-14} s^{-1}$ and
an average density of the molecular Aquila complex of $10^{4}$ cm$^{-3}$ \citep{2013A&A...553A..58L}. Then $\beta$ $\lesssim$
0.003, 0.01, and 0.02, respectively. It is clear that in the entire observed region, Serpens South, and W\,40, rotation is
presently playing a negligible role. The rotational energy is a very small fraction of the gravitational energy.

\citet{2017ApJ...834..143O,2018ApJ...869L..33O} obtained with parallax measurements using the Very Large Baseline Array
(VLBA) distances slightly in excess of 400\,pc for both Serpens South and the W\,40 region, which show (see Figs.\,\ref{fg1},
\ref{fg3}, \ref{FgB.1}) together an arc-like morphology. However, including the weaker region 5 (Fig.\,\ref{fg1}) and the high
and low Galactic latitude regions of region 4 (again Fig.\,\ref{fg1}), we encounter a ring-like morphology, possibly forming a
shell of gas. The basic question is whether the weakly emitting high Galactic longitude part of this ring is located at
a similar distance to that obtained by \citet{2017ApJ...834..143O,2018ApJ...869L..33O}. While this remains an open question,
we note that this high longitude part of the putative shell shows about the same radial velocities as Serpens South and W\,40
(Sect.\,\ref{sect-3-1}). With respect to any Galactic kinematical model, clouds with similar radial velocity, being not more
than one degree apart, should be located at similar distances.

While we cannot prove that similar kinematic distances correspond in our case to a physical connection, it is nevertheless worthwhile
to analyse the observed structure with this assumption, noting that a similar morphology is also seen in $^{13}$CO
(which likely traces lower density gas than NH$_{3}$) in Fig.\,10 of \citet{2020ApJ...893...91S} and in the H$_{2}$ column
density map of \citet{2010A&A...518L.106K}. The ring-like large scale morphology of both the dust (Fig.\,\ref{fg1}) and
the $^{13}$CO and NH$_{3}$ emitting gas (Fig.\,15 of \citealt{2019ApJS..240....9S} and our Figs.\,\ref{fg3} and \ref{FgB.1})
is reminiscent of an expanding interstellar bubble with large amounts of dense gas on one and lesser amounts of such gas on
the other side.

Since velocities in regions 1 and 5 (see Fig.\,\ref{fg1}) are similar (Sect. \ref{sect-3-1}), they might represent those parts of
the putative shell, which are expanding parallel to the plane of the sky. Then, however, velocities inside the rim of this structure
should be different, more redshifted on the back and more blueshifted on the front side. For all three rotation axes crossing the
low Galactic longitude side of the shell-like structure, lower velocities and blueshifted areas are located at lower, while higher
velocities and redshifted areas are located at higher Galactic longitudes. This is compatible with a radial motion of the entire
shell-like structure. It would indicate slow expansion, if the gas at higher longitudes forms a part of the back side of the shell,
or slow contraction, if the gas is part of the front side of the shell. The velocity difference is about 0.7\,km\,s$^{-1}$.

\begin{table*}
\caption{Parameters obtained from Serpens South and W\,40.}
\centering
\begin{tabular}
{lccccccc}
\hline\hline
Parameter & \multicolumn{2}{c}{Serpens South} & & \multicolumn{2}{c}{W\,40 (H\,{\scriptsize II} region)}&\\
\cline{2-3} \cline{5-6}
& Range & Mean & & Range & Mean & \\
\hline
total-$N(\rm NH_{3})$/$10^{15}$ cm$^{-2}$ & 0.03--6.4 &2.6 $\pm$ 1.4  & & 0.02--0.8  & 0.3 $\pm$ 0.2&\\
$^{a}N(\rm H_{2})$/$10^{22}$ cm$^{-2}$ &0.9--6.3  & 3.2 $\pm$ 1.4 & & 0.9--2.7 & 1.7 $\pm$ 0.5&\\
$\chi$(total-NH$_3$)/$10^{-8}$         & 0.2--21.2 & 8.6 $\pm$ 3.8 & & 0.1--4.3 & 1.6 $\pm$ 1.4 &\\
$T_{\rm gas}$/K                    & 8.9--16.8 &   12.3 $\pm$ 1.7 & & 17.7--35.5    & 25.1 $\pm$ 4.9   &  \\
$^{b}T_{\rm dust}$/K            & 11.9--15.5 &   13.4 $\pm$ 0.9 & & 16.8--23.6    & 19.1 $\pm$ 2.2  & \\
$v_{\rm th}$/\,km\,s$^{-1}$   & 0.09--0.12  &   0.10 $\pm$ 0.01  & & 0.13--0.18    & 0.15 $\pm$ 0.01 &  \\
$\sigma_{\rm NT}$/\,km\,s$^{-1}$   & 0.07--0.54  &   0.32 $\pm$ 0.12 & & 0.23--0.55    & 0.41 $\pm$ 0.08    &  \\
$c_{\rm s}$/\,km\,s$^{-1}$         &         0.18--0.25   &          0.21 $\pm$ 0.01   & &         0.25--0.36    & 0.29 $\pm$ 0.03 &  \\
$R_{\rm P}$                        & 0.16--6.07 &   0.70 $\pm$ 0.90  & & 0.31--1.43    & 0.57 $\pm$ 0.27 &  \\
$M$                                & 0.41--2.47 &   1.51 $\pm$ 0.50  & & 0.83--1.78    & 1.39 $\pm$ 0.23 &  \\
\hline 
\end{tabular}
\label{table2}
\tablefoot{$^{a}$Molecular hydrogen column densities and $^{b}$dust temperatures are taken from
\cite{2010A&A...518L..85B} and \cite{2015A&A...584A..91K}. The errors shown are the standard deviations of the mean.
For the areas adopted for Serpens South and W\,40, see Table\,\ref{table:parameters}.}
\end{table*}

For the kinematic age of the putative expanding shell we obtain with
\begin{equation}
t\,=\,\sqrt{(r_{x}\,\times\,r_{y})\,/\,v_{\rm exp}}\ \ ,
\end{equation}
$r_{x}$ and $r_{y}$ $\sim$ 3.5\,pc, and $v_{\rm exp}$\,=\,0.7\,km\,s$^{-1}$ a time scale of several million years. We note,
however, that the 0.7\,km\,s$^{-1}$ are for this approach only a lower limit: (1) We do not see gas along the center of
the putative shell, where the discrepancy in velocity relative to its rim should be largest; (2) the expansion of the gas may
have significantly slowed down due to entrainment of ambient gas originally not participating in this expansion.

In this context it may be worth noting that the center of the putative shell is not known to host a supernova remnant (W. Reich,
prov. comm.). More sensitive data, in particular from the high Galactic longitude parts of the ring-like structure (see Figs.\,\ref{fg1}
and \ref{fg3}) would be helpful to further investigate related scenarios.

\section{Summary}
\label{sect:summary}
We have mapped the western part of the Aquila Rift cloud complex in the NH$_3$\,(1,1) and (2,2) transitions, which
includes the densest regions of Serpens South and W\,40. The main results of this work are the following:
\begin{enumerate}
\item
The NH$_3$ morphology, revealing the distribution of the dense gas ($n$(H$_{2}$)\,$\gtrsim$\,10$^{-3}$\,cm$^{-3}$),
is correlated with that of the $Herschel$ infrared dust emission and shows an overall similar structure with two exceptions:
There is, in Galactic coordinates, the well known NE-SW ridge of dense gas, the core of the Serpens South region,
that is much more dominant in NH$_3$ than in the dust emission. W\,40 dominates the dust
emission but is much less pronounced in NH$_3$. The ridge with strong NH$_3$\,(1,1) emission
contains several cores along a total length of about 15$\arcmin$. Weaker emission extends further
to lower and higher Galactic latitudes and toward regions at larger longitudes. Overall, the
NH$_3$ distribution forms like the dust a circle with a diameter of 50$\arcmin$ --
60$\arcmin$ ($\sim$ 7\,pc), with comparatively strong emission at low and weak emission at high longitudes.

\item
The kinetic temperature of the dense gas in the Aquila rift cloud complex measured by
NH$_3$\,(2,2)/(1,1) line ratios ranges from 8.9 to 35.0\,K with an average of 15.3\,$\pm$\,6.1\,K.
The high mass star-forming region of W\,40 has a gas kinetic temperature $\sim$25\,K,
which is twice that ($\sim$12\,K) in the low mass star formation region of Serpens South.

\item
Fractional abundances of total-NH$_3$ vary from 0.1\,$\times$\,$10^{-8}$ to 2.1\,$\times$\,$10^{-7}$ with an
average of 6.9\,($\pm$4.5)\,$\times$\,10$^{-8}$. Serpens South has total NH$_3$ fractional abundances
also ranging from 0.2\,$\times$\,$10^{-8}$ to 2.1\,$\times$\,$10^{-7}$ with an average of
8.6\,($\pm$3.8)\,$\times$\,10$^{-8}$. Lower values of 0.1\,--\,4.3\,$\times$\,$10^{-8}$
with an average of 1.6\,($\pm$1.4)$\,\times$\,10$^{-8}$ characterize W\,40, which is a factor
of $\sim$5 below the result from Serpens South.

\item
A comparison of kinetic temperatures derived from NH$_3$ and dust emission indicates that gas
and dust temperatures are in agreement in Serpens South, but gas temperatures are higher than
those of the dust in W\,40. This suggests that gas and dust are coupled in Serpens South,
but not in W\,40.

\item
Dense gas traced by NH$_3$ is significantly influenced by supersonic non-thermal motions.
Similar levels of non-thermal turbulence are encountered in W\,40 and Serpens South. This
may either be caused by the fact that the entire region is already strongly affected by the
consequences of massive star formation based on activity related to W\,40 or that a similar
amount of turbulence is triggered by the large number of forming low mass stars in Serpens
South. W\,40 appears to have dissociated or expelled most dense molecular gas in its vicinity.

\item
The non-thermal velocity dispersion of NH$_3$\,(1,1) is positively correlated with
the gas temperature, which indicates that the dense gas may be heated by dissipation
of turbulent energy.

\item
Velocity gradients of 0.16\,$\pm\,0.01$, 0.27\,$\pm\,0.01$, and 0.38\,$\pm\,0.01$\,km\,s$^{-1}$\,pc$^{-1}$ appear to
be present in the observed region shown in Fig.\,\ref{FgD.1}, Serpens South alone, and W\,40 alone, respectively.
The rotational and gravitational energy of the Aquila region are compared by using those velocity gradients. For
the entire observed region, Serpens South, and W\,40, ratios are about $\beta$ $\lesssim$ 0.003, 0.01, and 0.02,
respectively. This demonstrates that the rotational energy is a negligible fraction of the gravitational energy.

\item
The morphology of the entire studied region can be described by a ring or shell with a diameter of about 7\,pc and
strong emission at lower and weak emission at higher Galactic longitudes. However, this only holds in a physical sense,
if all parts are located at approximately the same distance. This is presently only known for the lower Galactic longitude
part with its strong emission. While we find velocity gradients in radial direction at the low longitude side of the putative
shell, it is therefore not yet clear whether this indicates a systematic expansion (or contraction) of the entire structure.
\end{enumerate}

\begin{acknowledgements}
We would like to thank the anonymous referee for the
useful suggestions that improved this study. The authors are thankful for helpful comments by Wolfgang Reich.
We thank the staff of the Nanshan 26-m radio telescope for their assistance during the observations.
This work is based on measurements made with the Nanshan 26-m radio telescope, which is
operated by the Key Laboratory of Radio Astronomy, Chinese Academy of Sciences. This work was funded by
the National Natural Science foundation of China under grant 11433008, 11903070, 11603063,
11703074, 11703073, and 11973076, the Heaven Lake Hundred-Talent Program of Xinjiang
Uygur Autonomous Region of China, and the CAS "Light of West China" Program under Grant 2018-XBQNXZ-B-024,
2016-QNXZ-B-23, and 2016-QNXZ-B-22. C. H. acknowledges support by a Chinese Academy
of Sciences President's International Fellowship Initiative for visiting scientists (2021VMA0009).
This research has used NASA's Astrophysical Data System (ADS).
\end{acknowledgements}

\Online
\begin{appendix} 
\newpage
\twocolumn
\section{Calibration stability}
\label{Appendix A}
\begin{figure}[h]
\includegraphics[width=0.48\textwidth]{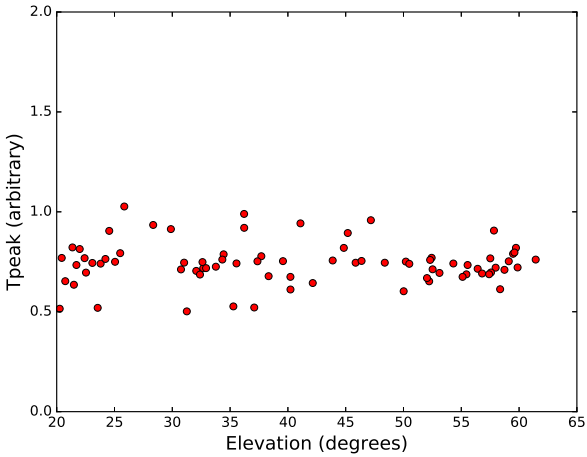}
\caption[]{Uncorrected NH$_3$\,(1,1) main line intensities against elevation of repeated observations
toward the reference position. The position corresponds to RA: 00:36:47.51, DEC: +63:29:02.1 (J2000).
The standard deviations of the mean of the flux is about $10\%$.}
\label{FgA.1}
\end{figure}

The system temperature was calibrated against a signal injected by a noise diode. Hot (ambient temperature)
and cold (liquid nitrogen) loads determine the temperature of the noise diode. We observed the reference
position (RA: 00:36:47.51, DEC: +63:29:02.1 with (0,0) offset, J2000) to check the calibration stability
every 2-3 hours. All reference position observations were made in the OTF mode of a small area of
$6\arcmin \times 6\arcmin$. To present the peak distribution against elevation (Fig.\,\ref{FgA.1}),
we regridded the data and then fitted the NH$_3$\,(1,1) main lines (the central group of NH$_3$\,(1,1)
hyperfine components). From Fig.\,\ref{FgA.1}, we clearly see that there is no significant systematic
variation. The standard deviations of the mean of the peak intensities is about $10\%$, thus
the observational system of the Nanshan observatory is stable.

\begin{figure}[h]
\includegraphics[width=0.48\textwidth]{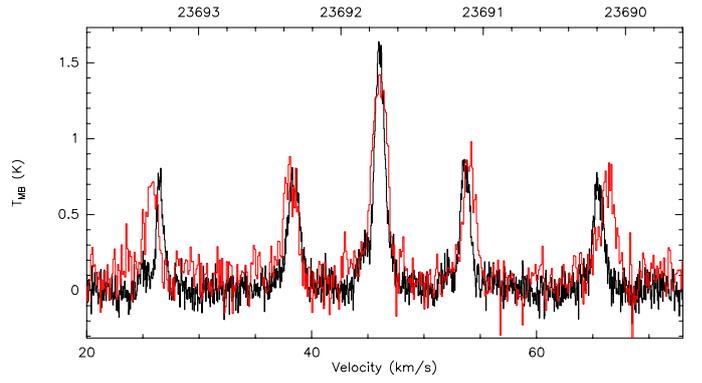}
\caption[]{The NH$_3$\,(1,1) spectra from the GBT (black) and our data (red) toward the reference position RA: 18:57:07.94, DEC: 02:10:51.40 (J2000).}
\label{FgA.2}
\end{figure}

To further check our calibration stability, the NH$_3$\,(1,1) data of G035.39-0.33 observed by the GBT
\citep{2017A&A...606A.133S} were used as a comparision with our NH$_3$ data. For the process, the GBT
data were smoothed to our beam size using the `XY\_MAP' routine in GILDAS. In Fig.\,\ref{FgA.2}, the GBT
spectrum (black) and our spectrum (red) of the reference position (RA: 18:57:07.94, DEC: 02:10:51.40, J2000)
are displayed. It is obvious that the two spectra match each other well and that the checking results are reliable.

\begin{table}[h]
\centering
\caption{Area name, central position, average noise level and covered size of the mapped regions (see also Fig.\,\ref{fg1}).}
\begin{tabular}{cccc}
\hline \hline
Area Name & Central position & Area & Noise level\\
& ($l$, $b$)  & $\rm degree \times \rm degree$ & K\\
\hline
1 &$(28.59\degr$, $3.55\degr)$ & $0.7 \times 1.1 $ & 0.05 \\
2 &$(28.75\degr$, $4.30\degr)$ & $0.4 \times 0.4 $ & 0.03 \\
3 &$(28.39\degr$, $2.81\degr)$ & $0.5 \times 0.4 $ & 0.04 \\
4 &$(29.19\degr$, $3.79\degr)$ & $0.5 \times 1.4 $ & 0.04 \\
5 &$(29.55\degr$, $3.54\degr)$ & $0.2 \times 1.1$  & 0.05 \\
6 &$(28.87\degr$, $4.65\degr)$ & $0.6 \times 0.3$  & 0.05 \\
\hline
\end{tabular}
\label{table:A.1}
\end{table}

\Online
\onecolumn
\section{Prestellar and protostellar cores and NH$_3$ spectra toward the Aquila Rift cloud complex and derived physical parameters}
\label{Appendix B}
\begin{figure*}[h]
\centerline{\hbox{
\includegraphics[width=23.0cm,height=10.6cm,angle=0]{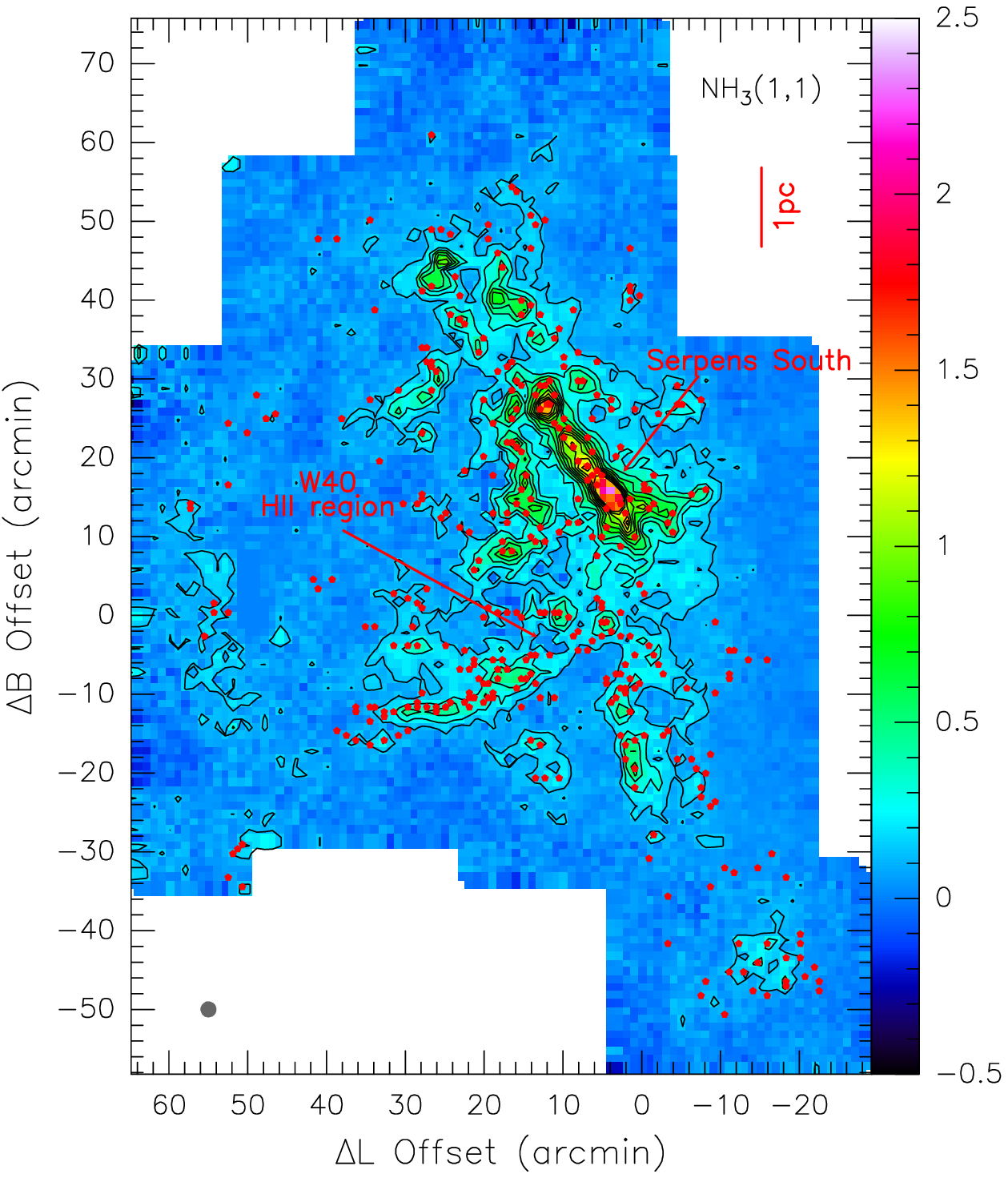}
\hspace{-13.2cm}
\includegraphics[width=8.4cm,height=10.6cm,angle=0]{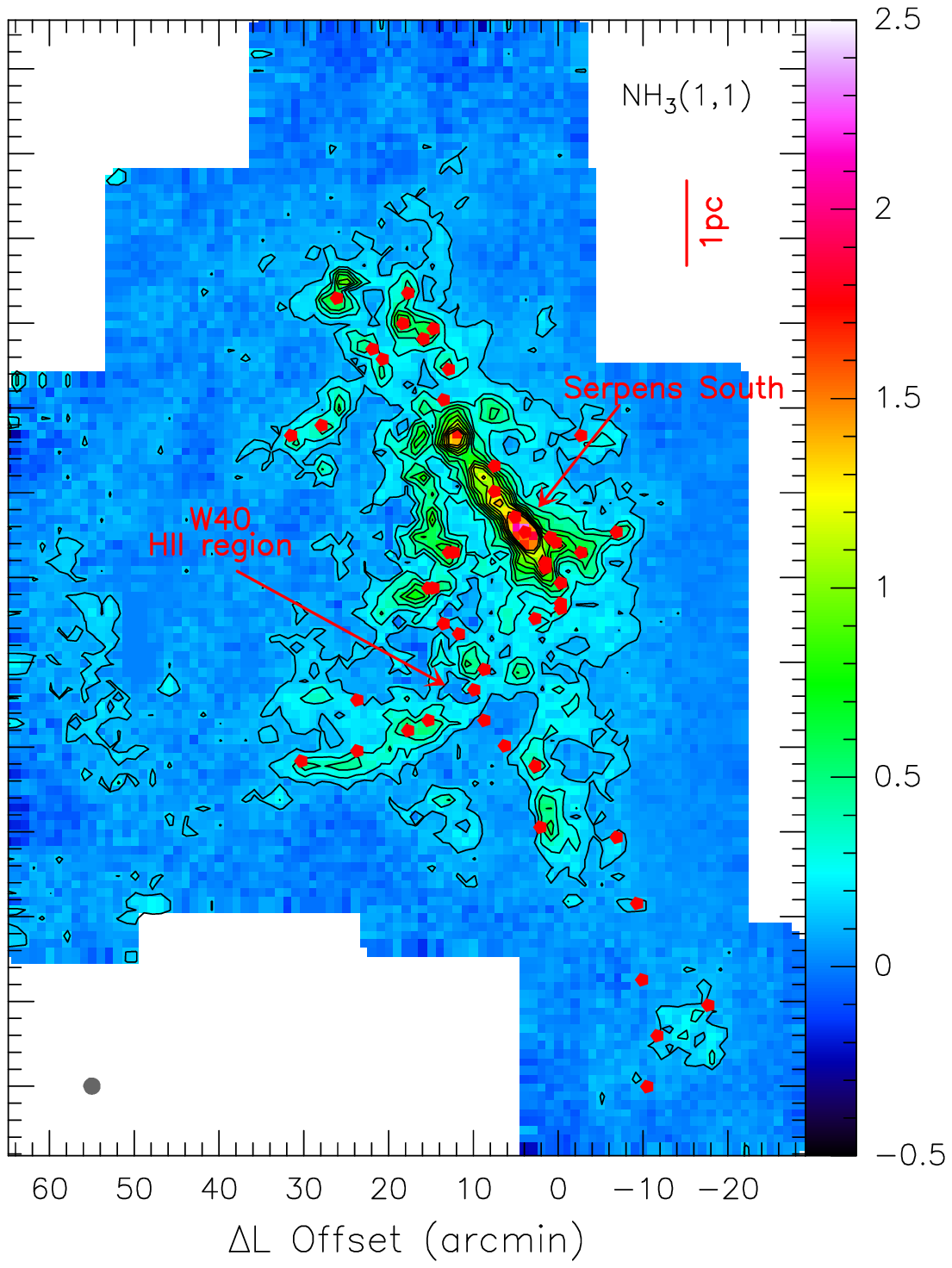}}}
\caption[]{Integrated intensity maps of NH$_3$\,(1,1) (\textit{left} and
\textit{right}), the reference position is $l$\,=\,28.59$\degr$, $b$\,=\,3.55$\degr$. The integration
range is 4\,<\,$V_{\rm LSR}$\,<\,10\,km\,s$^{-1}$. Contours start at 0.13\,K\,km\,s$^{-1}$ (3$\sigma$)
on a main beam brightness temperature scale and go up in steps of 0.13\,K\,km\,s$^{-1}$.
The unit of the color bars is K\,km\,s$^{-1}$. The half-power beam width is illustrated as a black
filled circle in the lower left corners of the images. In the left and right panel the red points show
the positions of the 362 candidate prestellar cores and 49 protostellar cores taken from \citet{2015A&A...584A..91K}, respectively.
The red line in the top right of each map illustrates the 1\,pc scale at a distance of 436\,pc \citep{2017ApJ...834..143O,2018ApJ...869L..33O}.}
\label{FgB.1}
\end{figure*}

\begin{figure*}[h]
\centering
\includegraphics[width=0.49\textwidth,angle=0]{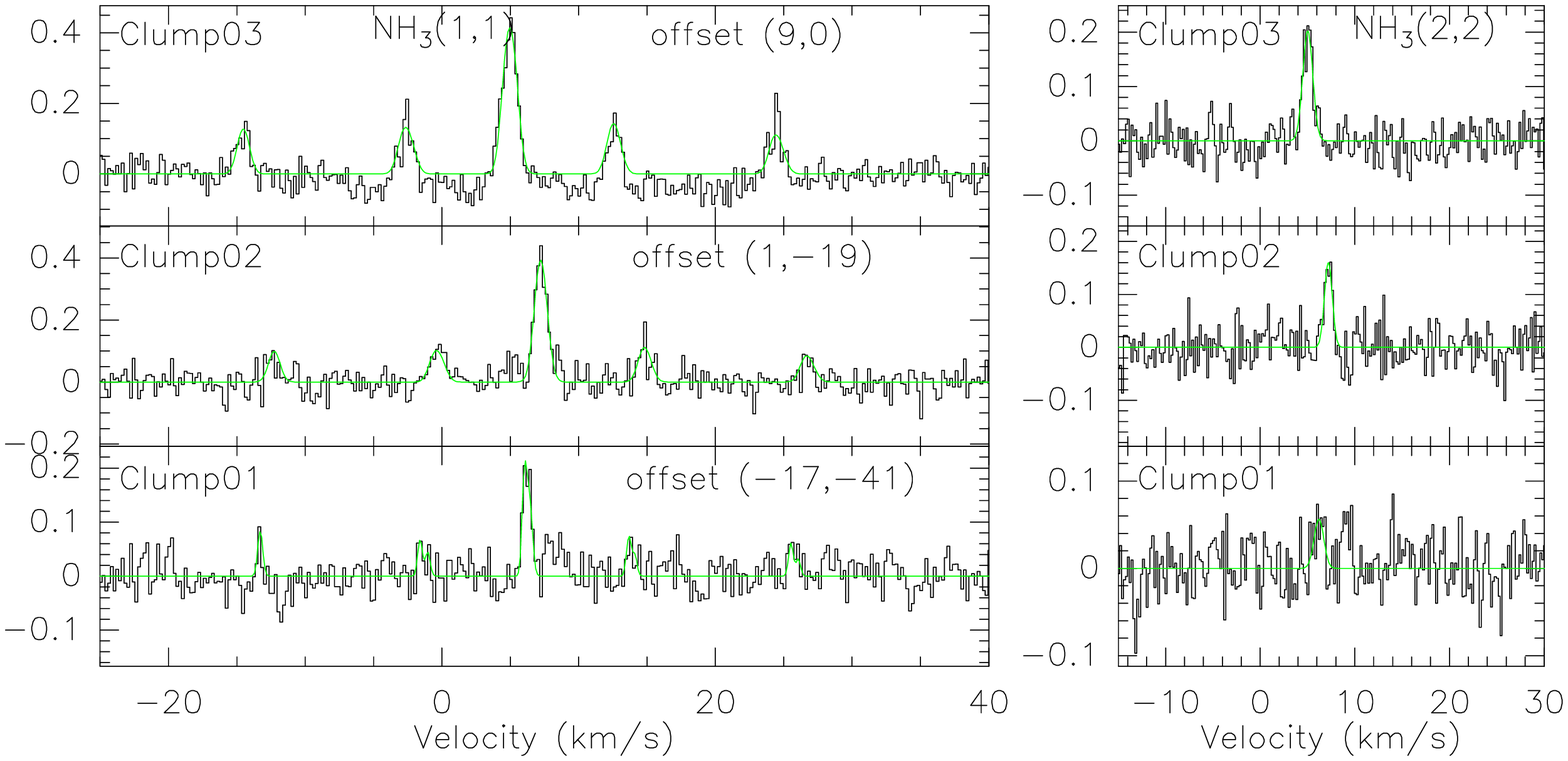}
\includegraphics[width=0.49\textwidth,angle=0]{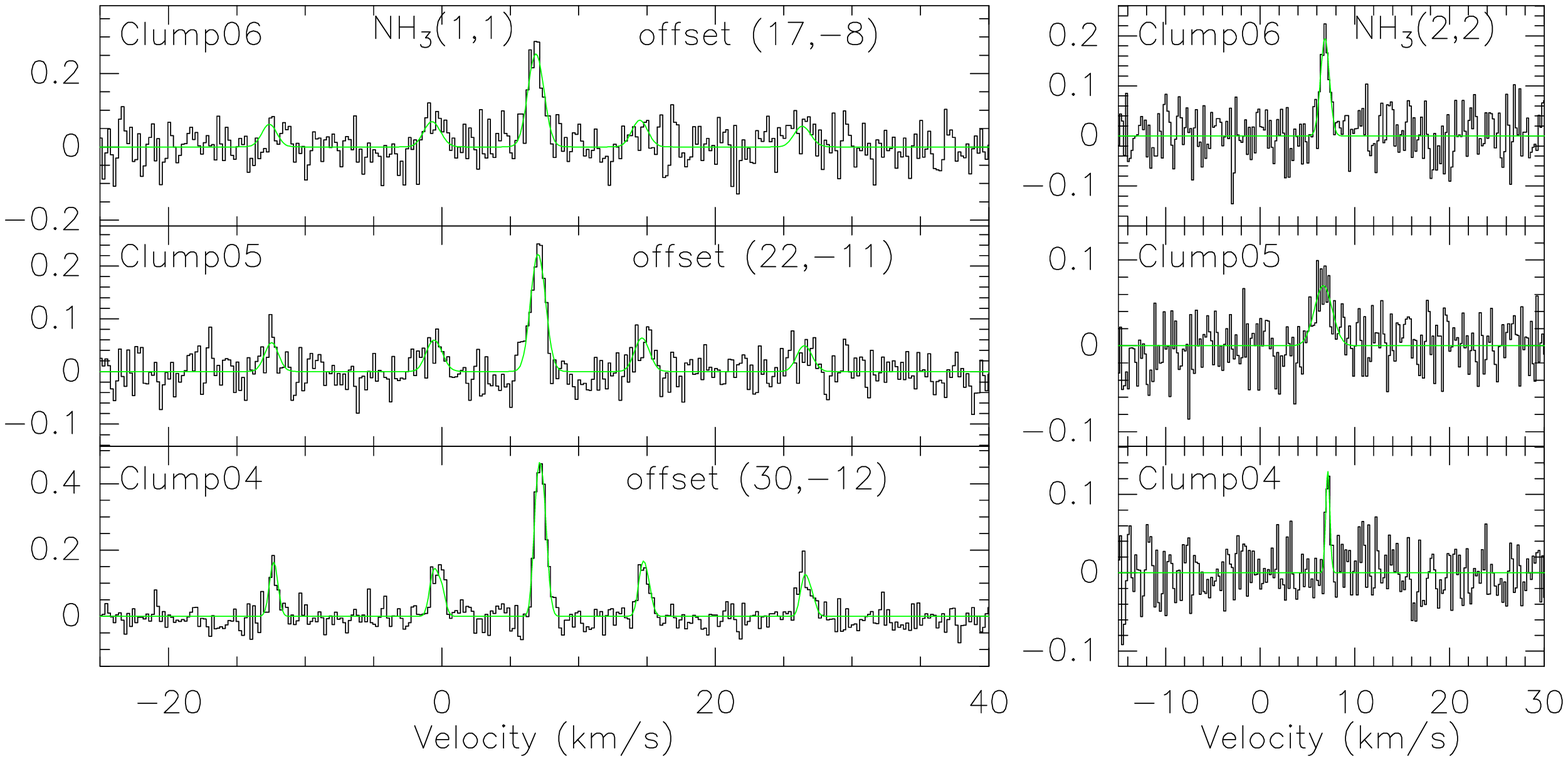}
\includegraphics[width=0.49\textwidth,angle=0]{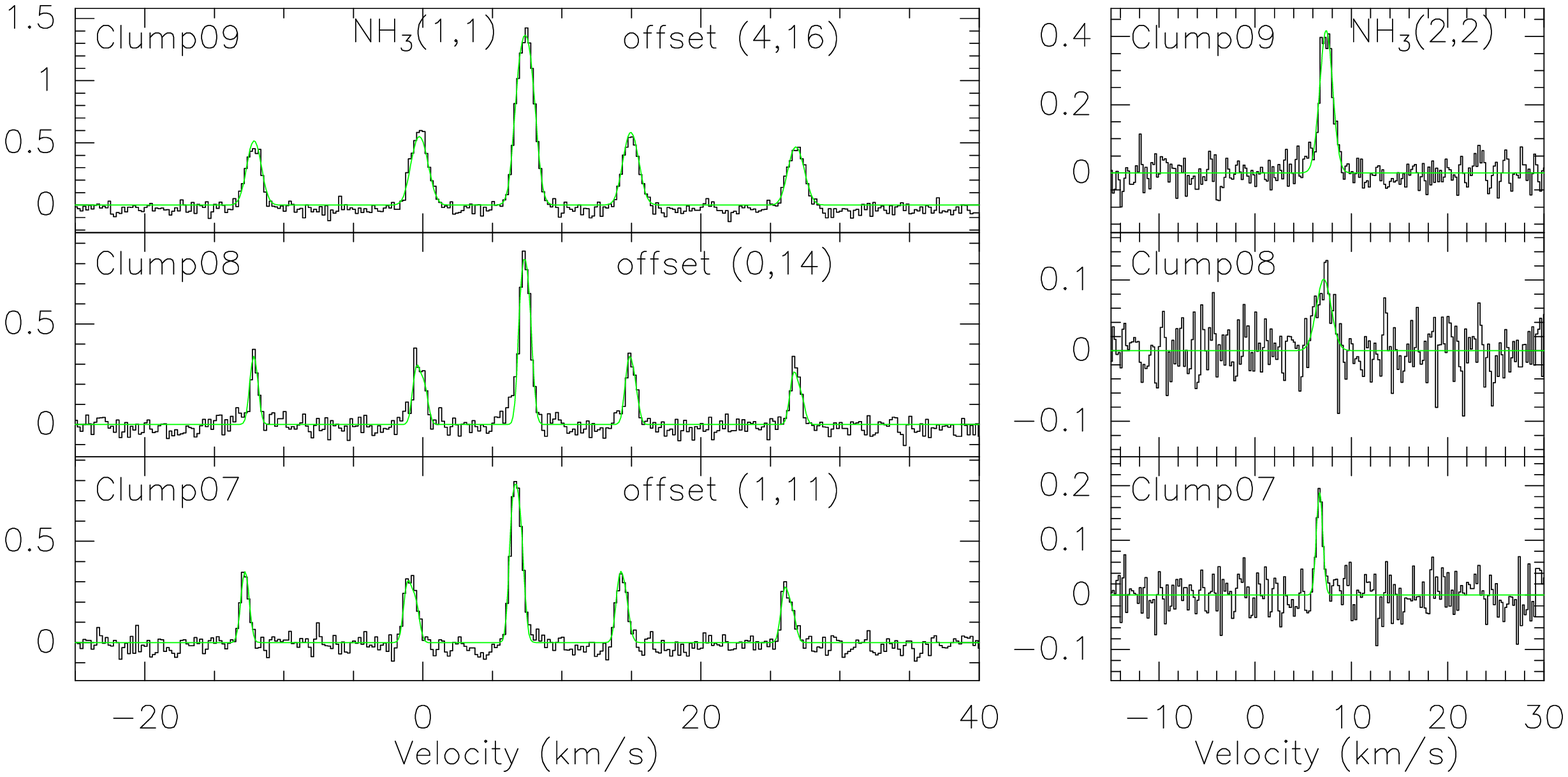}
\includegraphics[width=0.49\textwidth,angle=0]{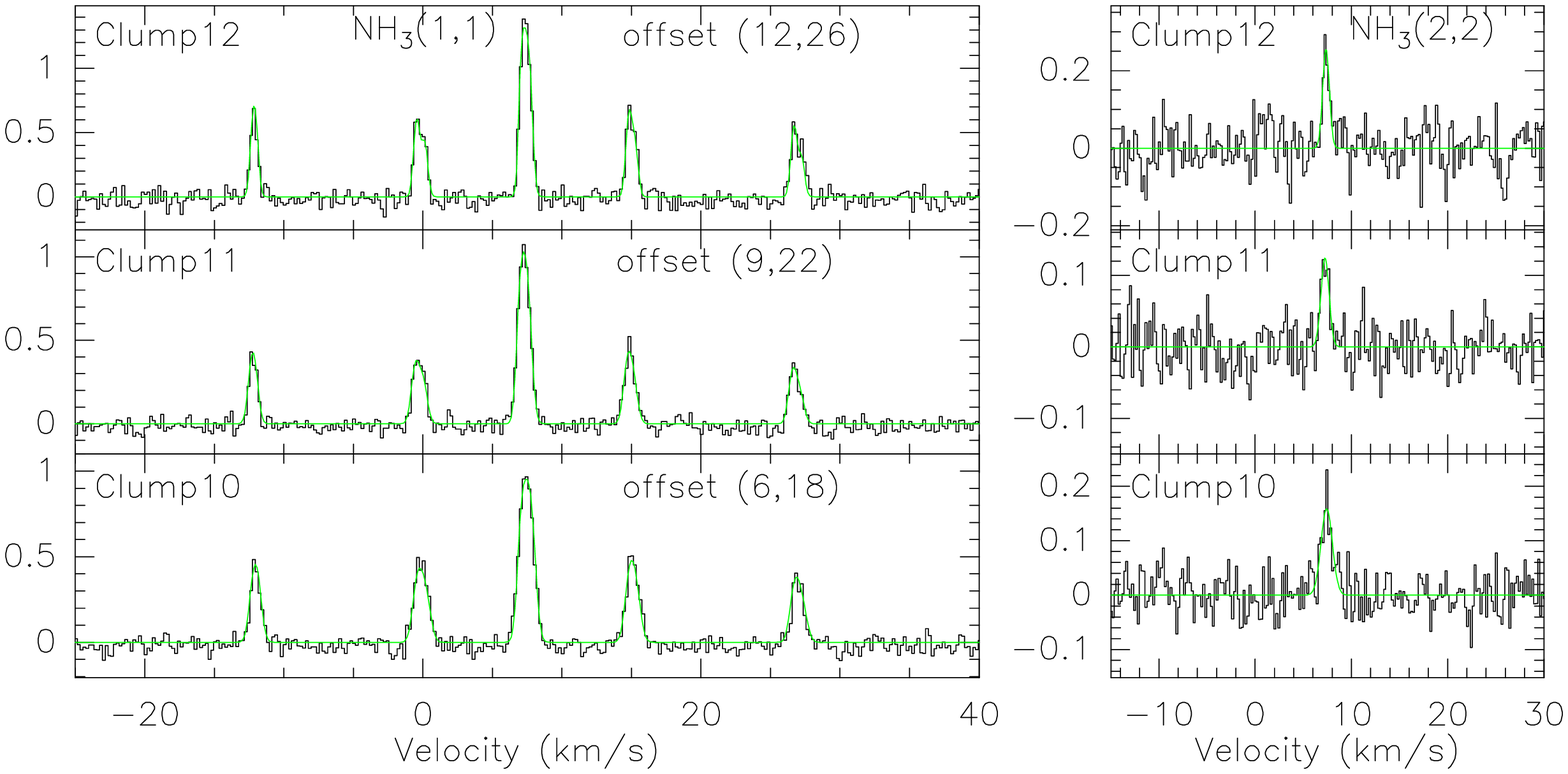}
\includegraphics[width=0.49\textwidth,angle=0]{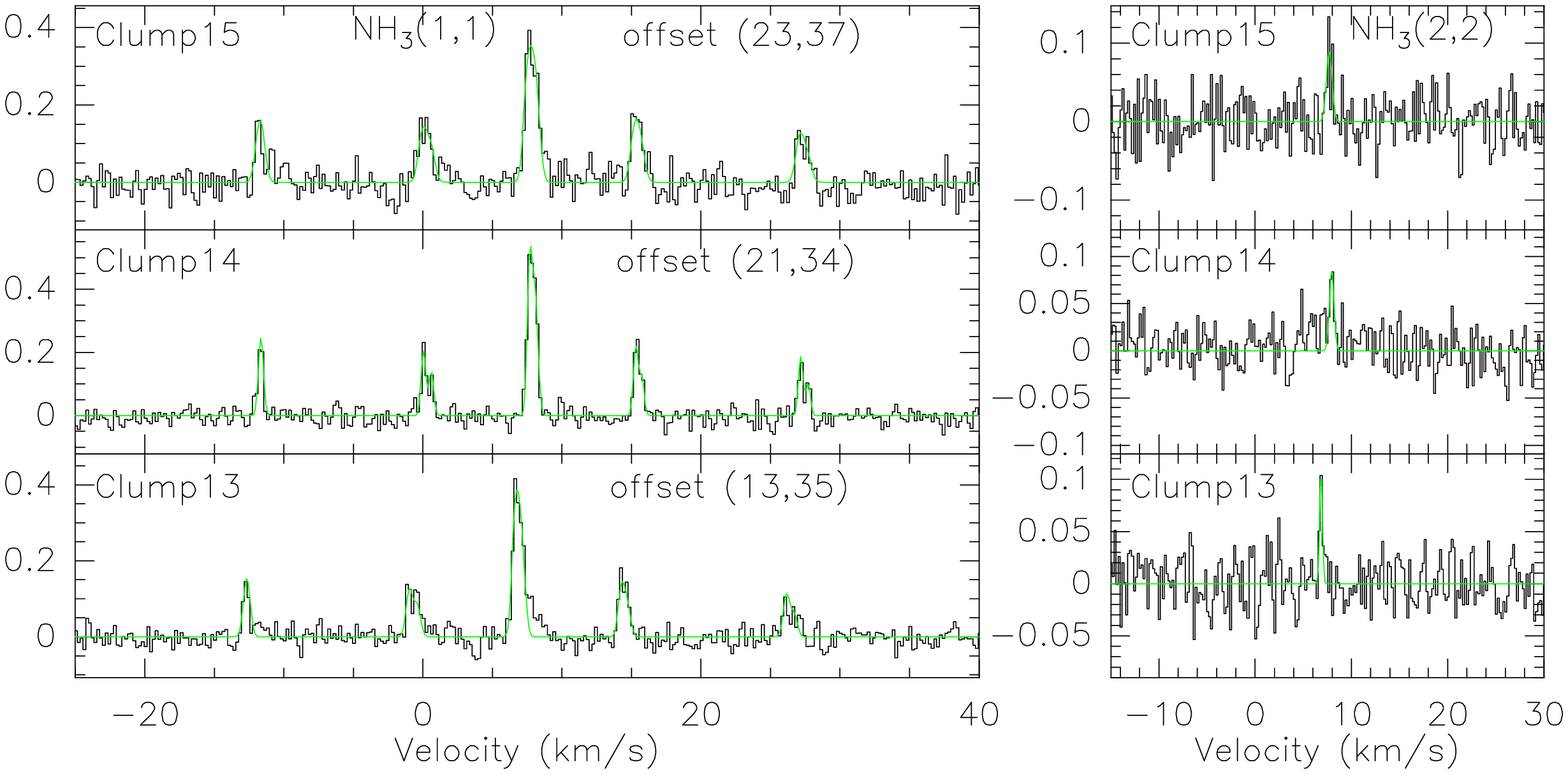}
\includegraphics[width=0.49\textwidth,angle=0]{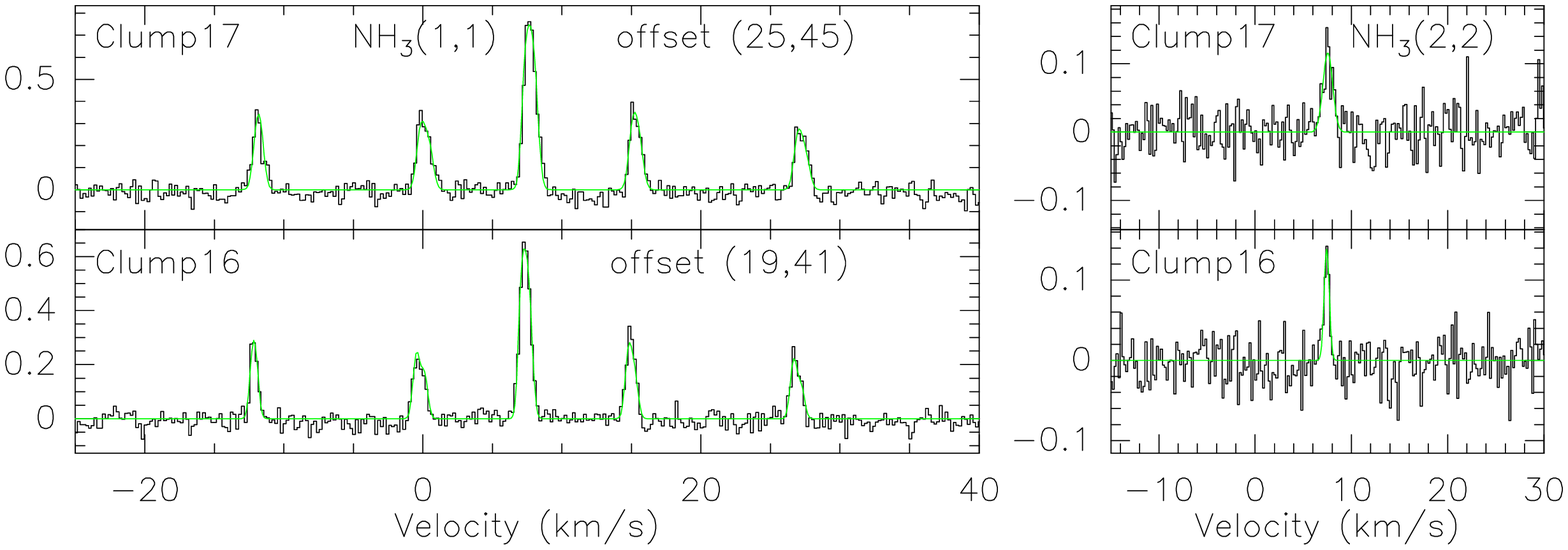}
\caption{NH$_3$\,(1,1) and (2,2) spectra towards clumps 01 to 17. Green colour
indicates the NH$_3$\,(1,1) fitting and Gaussian fitting of the NH$_3$\,(2,2) lines (see Sect. \ref{sect-2-2}).
The central position of this area is at \textbf{($l, b$)\,=\,($28.59\degr, 3.55\degr$)}. Offsets in
Galactic coordinates (unit: arcmin) are shown in the top right corner of each NH$_3$\,(1,1) panel.}
\label{FgB.2}
\end{figure*}

\begin{figure*}[h]
\centering
\includegraphics[width=0.33\textwidth,angle=0]{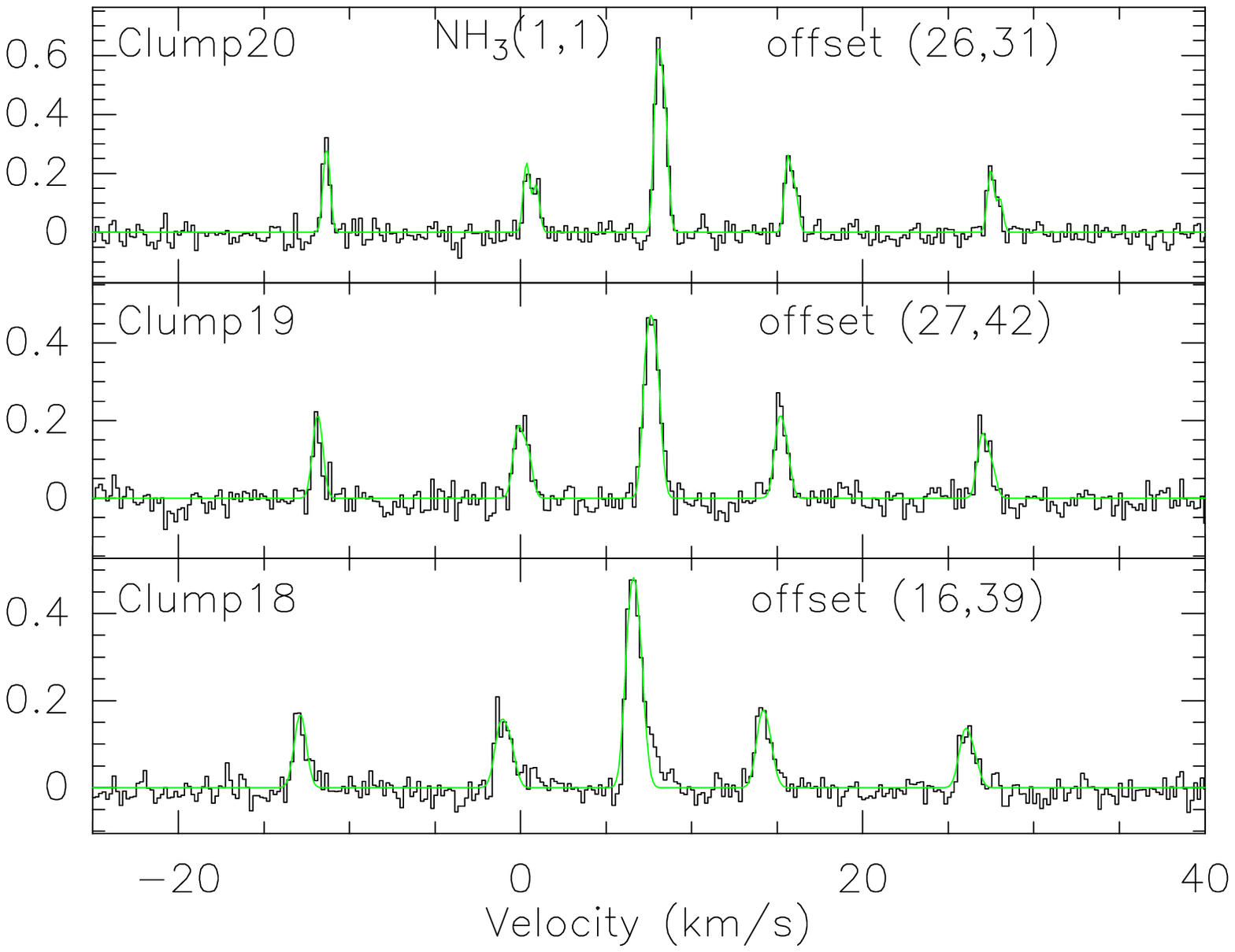}
\includegraphics[width=0.33\textwidth,angle=0]{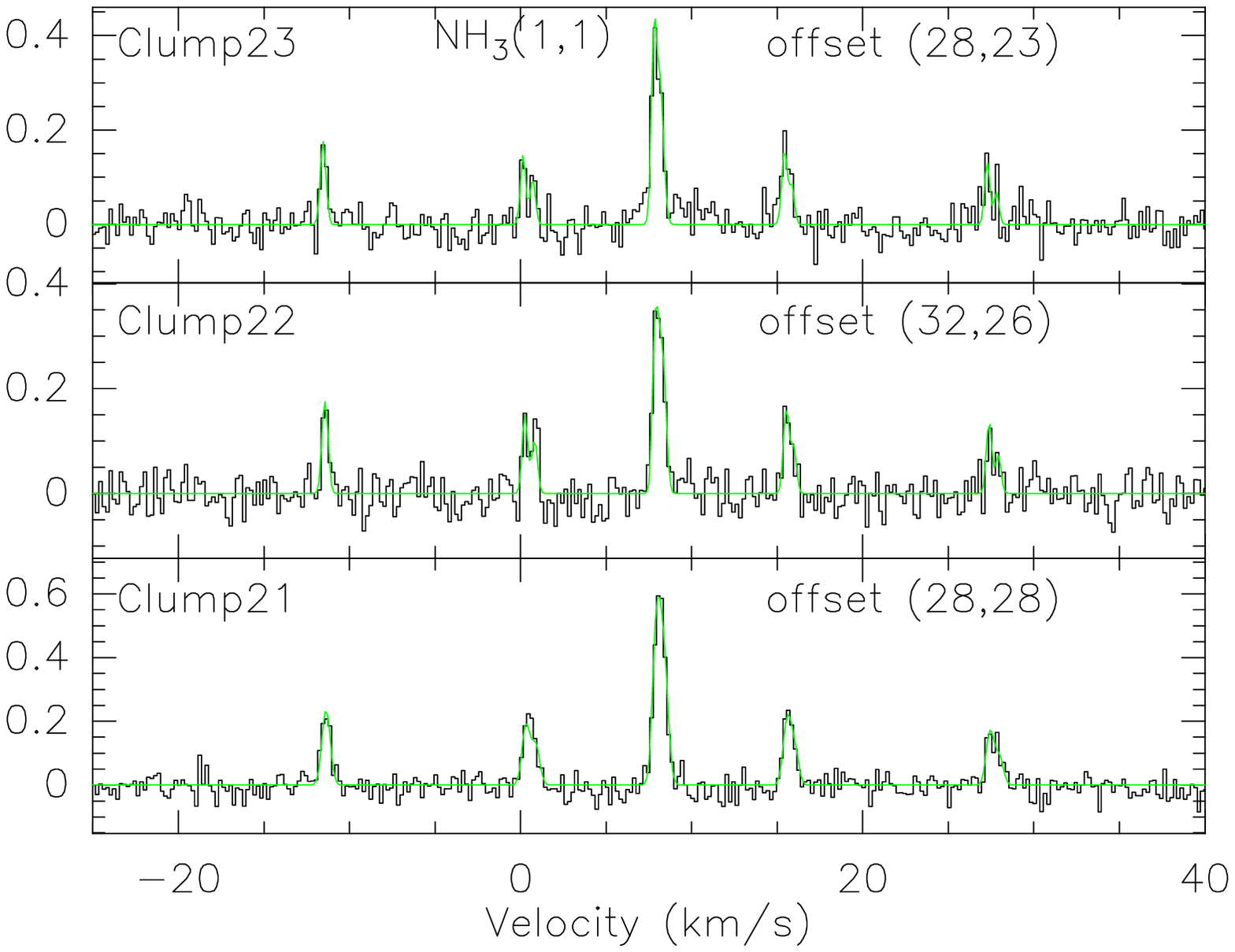}
\includegraphics[width=0.33\textwidth,angle=0]{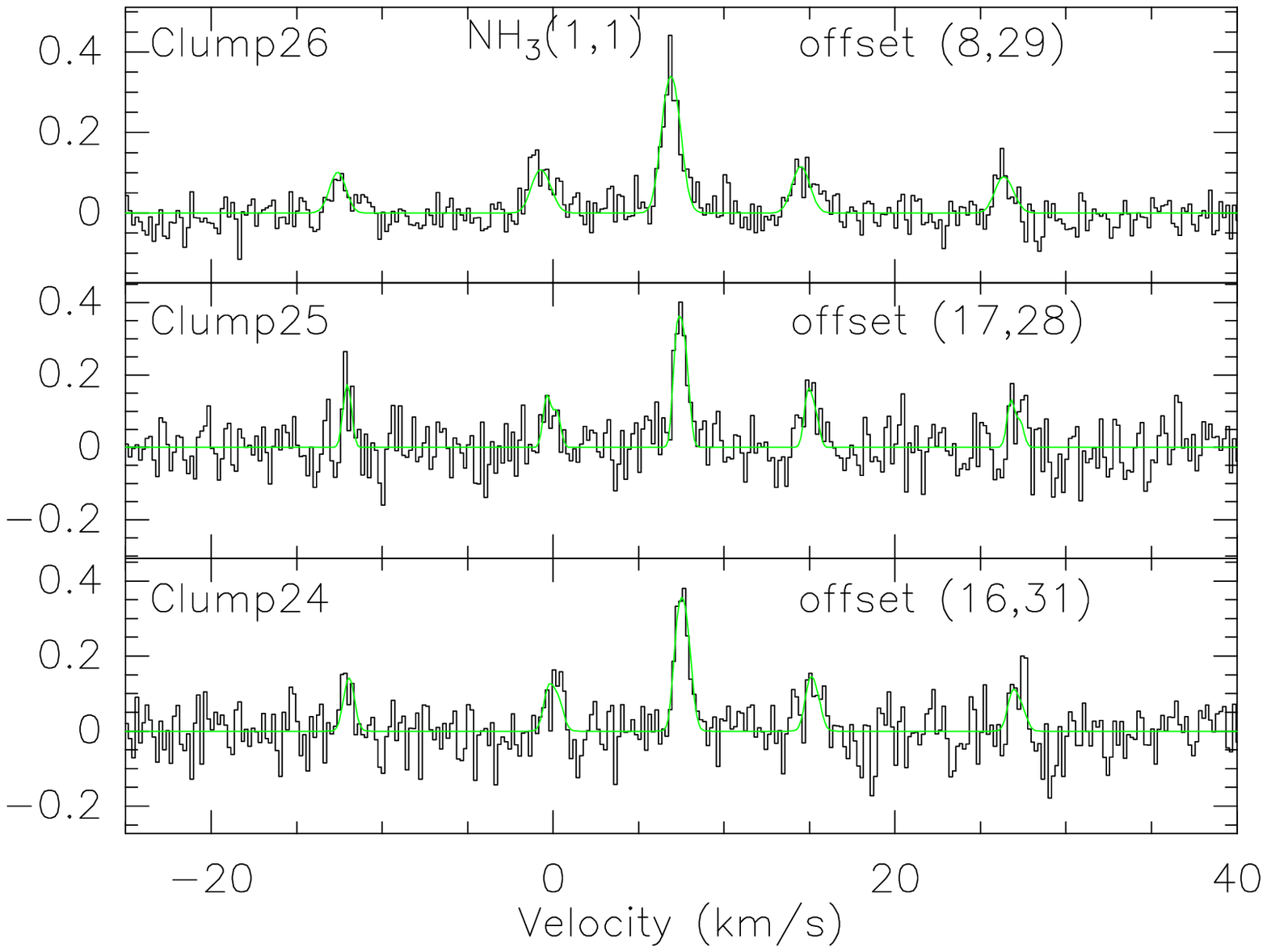}
\includegraphics[width=0.33\textwidth,angle=0]{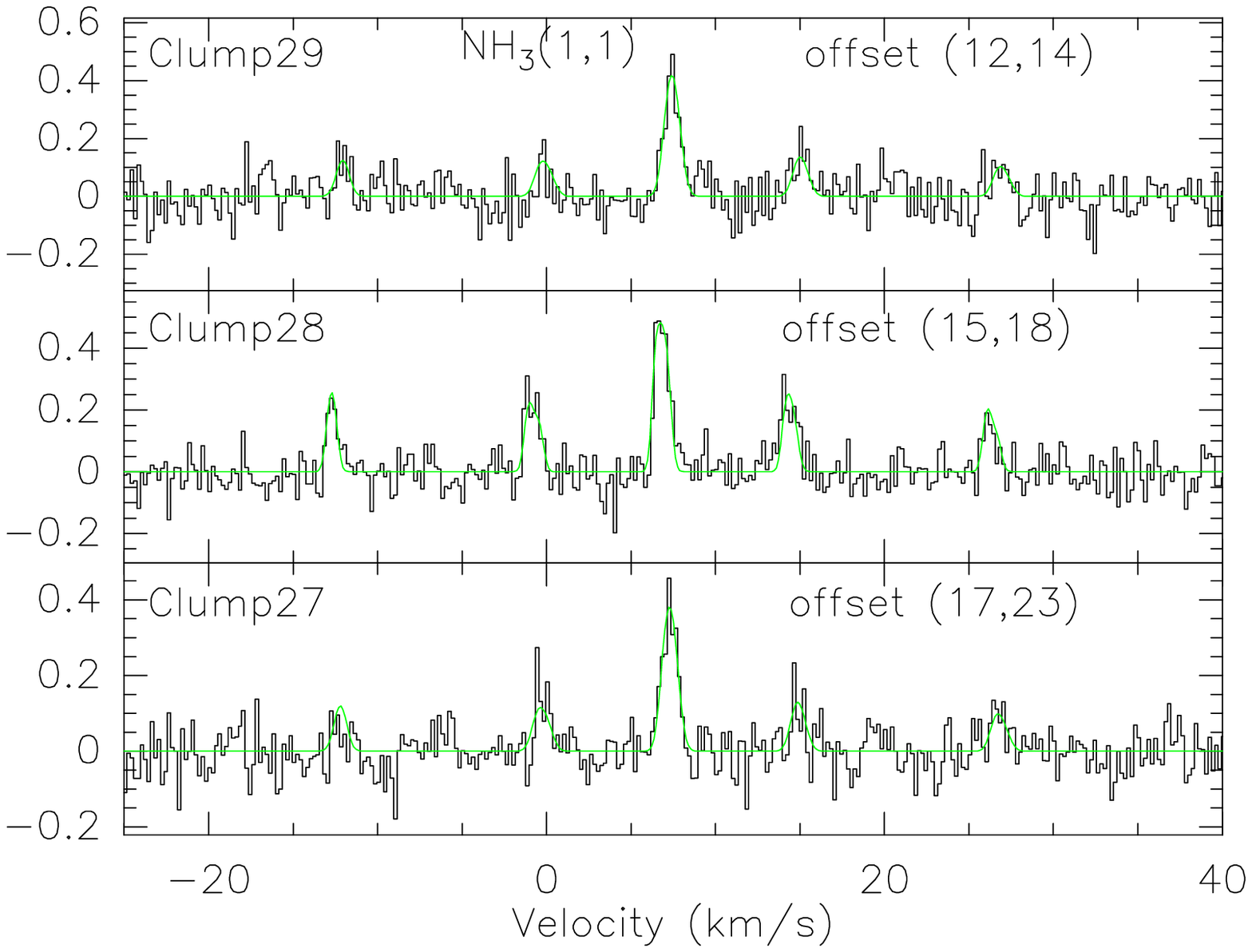}
\includegraphics[width=0.33\textwidth,angle=0]{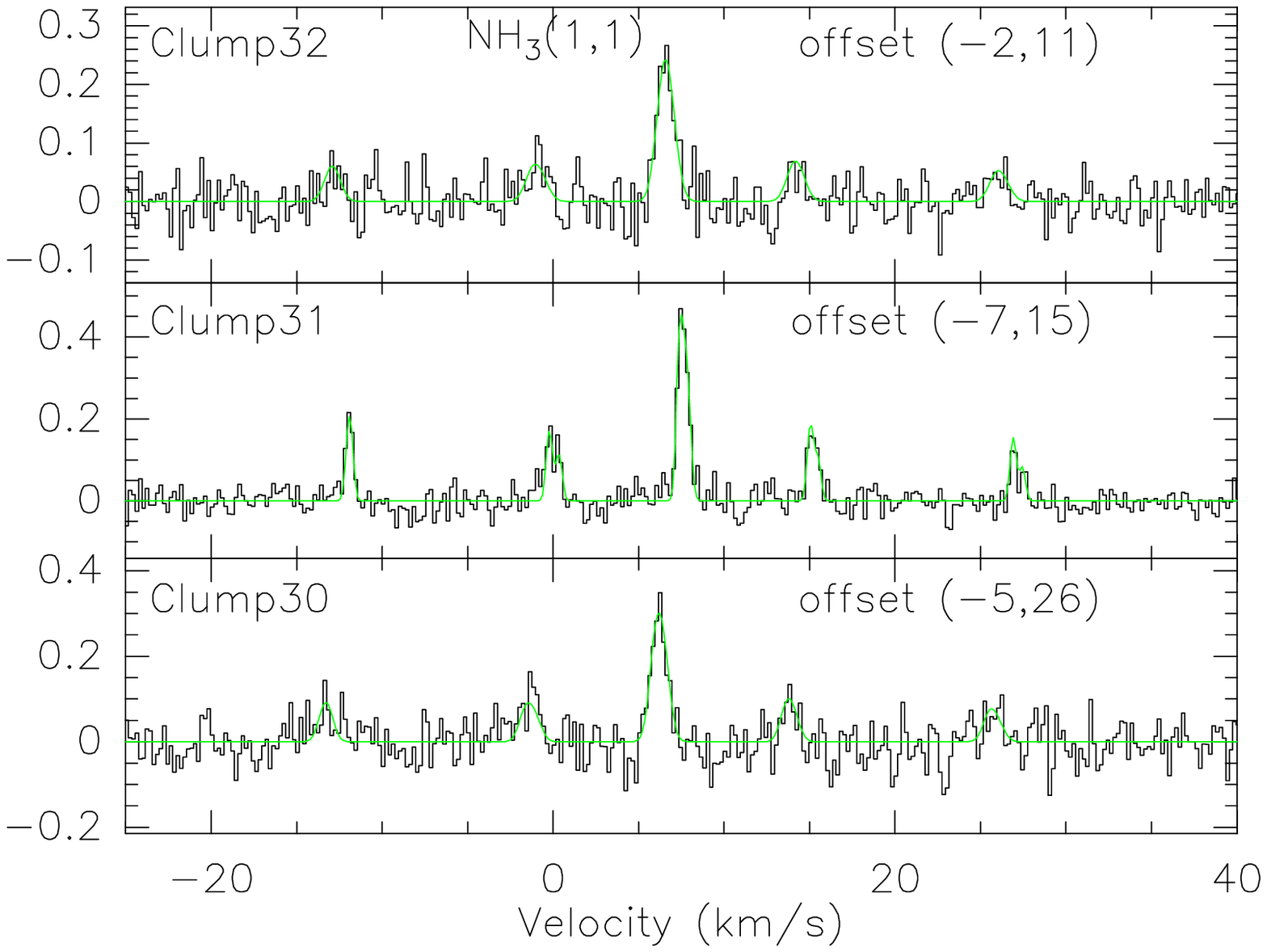}
\includegraphics[width=0.33\textwidth,angle=0]{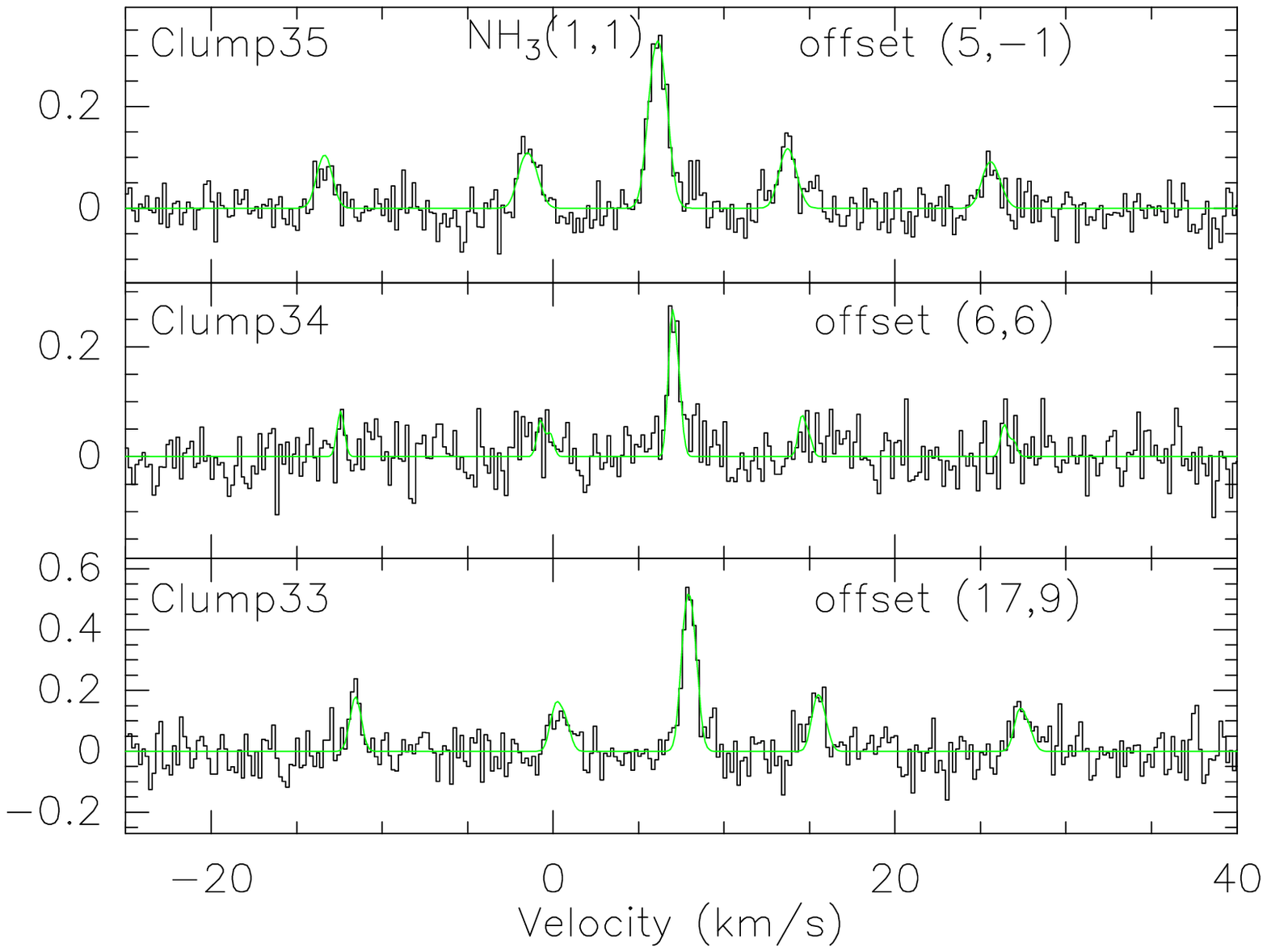}
\includegraphics[width=0.33\textwidth,angle=0]{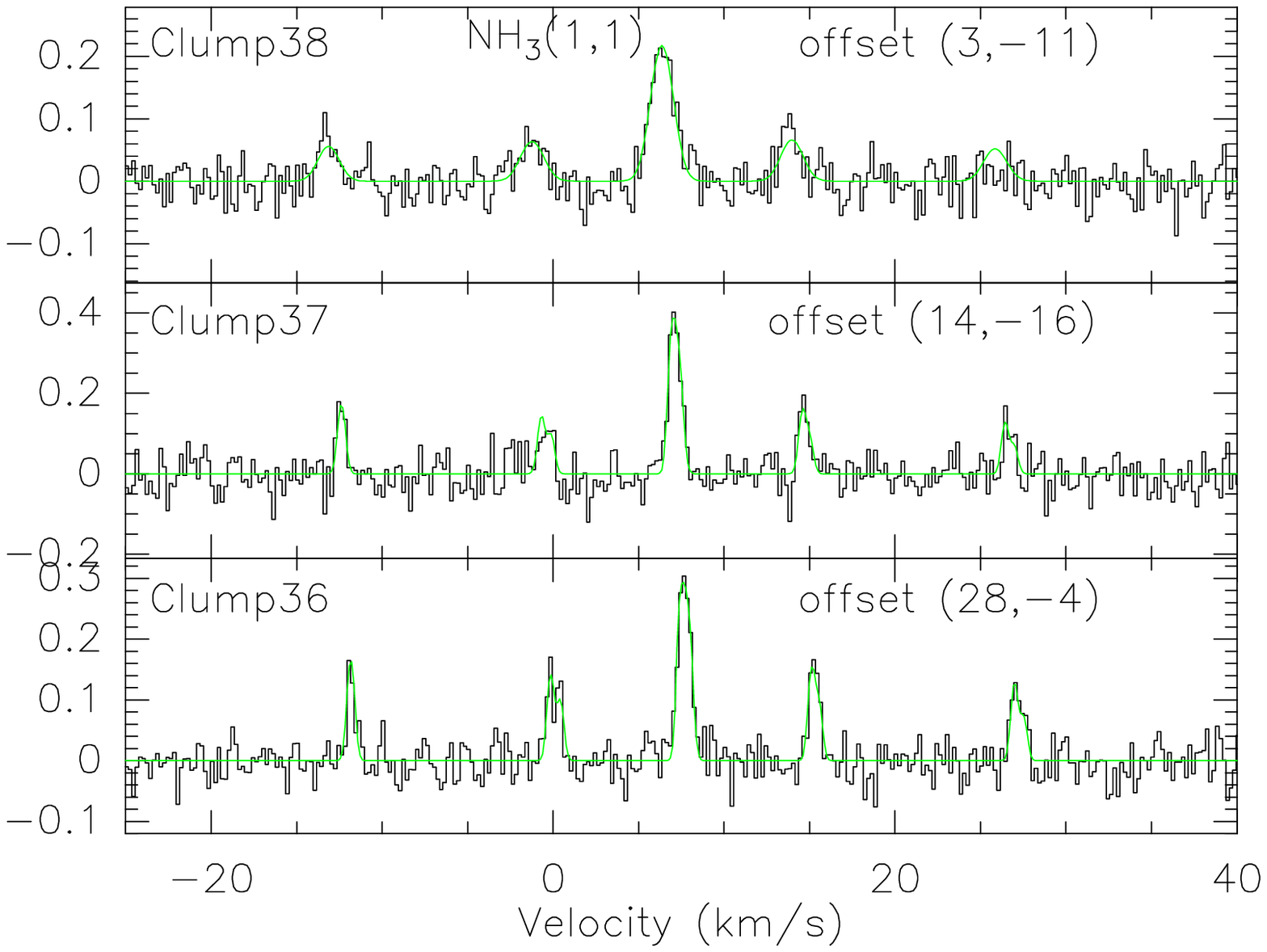}
\caption{NH$_3$\,(1,1) spectra towards clumps 18 to 38. Green colour indicates
the NH$_3$\,(1,1) fitting of the NH$_3$\,(1,1) lines. The central position of this area
 is at \textbf{($l, b$)\,=\,($28.59\degr, 3.55\degr$)}. Offsets in Galactic coordinates (unit: arcmin) are
shown in the top right corner of each NH$_3$\,(1,1) panel.}
\label{FgB.3}
\end{figure*}

In this Appendix we present prestellar and protostellar cores taken from \citet{2015A&A...584A..91K} (see Fig.\,\ref{FgB.1}),
and show the NH$_3$ spectra for the 38 Clumps identified with Clumpfind2d in Sect. \ref{sect-3-2}. The observed spectra of the
NH$_3$\,(1,1) and (2,2) transitions detected toward their peak positions are shown in Figs.\,\ref{FgB.2} and \ref{FgB.3}.
Measured physical parameters are listed in Tables\,\ref{table:B.1} to \ref{table:B.3}.

\newpage

\begin{table*}[t]
\centering
\caption{Observed parameters of the NH$_3$\,(1,1) emission lines detected in the 38 identified clumps (see Sect. \ref{sect-3-2}). }
\begin{tabular}{cccccccccc}
\hline \hline
Clump & Offset & $\int$$T_{\rm MB}$d$v$ &$V_{\rm LSR}$ & $\Delta v$ & $T_{\rm MB}$ & $\tau_{\rm}$ & $T_{\rm ex}$\\
 & (\arcmin, \arcmin) & K\,km\,s$^{-1}$ & km\,s$^{-1}$ & km\,s$^{-1}$ & K & & K  \\
\hline
01  & (-17, -41) & 0.55(0.11)& 6.21(0.02) &0.34(0.05)& 0.21(0.03)&1.50(0.83) &3.0(0.1)\\
02  & (1, -19) &0.45(0.03)&7.24(0.03)&1.04(0.07)& 0.40(0.04)& 0.10(0.14)  &...   \\
03  & (9, 0) &0.61(0.06) &4.98(0.02) &1.01(0.05)& 0.41(0.04) &0.68(0.26)&3.7(0.1) \\
04  & (30, -12) & 0.88(0.07)&7.19(0.01)&0.71(0.04)&0.48(0.03) &1.15(0.25)&3.4(0.1) \\
05  & (22, -11) &0.26(0.02)&7.04(0.04)&1.22(0.09) &0.24(0.03)& 0.10(0.21)   &...  \\
06  & (17, -8) &0.29(0.03)&6.89(0.05)&1.32(0.14)&0.28(0.05)& 0.10(0.20)   &...  \\
07  & (1, 11) & 2.15(0.09) &6.71(0.01)&0.65(0.02)&0.83(0.05)&2.29(0.19)&3.7(0.1) \\
08  & (0, 14) & 2.03(0.11)&7.35(0.01)&0.63(0.03)&0.86(0.07)&1.84(0.19)&3.7(0.1) \\
09  & (4, 16) &2.98(0.06)&7.36(0.01)&1.18(0.01)&1.46(0.04)&1.71(0.07)&4.5(0.1) \\
10  & (6, 18) &2.86(0.09)&7.46(0.01)&0.81(0.02)&1.02(0.03)&2.66(0.15) &3.8(0.1)\\
11  & (9, 22) &2.49(0.09)&7.29(0.01)&0.69(0.02)&1.08(0.03)&1.87(0.14)&4.0(0.1) \\
12  & (12, 26) &1.85(0.20)&7.38(0.01)&0.52(0.02)&1.46(0.06)&3.26(0.23)&4.3(0.1) \\
13  & (13, 35) &0.86(0.09)&6.84(0.01)&0.57(0.06)&0.38(0.06)&1.47(0.30)&3.2(0.1) \\
14  & (21, 34) & 1.59(0.09) &7.86(0.01)&0.36(0.01)&0.54(0.05)&2.09(0.25)&3.4(0.1) \\
15  & (23, 37) &0.94(0.09)&7.81(0.02)&0.72(0.06)&0.37(0.06)&2.19(0.36)&3.1(0.1)  \\
16  & (19, 41) &1.83(0.09)&7.36(0.01)&0.56(0.02)&0.67(0.03)&2.38(0.20)&3.5(0.1) \\
17  & (25, 45) & 2.08(0.08) & 7.69(0.01)&0.72(0.02)&0.80(0.05)&2.36(0.16)&3.6(0.1)  \\
18  & (16, 39)&0.91(0.05)&6.64(0.01)&0.81(0.03)&0.49(0.06)&1.20(0.16)& 3.4(0.1)  \\
19  & (27, 42) &1.30(0.09)&7.65(0.01) &0.64(0.03)& 0.49(0.03)&2.26(0.26)& 3.3(0.1) \\
20  &(26, 31) &1.81(0.09)&8.16(0.01)&0.42(0.01)&0.67(0.05)&2.12(0.25)&3.5(0.1) \\
21  & (28, 28) &1.33(0.09)&8.13(0.01)&0.54(0.02)& 0.64(0.03) &1.46(0.24)& 3.6(0.1) \\
22  & (32, 26) &1.09(0.11)&8.08(0.01)&0.38(0.03)&0.37(0.03)&2.27(0.47) & 3.2(0.1)\\
23  &(28, 23)&1.32(0.13)&7.96(0.01)&0.33(0.02)& 0.39(0.04) &2.47(0.51)& 3.2(0.1) \\
24  &(16, 31)&0.81(0.15)&7.56(0.03)&0.68(0.09)& 0.39(0.05) &1.69(0.69)& 3.2(0.1) \\
25  &(17, 28)&1.05(0.20)&7.46(0.03)&0.52(0.08)& 0.39(0.05) &2.33(0.84)& 3.2(0.1) \\
26  & (8, 29) &0.53(0.06)&6.91(0.03)&1.11(0.08)&0.36(0.05)&0.83(0.32)& 3.4(0.1) \\
27  &(17, 23)&0.64(0.12)&7.29(0.03)&0.83(0.10)& 0.39(0.05) &0.89(0.54)& 3.4(0.2) \\
28  &(15, 18)&1.81(0.21)&6.79(0.02)&0.59(0.05)& 0.53(0.07) &3.39(0.59)& 3.3(0.1) \\
29  & (12, 14)& 0.67(0.13) &7.43(0.04)&0.89(0.10)&0.42(0.06)&0.83(0.55)& 3.5(0.2) \\
30  & (-5, 26)& 0.48(0.08)&6.21(0.03)&0.93(0.10) & 0.31(0.03) &0.82(0.48)& 3.3(0.1) \\
31  & (-7, 15) &1.29(0.10)&7.59(0.01)&0.41(0.02)& 0.48(0.04) &2.03(0.35)& 3.3(0.1) \\
32  & (-2, 11) &0.29(0.02)&6.59(0.03)&1.08(0.08)&0.25(0.05)&  0.10(0.05)  &... \\
33  & (17, 9) &0.95(0.13)& 7.95(0.02) &0.77(0.08)&0.55(0.05)&1.09(0.39)& 3.6(0.1) \\
34  & (6, 6) & 0.38(0.07)&7.07(0.03)&0.54(0.14)&0.27(0.06)& 0.10(0.38)  &... \\
35  & (5, -1) &0.56(0.05)&6.13(0.02)&1.05(0.06)& 0.34(0.04) &1.04(0.28)& 3.3(0.1) \\
36  & (28, -4) &1.19(0.14)&7.67(0.01)&0.46(0.04)& 0.32(0.02) &3.67(0.63) & 3.1(0.1)\\
37  & (14, -16) &1.07(0.13)&7.12(0.02)&0.49(0.04)& 0.40(0.03)&2.09(0.54)& 3.2(0.1) \\
38  & (3, -11) &0.28(0.04)&6.37(0.04)&1.47(0.10)& 0.22(0.03) &0.42(0.32) & 3.4(0.1)\\
\hline
\end{tabular}
\tablefoot{The central position is \textbf{($l, b$)\,=\,($28.59\degr, 3.55\degr$)}. Offsets are given in Galactic coordinates. The errors shown in  parentheses are fitting uncertainties. In cases where no excitation temperature is given, the NH$_3$\,(1,1) line is optically
thin and does not allow for a determination of this parameter.}
\label{table:B.1}
\end{table*}

\begin{table*}[t]
\centering
\caption{Observed parameters of the NH$_3$\,(2,2) emission lines detected in 17 of the 38 NH$_3$\,(1,1) clumps (see Sect. \ref{sect-3-2}).}
\begin{tabular}{c  c c c c c c c}
\hline \hline
 Clump & Offset & $\int$$T_{\rm MB}$d$v$ & $V_{\rm LSR}$ & $\Delta v$ & $T_{\rm MB}$   \\
  & (\arcmin, \arcmin) & K\,km\,s$^{-1}$ & km\,s$^{-1}$ & km\,s$^{-1}$ & K   \\
\hline
01  & (-17, -41) &0.07(0.02)&6.19(0.15)&1.17(0.29)&0.06(0.03)    \\
02  & (1, -19) &0.17(0.02)&7.21(0.07)&0.99(0.14) & 0.16(0.04)    \\
03 & (9, 0) &0.26(0.02)&5.024(0.05)&1.23(0.11)&0.20(0.04)    \\
04 & (30, -12) & 0.07(0.01) &7.15(0.04)&0.49(0.10)&0.13(0.04)    \\
05 & (22, -11) &0.02(0.01)&10.41(0.08)&0.28(0.12)& 0.05(0.01)     \\
06 & (17, -8)&0.24(0.03) &6.85(0.07)&1.15(0.19)&0.19(0.04)    \\
07  & (1, 11) & 0.15(0.02)&6.68(0.04)&0.75(0.11)&0.19(0.03)     \\
08  & (0, 14) & 0.19(0.03)&7.13(0.12)&1.82(0.25) & 0.10(0.05)     \\
09  &(4, 16)&0.70(0.03)&7.38(0.03)&1.57(0.07)&0.42(0.05)    \\
10  &(6, 18) &0.23(0.03)&7.43(0.07)&1.36(0.25)&0.16(0.04)     \\
11  &(9, 22) & 0.13(0.02)&7.25(0.07)&0.99(0.14)&0.13(0.03)    \\
12  &(12, 26) &0.22 (0.03)&7.35(0.05)&0.79(0.13)&0.26(0.04)    \\
13  &(13, 35)&0.04(0.01) &6.84(0.03)&0.33(0.07)&0.11(0.01)    \\
14  &(21, 34)  &0.15(0.03)&7.26(0.32)&0.23(0.09) & 0.04(0.03)     \\
15  &(23, 37)& 0.07(0.01) &7.69(0.08)&0.70(0.16)  &0.09(0.04)    \\
16  &(19, 41)&0.09(0.01)& 7.42(0.04) &0.63(0.10)& 0.14(0.02)    \\
17  &(25, 45) &0.15(0.02)&7.54(0.07)&1.21(0.17)&0.12(0.03)    \\
\hline
\end{tabular}
\tablefoot{The central position is \textbf{($l, b$)\,=\,($28.59\degr, 3.55\degr$)}. Offsets are given in Galactic coordinates.
Parameters are derived from Gassian fits to the spectra.}
\label{table:B.2}
\end{table*}

\begin{table*}[t!]
\centering
\caption{Calculated model parameter of NH$_3$\,(1,1) and NH$_3$\,(2,2) emission lines detected in seventeen clumps.}
\begin{tabular}{c c c c c c c c c}
\hline \hline
Clump & Offset& $T_{\rm rot}$ & $T_{\rm kin}$ & $N(1,1)$ &total-$N{(\rm NH_3)}$$^{a}$ & $N(\rm H_{2})^{\it b}$ & $\chi$(total-NH$_3$) \\
 & (\arcmin, \arcmin) & K & K & cm$^{-2}$  &  cm$^{-2}$ &  cm$^{-2}$ &  \\
\hline
01 &(-17, -41)& 13.6 $\pm$ 2.9 & 14.9 $\pm$ 3.8  & $0.4\times10^{14}$ & $3.8\times10^{14}$ & $0.9\times10^{22}$ & $4.3\times10^{-8}$\\
02 &(1, -19)& 18.3 $\pm$ 2.4& 22.5 $\pm$ 3.9  & $0.1\times10^{14}$ & $0.3\times10^{14}$ & $1.4\times10^{22}$ & $0.2\times10^{-8}$\\
03 & (9, 0)& 19.1 $\pm$ 2.1 &23.2 $\pm$ 3.5  & $0.7\times10^{14}$ & $3.6\times10^{14}$ & $2.5\times10^{22}$ & $1.4\times10^{-8}$\\
04 &(30, -12)& 13.9 $\pm$ 1.6&15.5 $\pm$ 2.1 & $0.8\times10^{14}$ & $6.5\times10^{14}$ & $2.1\times10^{22}$ & $3.1\times10^{-8}$\\
05 &(22, -11)&14.8 $\pm$ 1.2&16.8 $\pm$ 1.7  & $0.4\times10^{13}$ & $0.3\times10^{14}$ & $1.8\times10^{22}$ & $0.2\times10^{-8}$\\
06 & (17, -8)& 25.2 $\pm$ 4.3&34.9 $\pm$ 9.2 & $0.5\times10^{13}$ & $0.2\times10^{14}$ & $1.9\times10^{22}$ & $0.1\times10^{-8}$\\
07 &(1, 11)& 11.7 $\pm$ 0.6 &12.5 $\pm$ 0.8  & $1.5\times10^{14}$ & $1.9\times10^{15}$ & $3.2\times10^{22}$ & $5.9\times10^{-8}$\\
08 &(0, 14)& 10.1 $\pm$ 1.2 &10.6 $\pm$ 1.4 & $1.2\times10^{14}$ & $2.3\times10^{15}$ & $2.7\times10^{22}$ & $8.9\times10^{-8}$\\
09 &(4, 16)& 13.3 $\pm$ 0.6 &14.6 $\pm$ 0.8  & $2.6\times10^{14}$ & $2.3\times10^{15}$ & $6.3\times10^{22}$ & $3.7\times10^{-8}$\\
10 &(6, 18)& 10.2 $\pm$ 0.7&10.7 $\pm$ 0.8  & $2.3\times10^{14}$ & $4.3\times10^{15}$ & $5.2\times10^{22}$ & $8.3\times10^{-8}$\\
11 &(9, 22)& 9.9 $\pm$ 0.6&10.5 $\pm$ 0.7  & $1.5\times10^{14}$ & $2.9\times10^{15}$ & $2.9\times10^{22}$ & $9.9\times10^{-8}$\\
12 &(12, 26)&10.1 $\pm$ 0.4 &10.6 $\pm$ 0.5  & $1.9\times10^{14}$ & $3.8\times10^{15}$ & $3.6\times10^{22}$ & $1.1\times10^{-7}$\\
13 &(13, 35)& 13.8 $\pm$ 0.9 &15.2 $\pm$ 1.3 & $0.8\times10^{14}$ & $6.4\times10^{14}$ & $1.6\times10^{22}$ & $4.1\times10^{-8}$\\
14 &(21, 34)& 8.6 $\pm$ 1.6 &8.9 $\pm$ 1.9 & $0.7\times10^{14}$ & $2.5\times10^{15}$ & $1.2\times10^{22}$ & $2.1\times10^{-7}$\\
15 &(23, 37)& 12.1 $\pm$ 1.9&13.0 $\pm$ 2.5  & $1.4\times10^{14}$ & $1.6\times10^{15}$ & $1.3\times10^{22}$ & $1.2\times10^{-7}$\\
16 &(19, 41)& 11.2 $\pm$ 0.4 &11.9 $\pm$ 0.5 & $1.3\times10^{14}$ & $1.8\times10^{15}$& $2.1\times10^{22}$ & $8.5\times10^{-8}$ \\
17 &(25, 45)& 10.2 $\pm$ 0.7&10.7 $\pm$ 0.9 & $1.7\times10^{14}$  & $3.2\times10^{15}$& $2.7\times10^{22}$ & $1.2\times10^{-7}$\\
\hline
\end{tabular}
\tablefoot{$^{a}$Total(para+ortho) column densities of NH$_3$, see Sect. \ref{sect-3-4}. $^{b}$H$_2$ column densities are
taken from \cite{2010A&A...518L..85B} and \cite{2015A&A...584A..91K}. The central position is
\textbf{($l, b$)\,=\,($28.59\degr, 3.55\degr$)}. Offsets are based on Galactic coordinates.}
\label{table:B.3}
\end{table*}

\Online
\twocolumn
\section{Uncertainty estimation and derivation of the physical parameters}
\label{Appendix-C}

The uncertainty of $T_{\rm rot}$ in Eq.\,(\ref{Eq1}) is

\begin{eqnarray}
\Delta T_{\rm rot} &=& \frac{\partial T_{\rm rot}}{\partial \tau_m(1,1)} \Delta \tau_m(1,1) + \frac{\partial T_{\rm rot}}{\partial T_{\rm MB}(2,2)} \Delta T_{\rm MB}(2,2) \\ \nonumber
\ &+& \frac{\partial T_{\rm rot}}{\partial T_{\rm MB}(1,1)} \Delta T_{\rm MB}(1,1),
\end{eqnarray}
where $\Delta \tau_m(1,1)$, $\Delta T_{\rm MB}(2,2)$ and $\Delta T_{\rm MB}(1,1)$ are uncertainties of $\tau_m(1,1)$, $T_{\rm MB}(2,2)$ and $T_{\rm MB}(1,1)$, respectively.

The error of $T_{\rm kin}$ in Eq.\,(\ref{Eq2}) is

\begin{equation}
  \Delta T_{\rm kin} = \frac{\partial T_{\rm kin}}{\partial T_{\rm rot}} \Delta T_{\rm rot}.
\end{equation}

Uncertainties of $N_{\rm tot}$ in Eq.\,(\ref{Eq3}) are

\begin{equation}
  \Delta N_{\rm tot} = \frac{\partial N_{\rm tot}}{\partial T_{\rm rot}} \Delta T_{\rm rot} + \frac{\partial N_{\rm tot}}{\partial N(1,1)} \Delta N(1,1),
\end{equation}
where $\Delta N(1,1)$ is the uncertainty of $N(1,1)$ in Eq.\,(\ref{Eq4}).

When $\tau \ll 1$, $N(1,1)$ is given by

\begin{eqnarray}
  && T_{\rm MB} \varpropto T_{\rm ex} \tau \, \, \rm and \\
  && N(1,1) = \frac{1.65 \times 10^{14}}{v} \frac{J(J+1)}{K^2} \Delta v T_{\rm MB},
\end{eqnarray}
The uncertainty of N(1,1) then becomes

\begin{eqnarray}
  &&  \Delta N(1,1) = \frac{\partial N(1,1)}{\partial \Delta v} \Delta \Delta v + \frac{\partial N(1,1)}{\partial T_{\rm MB}} \Delta T_{\rm MB} \\
  &&  \ \ \ \ = \frac{1.65 \times 10^{14}}{v} \frac{J(J+1)}{K^2} \left( T_{\rm MB} \Delta \Delta v + \Delta v \Delta T_{\rm MB} \right),
\end{eqnarray}
$\Delta \Delta v$ represents the error in the line width $\Delta v$. If $\tau \gtrsim 1$, $T_{\rm ex}$ is obtained from Eq.\,(\ref{Eq5}) as a function of $T_{\rm MB}$ and $\tau$. Then the uncertainty of $N(1,1)$ is defined by

\begin{eqnarray}
  \Delta N(1,1) = \frac{\partial N(1,1)}{\partial \Delta v} \Delta \Delta v + \frac{\partial N(1,1)}{\partial \tau_{\rm tot}} \Delta \tau_{\rm tot} + \frac{\partial N(1,1)}{\partial T_{\rm MB}} \Delta T_{\rm MB}.
\end{eqnarray}

\textit{Conversion of measured full width to half maximum (FWHM) line widths into velocity dispersion:} The intensity of a line with Gaussian distribution at radial velocity $v_{\rm r}$ is

\begin{eqnarray}
\rm P(v_{\rm r}) = \frac{1}{\sqrt{2\pi}\sigma}e^{-\frac{1}{2} \left( \frac{v_{\rm r} - v_p}{\sigma} \right)^{2}},
\end{eqnarray}
where $v_{\rm p}$ is the peak velocity, and $\sigma$ is the velocity dispersion. From the definition of the FWHM, we can infer

\begin{eqnarray}
e^{-\frac{1}{2}(\Delta v / 2 \sigma)^{2}} = 1/2 \ \ \Rightarrow \sigma = \Delta v / \left(2 \sqrt{2 \ln{2}} \right) = \Delta v / 2.35,
\end{eqnarray}
where $\Delta v$ is the FWHM.

\textit{Conversion of velocity dispersion into thermal contribution to the observed line width (see Eq.\,(\ref{Eq8}))}:
\begin{eqnarray}
v_{\rm obs} = \sqrt{2} \sigma
\end{eqnarray}

\Online
\onecolumn
\section{Fitted velocity and residuals of the velocity fitting}
\label{Appendix D}
\begin{figure*}[h]
\includegraphics[width=0.49\textwidth]{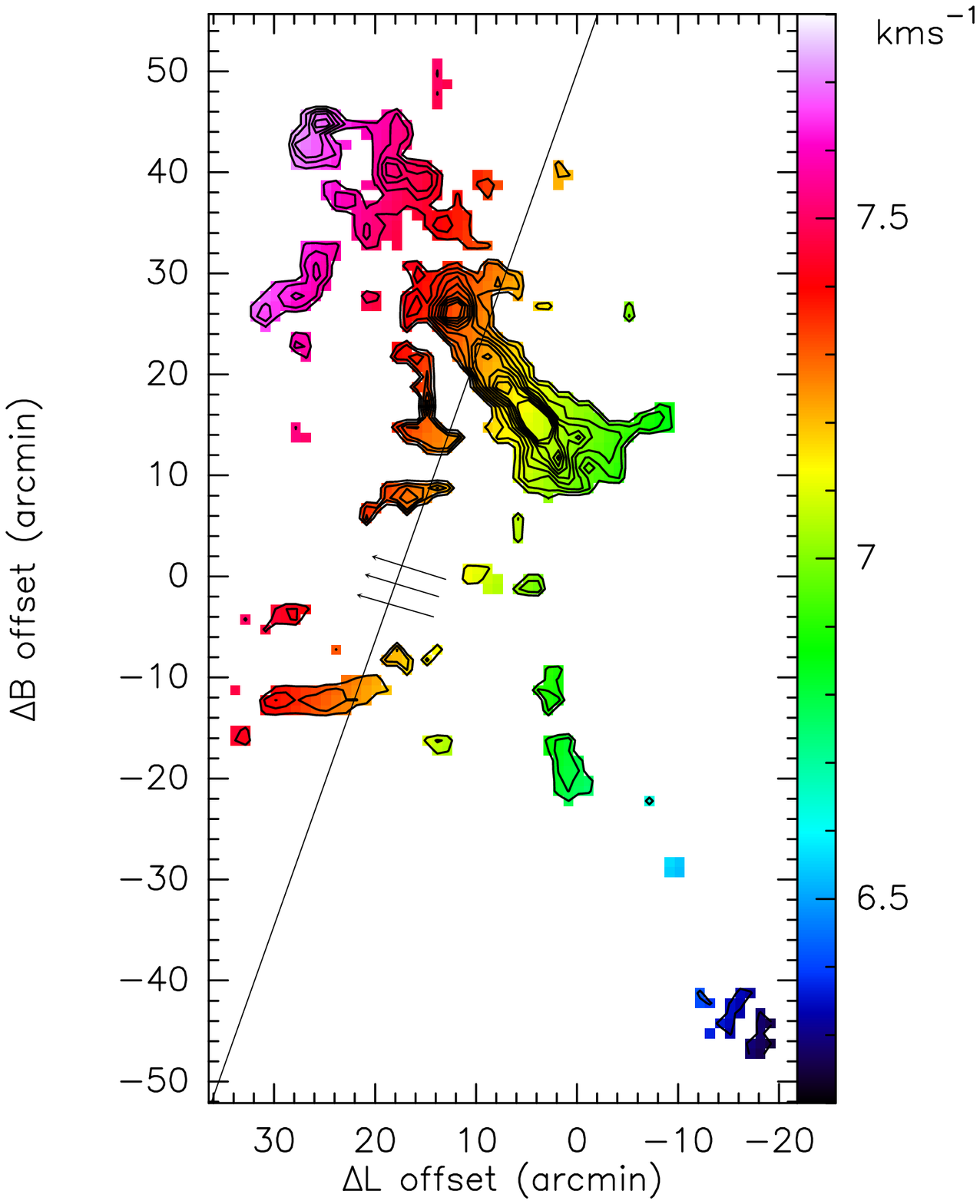}
\includegraphics[width=0.49\textwidth]{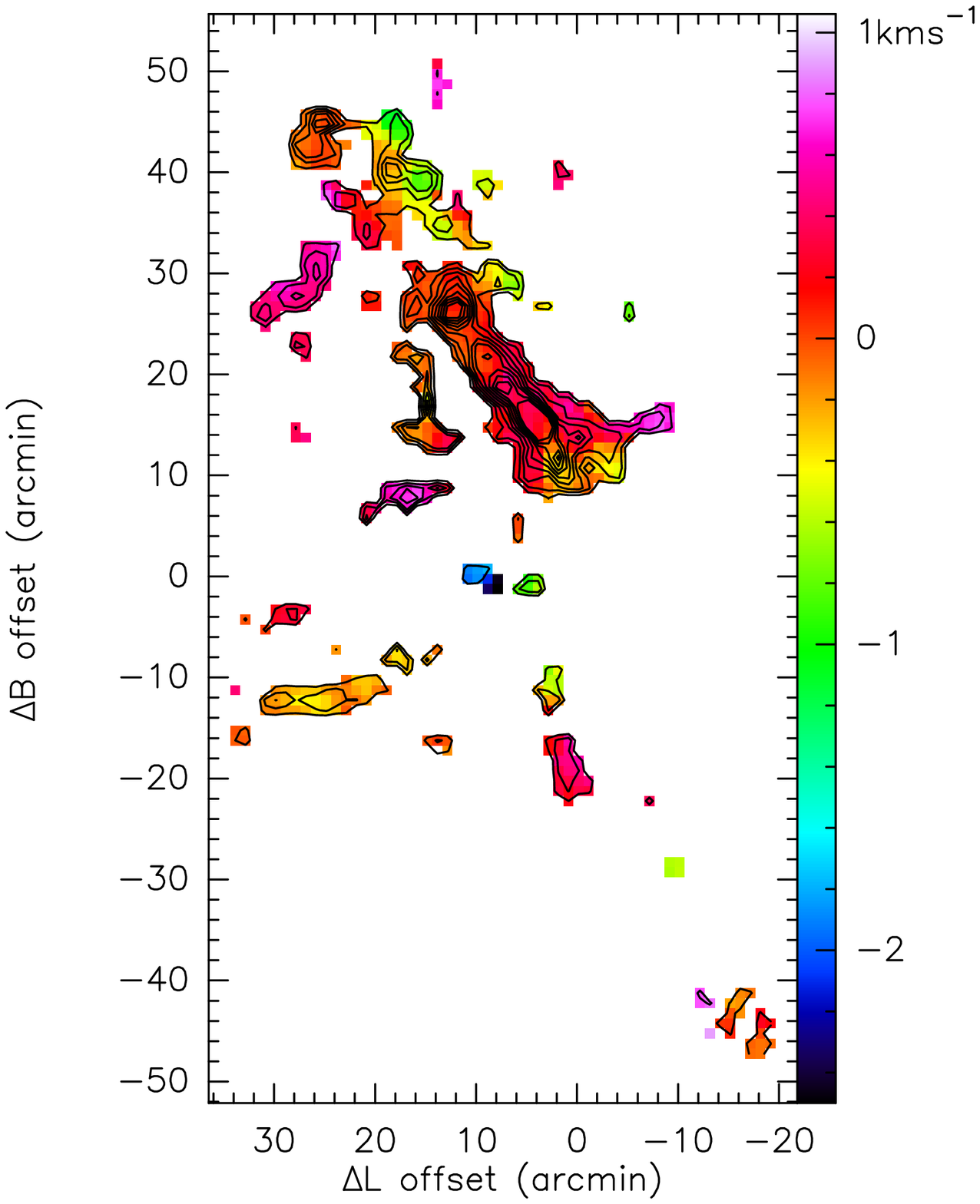}
\caption[]{Fitted velocity (\textit{left}) and velocity residual ($V_{\rm obs}$ -- $V_{\rm fit}$)\,(\textit{right}) maps of the NH$_3$\,(1,1) lines with signal-to-noise ratios $>$5$\sigma$. Contours are as in Figs.\,\ref{fg4} and \,\ref{fg6}. The black solid polyline in the left panel indicates a potential rotation axis, and three black parallel lines perpendicular to the rotation axis show the direction of the velocity gradient.}
\label{FgD.1}
\end{figure*}

\begin{figure*}[h]
\includegraphics[width=0.48\textwidth]{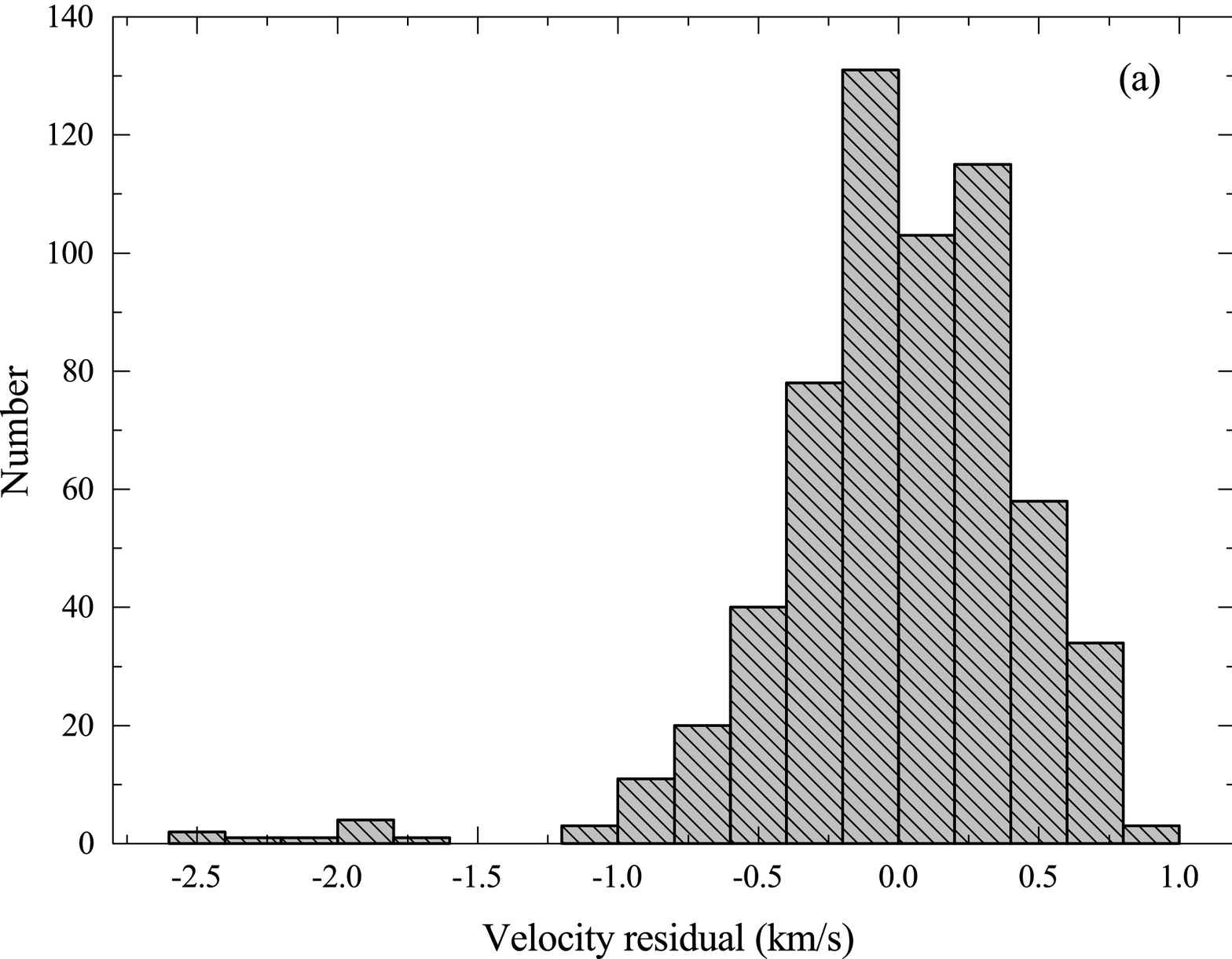}
\includegraphics[width=0.48\textwidth]{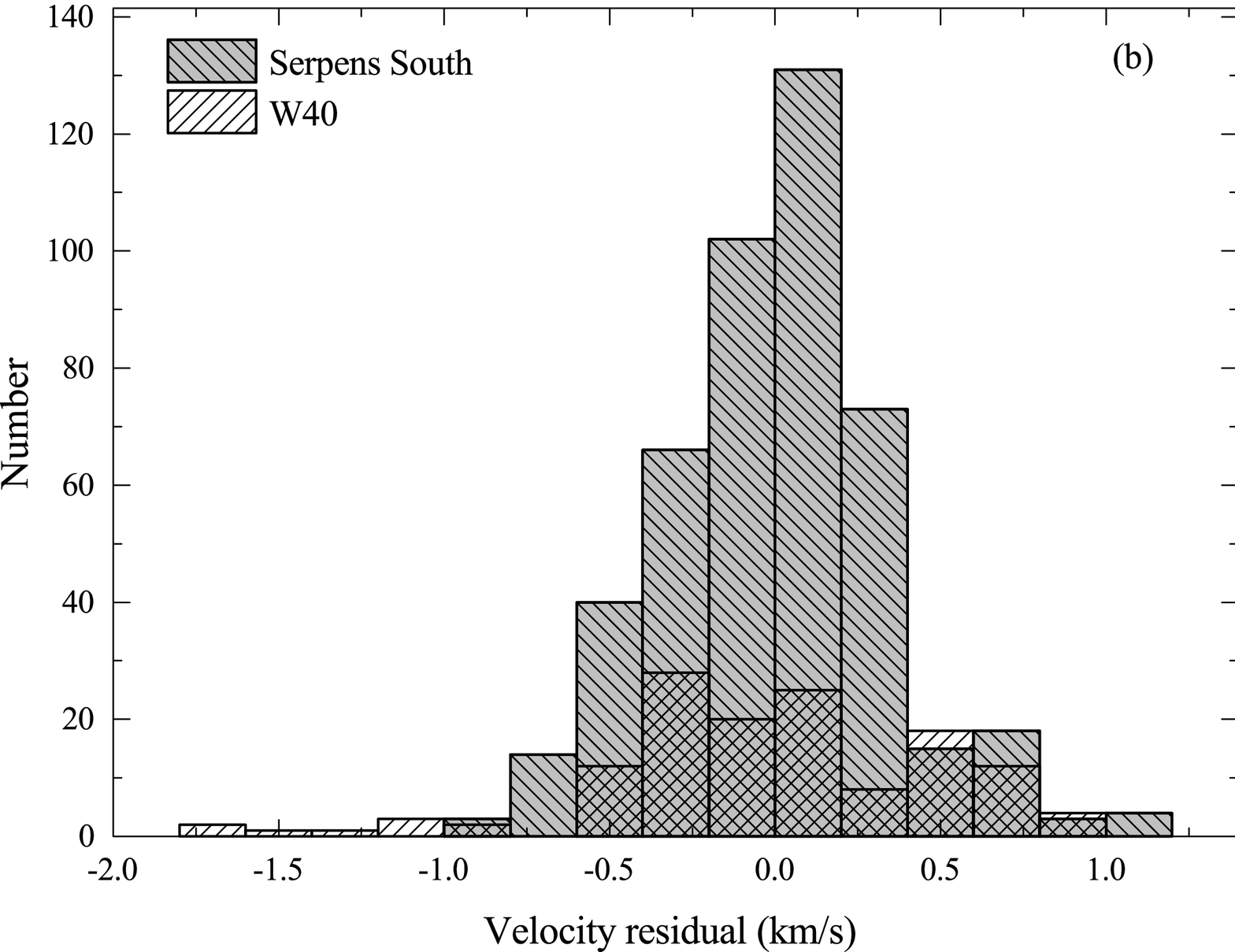}
\caption[]{Histograms of the velocity residuals ($V_{\rm obs}$ -- $V_{\rm fit}$) derived from our NH$_3$\,(1,1) data with signal-to-noise ratios $>$5$\sigma$: (a) entire observed region, (b) Serpens South and W\,40.}
\label{FgD.2}
\end{figure*}

We have assumed that the observed region shown in Fig.\,\ref{FgD.1} is a rigid body and fitted its
velocity field in a linear form. To check this assumption, we present the distribution of the fitted
velocity (Fig.\,\ref{FgD.1} left panel) and velocity residuals (Fig.\,\ref{FgD.1} right
panel) between the observed velocity and the fitted velocity. Statistics of the velocity residuals are
shown in Fig.\,\ref{FgD.2}. From the right panel of Fig.\,\ref{FgD.1}, we can see that
most of our velocity residuals are in the range of $V_{\rm obs}$ -- \textbf{$V_{\rm fit}$} = --1 to 1 km\,s$^{-1}$,
which can be even more clearly seen in Fig.\,\ref{FgD.2}. To be specific, $90\%$ of the velocity residuals
of the shown region, Serpens South alone and W\,40 alone are in the range $V_{\rm obs} - V_{\rm fit}$\,<\,
0.65, 0.56, and 0.79 km\,s$^{-1}$, respectively. Larger velocity residuals are mainly
located, in Galactic coordinates, at the western and northeastern part of the Serpens South and the southern part of the W\,40.

\end{appendix}

\begin{thebibliography}{}
\bibitem[Andr{\'e} et al.(2010)]{2010A&A...518L.102A} Andr{\'e}, P., Men'shchikov, A., Bontemps, S., et al. 2010, \aap, 518, L102
\bibitem[Barranco \& Goodman(1998)]{1998ApJ...504..207B} Barranco, J.~A., \& Goodman, A.~A.\ 1998, \apj, 504, 207
\bibitem[Batrla \& Wilson(2003)]{2003A&A...408..231B} Batrla, W., \& Wilson, T.~L.\ 2003, \aap, 408, 231
\bibitem[Battersby et al.(2014)]{2014ApJ...786..116B} Battersby, C., Bally, J., Dunham, M., et al.\ 2014, \apj, 786, 116
\bibitem[Benson \& Myers(1983)]{1983ApJ...270..589B} Benson, P.~J., \& Myers, P.~C.\ 1983, \apj, 270, 589
\bibitem[Bonnell et al.(1992)]{1992ApJ...400..579B} Bonnell, I., Arcoragi, J.-P., Martel, H., et al.\ 1992, \apj, 400, 579
\bibitem[Bontemps et al.(2010)]{2010A&A...518L..85B} Bontemps, S., Andr{\'e}, P., K{\"o}nyves, V., et al.\ 2010, \aap, 518, L85
\bibitem[Charnley et al.(1992)]{1992ApJ...399L..71C} Charnley, S.~B., Tielens, A.~G.~G.~M., \& Millar, T.~J.\ 1992, \apjl, 399, L71
\bibitem[Cheung et al.(1969)]{1969ApJ...157L..13C} Cheung, A.~C., Rank, D.~M., Townes, C.~H., et al.\ 1969, \apjl, 157, L13
\bibitem[Chira et al.(2013)]{2013A&A...552A..40C} Chira, R.-A., Beuther, H., Linz, H., et al.\ 2013, \aap, 552, A40
\bibitem[Danby et al.(1988)]{1988MNRAS.235..229D} Danby, G., Flower, D.~R., Valiron, P., et al. 1988, \mnras, 235, 229
\bibitem[Dewangan et al.(2016)]{2016ApJ...819...66D} Dewangan, L.~K., Ojha, D.~K., Luna, A., et al.\ 2016, \apj, 819, 66
\bibitem[Dunham et al.(2010)]{2010ApJ...717.1157D} Dunham, M.~K., Rosolowsky, E., Evans, N.~J., II, et al.\ 2010, \apj, 717, 1157
\bibitem[Dunham et al.(2011)]{2011ApJ...741..110D} Dunham, M.~K., Rosolowsky, E., Evans, N.~J., II, et al. 2011, \apj, 741, 110
\bibitem[Dobashi et al.(2005)]{2005PASJ...57S...1D} Dobashi, K., Uehara, H., Kandori, R., et al.\ 2005, \pasj, 57, S1
\bibitem[Evans(1999)]{1999ARA&A..37..311E} Evans, N.~J., II 1999, \araa, 37, 311
\bibitem[Faure et al.(2013)]{2013ApJ...770L...2F} Faure, A., Hily-Blant, P., Le Gal, R., et al.\ 2013, \apjl, 770, L2
\bibitem[Foster et al.(2009)]{2009ApJ...696..298F} Foster, J.~B., Rosolowsky, E.~W., Kauffmann, J., et al.\ 2009, \apj, 696, 298
\bibitem[Friesen et al.(2016)]{2016ApJ...833..204F} Friesen, R.~K., Bourke, T.~L., Di Francesco, J., Gutermuth, R., Myers, P.~C.\ 2016, \apj, 833, 204
\bibitem[Friesen et al.(2017)]{2017ApJ...843...63F} Friesen, R.~K., Pineda, J.~E., co-PIs, et al.\ 2017, \apj, 843, 63
\bibitem[Giannetti et al.(2013)]{2013A&A...556A..16G} Giannetti, A., Brand, J., S{\'a}nchez-Monge, {\'A}., et al.\ 2013, \aap, 556, A16
\bibitem[Goodman et al.(1993)]{1993ApJ...406..528G} Goodman, A.~A., Benson, P.~J., Fuller, G.~A., et al.\ 1993, \apj, 406, 528
\bibitem[Goldsmith(2001)]{2001ApJ...557..736G} Goldsmith, P.~F.\ 2001, \apj, 557, 736
\bibitem[Gutermuth et al.(2008)]{2008ApJ...673L.151G} Gutermuth, R.~A., Bourke, T.~L., Allen, L.~E., et al.\ 2008, \apjl, 673, L151
\bibitem[Ho \& Townes(1983)]{1983ARA&A..21..239H} Ho, P.~T.~P., \& Townes, C.~H.\ 1983, \araa, 21, 239
\bibitem[Hunter(2007)]{2007CSE.....9...90H} Hunter, J.~D.\ 2007, Computing in Science and Engineering, 9, 90
\bibitem[Jijina et al.(1999)]{1999ApJS..125..161J} Jijina, J., Myers, P.~C., \& Adams, F.~C.\ 1999, \apjs, 125, 161
\bibitem[Komesh et al.(2019)]{2019ApJ...874..172K} Komesh, T., Esimbek, J., Baan, W., et al.\ 2019, \apj, 874, 172
\bibitem[K{\"o}nyves et al.(2010)]{2010A&A...518L.106K} K{\"o}nyves, V., Andr{\'e}, P., Men'shchikov, A., et al.\ 2010, \aap, 518, L106
\bibitem[K{\"o}nyves et al.(2015)]{2015A&A...584A..91K} K{\"o}nyves, V., Andr{\'e}, P., Men'shchikov, A., et al.\ 2015, \aap, 584, A91
\bibitem[Koumpia et al.(2015)]{2015A&A...580A..68K} Koumpia, E., Harvey, P.~M., Ossenkopf, V., et al.\ 2015, \aap, 580, A68
\bibitem[Kuhn et al.(2010)]{2010ApJ...725.2485K} Kuhn, M.~A., Getman, K.~V., Feigelson, E.~D., et al.\ 2010, \apj, 725, 2485
\bibitem[Lada \& Lada(2003)]{2003ARA&A..41...57L} Lada, C.~J., \& Lada, E.~A.\ 2003, \araa, 41, 57
\bibitem[Lada et al.(2003)]{2003ApJ...586..286L} Lada, C.~J., Bergin, E.~A., Alves, J.~F., \& Huard, T.~L.\ 2003, \apj, 586, 286
\bibitem[Larsson et al.(2003)]{2003A&A...402L..69L} Larsson, B., Liseau, R., Bergman, P., et al.\ 2003, \aap, 402, L69
\bibitem[Levshakov et al.(2013)]{2013A&A...553A..58L} Levshakov, S.~A., Henkel, C., Reimers, D., et al.\ 2013, \aap, 553, A58
\bibitem[Levshakov et al.(2014)]{2014A&A...567A..78L} Levshakov, S.~A., Henkel, C., Reimers, D., \& Wang, M.\ 2014, \aap, 567, A78
\bibitem[Lu et al.(2014)]{2014ApJ...790...84L} Lu, X., Zhang, Q., Liu, H.~B., Wang, J., \& Gu, Q.\ 2014, \apj, 790, 84
\bibitem[Mallick et al.(2013)]{2013ApJ...779..113M} Mallick, K.~K., Kumar, M.~S.~N., Ojha, D.~K., et al.\ 2013, \apj, 779, 113
\bibitem[Mauersberger et al.(1986)]{1986A&A...162..199M} Mauersberger, R., Henkel, C., Wilson, T.~L., \& Walmsley, C.~M.\ 1986, \aap, 162, 199
\bibitem[Mauersberger et al.(1987)]{1987A&A...173..352M} Mauersberger, R., Henkel, C., \& Wilson, T.~L.\ 1987, \aap, 173, 352
\bibitem[Merello et al.(2019)]{2019MNRAS.483.5355M} Merello, M.; Molinari, S.; Rygl, K. L. J. et al.\  2019, \mnras, 483, 5355
\bibitem[Molinari et al.(1996)]{1996A&A...308..573M} Molinari, S., Brand, J., Cesaroni, R., \& Palla, F.\ 1996, \aap, 308, 573
\bibitem[Myers \& Benson(1983)]{1983ApJ...266..309M} Myers, P.~C., \& Benson, P.~J.\ 1983, \apj, 266, 309
\bibitem[Myers et al.(1986)]{1986ApJ...301..398M} Myers, P.~C., Dame, T.~M., Thaddeus, P., et al.\ 1986, \apj, 301, 398
\bibitem[Nakamura et al.(2017)]{2017ApJ...837..154N} Nakamura, F., Dobashi, K., Shimoikura, T., Tanaka, T., \& Onishi, T.\  2017, \apj, 837, 154
\bibitem[Ortiz-Le{\'o}n et al.(2017)]{2017ApJ...834..143O} Ortiz-Le{\'o}n, G.~N., Dzib, S.~A., Kounkel, M.~A., et al.\ 2017, \apj, 834, 143
\bibitem[Ortiz-Le{\'o}n et al.(2018)]{2018ApJ...869L..33O} Ortiz-Le{\'o}n, G.~N., Loinard, L., Dzib, S.~A., et al.\ 2018, \apjl, 869, L33
\bibitem[Ott et al.(2010)]{2010ApJ...710..105O} Ott, J., Henkel, C., Staveley-Smith, L., et al.\ 2010, \apj, 710, 105
\bibitem[Pandian et al.(2012)]{2012ApJ...753...50P} Pandian, J.~D., Wyrowski, F., \& Menten, K.~M.\ 2012, \apj, 753, 50
\bibitem[Persson et al.(2012)]{2012A&A...543A.145P} Persson, C.~M., De Luca, M., Mookerjea, B., et al.\ 2012, \aap, 543, A145
\bibitem[Rodr{\'{\i}}guez et al.(2010)]{2010AJ....140..968R} Rodr{\'{\i}}guez, L.~F., Rodney, S.~A., \& Reipurth, B.\ 2010, \aj, 140, 968
\bibitem[Rohlfs \& Wilson(2004)]{2004tra..book.....R} Rohlfs, K., \& Wilson, T.~L.\ 2004, Tools of radio astronomy, 4th rev.~and enl.~ed., by K.~Rohlfs and T.L.~Wilson.~ Berlin: Springer, 2004,
\bibitem[Rumble et al.(2016)]{2016MNRAS.460.4150R} Rumble, D., Hatchell, J., Pattle, K., et al.\ 2016, \mnras, 460, 4150
\bibitem[Schreyer et al.(1996)]{1996A&A...306..267S} Schreyer, K., Henning, T., Koempe, C., \& Harjunpaeae, P.\ 1996, \aap, 306, 267
\bibitem[Shirley(2015)]{2015PASP..127..299S} Shirley, Y.~L.\ 2015, \pasp, 127, 299
\bibitem[Smith et al.(1985)]{1985ApJ...291..571S} Smith, J., Bentley, A., Castelaz, M., et al.\ 1985, \apj, 291, 571
\bibitem[Sokolov et al.(2017)]{2017A&A...606A.133S} Sokolov, V., Wang, K., Pineda, J.~E., et al.\ 2017, \aap, 606, A133
\bibitem[Su et al.(2019)]{2019ApJS..240....9S} Su, Y., Yang, J., Zhang, S., et al.\ 2019, \apjs, 240, 9
\bibitem[Su et al.(2020)]{2020ApJ...893...91S} Su, Y., Yang, J., Yan, Q.-Z., et al.\ 2020, \apj, 893, 91
\bibitem[Suzuki et al.(1992)]{1992ApJ...392..551S} Suzuki, H., Yamamoto, S., Ohishi, M., et al.\ 1992, \apj, 392, 551
\bibitem[Tafalla et al.(2004)]{2004A&A...416..191T} Tafalla, M., Myers, P.~C., Caselli, P., \& Walmsley, C.~M.\ 2004, \aap, 416, 191
\bibitem[Tafalla et al.(2006)]{2006A&A...455..577T} Tafalla, M., Santiago-Garc{\'{\i}}a, J., Myers, P.~C., et al.\ 2006, \aap, 455, 577
\bibitem[Takano et al.(2002)]{2002PASJ...54..195T} Takano, S., Nakai, N., \& Kawaguchi, K.\ 2002, \pasj, 54, 195
\bibitem[Tan et al.(2014)]{2014prpl.conf..149T} Tan, J.~C., Beltr{\'a}n, M.~T., Caselli, P., et al.\ 2014, Protostars and Planets VI, 149
\bibitem[Tang et al.(2017)]{2017A&A...598A..30T} Tang, X.D., Henkel, C., Menten, K.~M., et al.\ 2017, \aap, 598, 30
\bibitem[Tang et al.(2018a)]{2018A&A...609A..16T} Tang, X. D., Henkel, C., Menten, K.~M., et al.\ 2018, \aap, 609, 16
\bibitem[Tang et al.(2018b)]{2018A&A...611A...6T} Tang, X. D., Henkel, C., Wyrowski, F., et al. 2018, \aap, 611, 6
\bibitem[Ungerechts et al.(1986)]{1986A&A...157..207U} Ungerechts, H., Walmsley, C.~M., \& Winnewisser, G.\ 1986, \aap, 157, 207
\bibitem[Urquhart et al.(2015)]{2015MNRAS.452.4029U} Urquhart, J.~S., Figura, C.~C., Moore, T.~J.~T., et al.\ 2015, \mnras, 452, 4029
\bibitem[Urquhart et al.(2011)]{2011MNRAS.418.1689U} Urquhart, J.~S., Morgan, L.~K., Figura, C.~C., et al.\ 2011, \mnras, 418, 1689
\bibitem[Vallee(1987)]{1987A&A...178..237V} Vallee, J.~P.\ 1987, \aap, 178, 237
\bibitem[Walmsley \& Ungerechts(1983)]{1983A&A...122..164W} Walmsley, C.~M., \& Ungerechts, H.\ 1983, \aap, 122, 164
\bibitem[Wei{\ss} et al.(2001)]{2001ApJ...554L.143W} Wei{\ss}, A., Neininger, N., Henkel, C., Stutzki, J., Klein, U.\ 2001, \apjl, 554, L143
\bibitem[Wienen et al.(2012)]{2012A&A...544A.146W} Wienen, M., Wyrowski, F., Schuller, F., et al.\ 2012, \aap, 544, A146
\bibitem[Williams et al.(1994)]{1994ApJ...428..693W} Williams, J.~P., de Geus, E.~J., \& Blitz, L.\ 1994, \apj, 428, 693
\bibitem[Wiseman \& Ho(1998)]{1998ApJ...502..676W} Wiseman, J.~J., \& Ho, P.~T.~P.\ 1998, \apj, 502, 676
\bibitem[Wouterloot et al.(1988)]{1988A&A...203..367W} Wouterloot, J.~G.~A., Walmsley, C.~M., \& Henkel, C.\ 1988, \aap, 203, 367
\bibitem[Wu et al.(2018)]{2018A&A...616A.111W} Wu, G., Qiu, K., Esimbek, J., et al.\ 2018, \aap, 616, A111
\bibitem[Wu et al.(2006)]{2006A&A...450..607W} Wu, Y., Zhang, Q., Yu, W., et al.\ 2006, \aap, 450, 607
\bibitem[Young et al.(2004)]{2004ApJ...614..252Y} Young, K.~E., Lee, J.-E., Evans, N.~J., II, et al. 2004, \apj, 614, 252
\bibitem[Zeilik \& Lada(1978)]{1978ApJ...222..896Z} Zeilik, M., II, \& Lada, C.~J.\ 1978, \apj, 222, 896
\bibitem[Zinnecker \& Yorke(2007)]{2007ARA&A..45..481Z} Zinnecker, H., \& Yorke, H.~W.\ 2007, \araa, 45, 481
\end{thebibliography}
\end{document}